\documentclass[12pt,preprint]{aastex}

\shorttitle{Complex chemistry in star-forming regions}
\shortauthors{Garrod, Widicus Weaver \& Herbst}

\begin{document}

\title{Complex Chemistry in Star-Forming Regions: An Expanded Gas-Grain Warm-up Chemical Model}

\author{Robin T. Garrod}
\affil{Max-Planck-Institut f{\"u}r Radioastronomie, Auf dem H{\"u}gel 69, Bonn, 53121, Germany}
\email{rgarrod@mpifr-bonn.mpg.de}

\author{Susanna L. Widicus Weaver}
\affil{Departments of Chemistry and Astronomy, University of Illinois at Urbana-Champaign, Urbana, IL 61801}
\email{slww@uiuc.edu}
\and

\author{Eric Herbst}
\affil{Departments of Physics, Chemistry and Astronomy, The Ohio State University, Columbus, OH 43210}

\begin{abstract}

Gas-phase processes were long thought to be the key formation mechanisms for complex organic molecules in star-forming regions. However, recent experimental and theoretical evidence has cast doubt on the efficiency of such processes. Grain-surface chemistry is frequently invoked as a solution, but until now there have been no quantitative models taking into account both the high degree of chemical complexity and the evolving physical conditions of star-forming regions. Here, we introduce a new gas-grain chemical network, wherein a wide array of complex species may be formed by reactions involving radicals. The radicals we consider (H, OH, CO, HCO, CH$_3$, CH$_3$O, CH$_2$OH, NH and NH$_2$) are produced primarily by cosmic ray-induced photodissociation of the granular ices formed during the colder, earlier stages of evolution. The gradual warm-up of the hot core is crucial to the formation of complex molecules, allowing the more strongly-bound radicals to become mobile on grain surfaces. This type of chemistry is capable of reproducing the high degree of complexity seen in Sgr B2(N), and can explain the observed abundances and temperatures of a variety of previously detected complex organic molecules, including structural isomers. Many other complex species are predicted by this model, and several of these species may be detectable in hot cores. Differences in the chemistry of high- and low-mass star-formation are also addressed; greater chemical complexity is expected where evolution timescales are longer.
\end{abstract}
\keywords{Astrochemistry, stars: formation, ISM: abundances, ISM: clouds, ISM: molecules, ISM: individual (Sagittarius
B2(N))}

\section{Introduction}
More than 140 molecules have been detected in the interstellar medium
(ISM) or in circumstellar environments. Many of these species have been
identified from their rotational emission spectra in hot cores --
molecular regions that are spatially associated with high-mass star
formation. Hot cores and their low-mass analogs, hot corinos, exhibit
high temperatures ($>$$100$ K) and densities
(10$^6$ -- 10$^8$ cm$^{-3}$), and are characterized by significant
abundances of large, complex organic molecules including methanol
(CH$_3$OH), formaldehyde (H$_2$CO), formic acid (HCOOH), methyl formate
(HCOOCH$_3$), and dimethyl ether (CH$_3$OCH$_3$)
\cite[]{hotcore,Bottinelli}. The Galactic-Center hot-core source
Sgr B2(N-LMH) exhibits the richest molecular inventory observed to date
\cite[e.g.][]{Nummelin}, showing spectral signatures of many additional
complex organics, including acetone \cite[CH$_3$COCH$_3$,][]{Snyder2002},
ethylene glycol [(CH$_2$OH)$_2$, Hollis et al. 2002],
and glycolaldehyde [CH$_2$(OH)CHO, Hollis et al. 2004]. The high degree
of chemical complexity of these molecules, and the structural
relationships between them, make their precise origins and formation
mechanisms the subject of debate.

In molecular clouds, cold dust grains build up molecular ice mantles
by the accretion and grain-surface hydrogenation of atoms and simple
molecules from the gas phase. In regions of star formation, the
gas and dust are heated to temperatures sufficient for these ices to
evaporate. Infrared observations indicate the presence of icy grain
mantles in the cold outer envelopes of protostellar objects
\cite[]{Gibb2004}; these mantles are typically comprised mainly of H$_2$O, CO
and CO$_2$. Methanol, methane, formaldehyde, and ammonia are also
observed in varying quantities, while the abundances of other, trace
species are less well determined.

The evaporation of granular ices in the hot, dense regions fuels a rich network of gas-phase chemistry, to which the
formation of many hot-core molecules has been attributed \cite[e.g.][]{Millar}. Ion--molecule reactions, especially
those involving evaporated formaldehyde and methanol, and the subsequent electronic recombination of protonated ions,
are viewed as potential routes to forming complex hot-core species \cite[]{Herbst,Charnley95}. However, recent
investigations call into doubt the efficiency of such gas-phase routes. The experimental results of \cite{Geppert06}
suggest that formation of saturated complex molecules by electronic recombination of their protonated precursors is
significantly less efficient than previously assumed; methanol and dimethyl ether are formed with respective
efficiencies of $\sim$3\% and $\sim$5\% (W. Geppert, private comm.), implying similar inefficiency for other complex
species. In addition, the quantum calculations of \cite{Horn} show that the ionic precursor of methyl formate cannot be
formed in the gas phase at temperatures appropriate for hot cores. Grain-surface mechanisms may therefore be important
to the formation of methyl formate, and perhaps many other observed hot-core molecules.  The detection of complex
molecules in hot corinos also points to grain-surface production, as gas-phase chemical timescales are greater than the
transit time of the infalling gas in these sources \cite[]{Schoier02,Bottinelli,Aikawa2007}.

Garrod \& Herbst (2006, hereafter GH06) used the OSU gas-grain chemical code to test grain-surface formation mechanisms
for methyl formate, as well as dimethyl ether and formic acid. GH06 employed three heavy-radical combination reactions
that were originally included in a large network developed by \cite{Allen} in the context of cold dark clouds. The
surface radicals OH, HCO, CH$_3$ and CH$_3$O were allowed to combine to produce the three target molecules.
\cite{HolChurch} pointed out the potential for complex-molecule formation from functional-group radicals such as these,
suggesting a cold chemistry in which reactive functional groups are built up from accreted atoms. However, GH06 found
that the complex molecules are formed predominantly at intermediate temperatures, when the grains become warm enough
for heavy radicals to become mobile on grain surfaces, during the gradual warm-up of the hot core from 10 K to
approximately 200 K. Further, the strongest source of heavy radicals was shown to be the cosmic ray-induced
photodissociation of the icy mantles formed in the earlier cold collapse phase. Hence, in this scenario, the complex
molecules are not themselves relics of the cold phase -- but the ice composition preserved from that phase of evolution
influences the chemistry by which they form at higher temperatures, as the hot core evolves.

The GH06 model demonstrated efficient means of forming complex molecules
on grain surfaces, and served to highlight the complexity of gas-grain
chemical interactions as temperatures evolve. It was shown that methyl
formate almost certainly has a grain-surface origin, and may be produced
in sufficient quantities to match observations, while dimethyl ether and
formic acid may equally be formed in the gas phase or on grain surfaces,
depending on the physical conditions and/or evolution timescale.

Here, we extend the reaction network to present a more complete study of
hot-core chemistry, incorporating all thermodynamically-viable surface
reactions between the radicals H, OH, HCO, CH$_3$, CH$_3$O, CH$_2$OH,
NH and NH$_2$, as well as reactions with the resultant products. The
radicals listed above may be produced by cosmic ray-induced
photodissociation of the most abundant ice-mantle constituents. The
majority of the new surface reactions form a subset of those suggested by
\cite{Allen}. We introduce 50 new grain surface-formed species, as well
as a number of gas-phase species related to these, and we include
gas-phase and grain-surface destruction mechanisms in keeping with
other species in the network. The individual treatment of the
structurally-isomeric radicals CH$_3$O and CH$_2$OH allows the
differentiation of such important hot-core species as
ethanol/dimethyl ether (C$_2$H$_5$OH, CH$_3$OCH$_3$), and
glycolaldehyde/methyl formate/acetic acid
[CH$_2$(OH)CHO, HCOOCH$_3$, CH$_3$COOH] -- an important point, given the
very different observational abundances of these molecules
\cite[]{Snyder2002}.

We employ the physical model adopted by GH06 (based on Viti et al. 2004),
which consists of an isothermal collapse, followed by a warm-up from
10 to 200 K. Warm-up timescales are chosen to approximate
low-, intermediate-, and high-mass star formation. We also adjust initial
solid-state abundances to explore the influence of ice composition on
observable hot-core species.  With such a large number of new species and
reactions, here we set up the basic network using plausible values for
chemical quantities and physical parameters, leaving the full parameter
space to be explored in future papers.

The chemical and physical model is presented in Section 2, and the results and discussion in Section 3. The influence of
granular ice-mantle composition is explored in Section 4. Conclusions are presented in Section 5. The Appendix gives
detailed chemical information regarding the formation and destruction mechanisms of all the new complex species
included in the model.

\section{Model}

\subsection{Chemical Network}

We have used as the basis for our extended chemistry the latest version of the Ohio State University gas-grain network \citep{Garrod07}, which is based on the {\em osu.2005} gas-phase network. To this set we have added 50 new neutral species, and a further 32 ionic species, along
with a large number of associated chemical reactions. In general, for every new grain surface-formed species included,
the following new reactions and mechanisms are added to the network: accretion; thermal evaporation; grain-surface
chemical reactions; gas-phase and grain-surface photodissociation, {\em via} both the cosmic ray-induced UV field
($\zeta=1.3 \times 10^{-17}$ s$^{-1}$) and the external interstellar radiation field (ISRF); and gas-phase
ion--molecule reactions. In most cases, ion--molecule reactions involving new complex molecules result in a protonated
complex ion. Dissociative recombination routes were added for those ions not already present in the network. A sticking coefficient of $S=0.5$ is assumed for all neutral species. The reactive desorption mechanism of \cite{Garrod07} is not included here; test runs show that its effects are generally small in this case. Evaporation due to cosmic ray-induced grain heating is included, but is generally weak, due to the ice-surface binding-energy values used here. The
complete reaction network includes approximately 7500 reactions, and is available online at {\em http://www.physics.ohio-state.edu/$\sim$eric/research.html}.

\subsubsection{New grain-surface chemistry}

We begin with the assumption that complex organic molecules may
be formed on grain surfaces from large radicals primarily
derived from the icy mantles. We consider reactions between the
radicals H, OH, CO, HCO, CH$_3$O, CH$_2$OH, CH$_3$, NH
and NH$_2$, all of which (except H and CO) are mainly derived from 
the photolysis of the major ice-mantle constituents H$_2$O, CH$_4$, H$_2$CO,
CH$_3$OH, or NH$_3$. We use the term `radical' loosely, either to mean an atom/molecule 
that has at least one unpaired electron, and which will therefore
readily react; or an unsaturated molecule, primarily CO -- whose multiple bonds may be broken 
according to some activation energy. Dependent on their state of 
hydrogenation, the radicals listed above may react together either to 
form other radicals, or stable (typically saturated) molecules. We designate as 
`primary radicals' those radicals (as listed above) that derive directly 
from photodissociation of ices or from accretion from the gas phase. 
We designate as `secondary radicals' those species with unpaired electrons 
that form by reactions between primary radicals. The primary and
secondary radicals may further react to produce stable (typically saturated)
species. In some cases, the reactions have activation energies; those
without barriers make up a subset of the reactions suggested by \cite{Allen}. 
We assume that reaction occurs only at the radical sites and that no
intramolecular rearrangement occurs upon reaction; hence the
structure of the component radicals is retained in the product. Many
structural isomers may thus be formed through the combination of
primary and secondary radicals.

Figures 1 \& 2 show all reactions between primary radicals, and between primary radicals and secondary radicals, a few
of which were already present in the network. The reaction CO + HCO $\rightarrow$ COCHO is excluded due to its
endothermicity \cite[based on the formation enthalpy of the COCHO radical as determined by][]{COCHOref}.
Primary-radical reactions with CO are assumed to have activation energies; the reactions H + CO $\rightarrow$ HCO and
OH + CO $\rightarrow$ CO$_2$ + H ($E_A$ = 2500 K, 80 K, respectively) are treated as in previous models \cite[]{Woon02,
Ruffle00, Ruffle2}. The CO + CO reaction is assumed to have a significantly higher barrier and is thus omitted.
Newly-added CO reactions are assigned an activation energy of 1500 K, based on values in the NIST Chemical Kinetics
Database (http://kinetics.nist.gov/kinetics/index.jsp). Some primary--secondary reactions involving NH-bearing species can result in unsaturated products; see Figure 2.
For simplicity, we allow these products to subsequently react only with hydrogen (not shown).

Diffusion energy barriers, $E_b$, for each radical are also noted in Figures 1 \& 2 to indicate the order in which
different radicals become mobile on the grain surface during the warm-up phase; see Section 2.1.2. The mobility of the
reacting species with the lower $E_b$ value dominates the reaction rate; however, the availability of the reactants is
also important, and varies in a more complex manner, according to both the gas-phase and grain-surface chemistry.

Besides the basic radical--radical addition reactions, we allow
reactions between primary radicals and species that contain an aldehyde 
functional group (-CHO), in cases where such reactions
are expected to be exothermic. Affected species include actual aldehydes like 
formaldehyde and acetaldehyde (CH$_3$CHO), but also formic acid (HCOOH), 
formamide (NH$_2$CHO), and the ester, methyl formate (HCOOCH$_3$). 
Figure 3 shows all of these reactions,
with their activation energies in K. These types of reactions have not
been considered in previous grain-surface models. {\it Ab initio}
studies show that the barriers to radical abstraction of an
aldehyde proton are much lower than the barriers to radical addition
to the aldehyde group \citep{Hippler}. Some complex organics observed
in hot cores may be formed from aldehyde group-bearing radicals produced in this way,
followed by further primary--secondary radical combination reactions, e.g.
\begin{eqnarray*}
 {\textrm{CH}_3\textrm{CHO} + \textrm{CH}_3} & \rightarrow & {\textrm{CH}_3\textrm{CO} + \textrm{CH}_4} \\
 {\textrm{CH}_3\textrm{CO} + \textrm{CH}_3}  & \rightarrow & {\textrm{CH}_3\textrm{COCH}_3}
\end{eqnarray*}
In some cases, instead of the abstraction of a hydrogen atom from
the aldehyde group, the substitution of the attacking radical for a
functional group on the saturated molecule may be possible, e.g.
\begin{eqnarray*}
 {\textrm{CH}_3\textrm{CHO} + \textrm{NH}_2} & \rightarrow & {\textrm{NH}_2\textrm{CH}_3 + \textrm{HCO}} \\
 {\textrm{CH}_3\textrm{CHO} + \textrm{NH}_2} & \rightarrow & {\textrm{NH}_2\textrm{CHO} + \textrm{CH}_3}
\end{eqnarray*}
Substitution reactions are assigned activation energies $E_{A}=2400$ K.
Abstraction reactions are assigned $E_{A}=2850$ K, with the
exception of OH reactions, for which we assume $E_{A}=1500$ K.
Estimates are based on similar reactions listed in the NIST online
Chemical Kinetics Database (http://kinetics.nist.gov/kinetics/index.jsp),
but measured values are used where available. Estimates are in
agreement with values determined in the {\it ab initio} studies of
\cite{Hippler}. The addition of hydrogen to formaldehyde, as to CO,
has a barrier of 2500 K. The two possible products, CH$_3$O and CH$_2$OH, form at equal rates.

It should be noted that those species already present in
previous gas-grain networks may be involved in other reactions
not listed in Figures 1 -- 3; we illustrate only those reactions
that involve exclusively the nine primary radicals we identify
above and their secondary-radical products.

\subsubsection{Surface-reaction rates and barriers}

Grain-surface reaction rates are treated in the same way as in
GH06. Surface-based species are assigned binding (desorption)
energies, $E_D$, and diffusion barrier energies, $E_b$. Reactions
occur {\em via} the Langmuir-Hinshelwood mechanism, i.e. reactants
migrate around the grain surface until they meet at a binding site.
Migration occurs by thermal hopping of reactants over the barrier
$E_b$ between sites; quantum-tunnelling effects are assumed to be
insignificant \cite[]{Katz}. Hence, the diffusion energy barriers
define the rates at which reactions take place. Activation-energy
barriers to reaction, $E_A$, may be overcome thermally (which introduces 
a simple Boltzmann factor to the reaction rate), or {\em via} quantum
tunneling, whichever is faster. Therefore, where $E_{A}/T$ is
`large', the activation barrier term loses its temperature
dependence. We refer the reader to \cite{Hasegawa} for a full
explanation of grain-surface reaction rates.

In this study, the modified rate treatment sometimes used for
hydrogen reaction rates \citep{Caselli98,Shal98} is discarded.
This treatment was employed to remedy inaccuracies when hydrogen
abundances fall to less than one atom per grain. While the method
has had moderate success in the low-temperature conditions of dark
clouds where hydrogen chemistry is dominant, its extension to the
wide range of temperatures used in this model is anomalous when so
many active surface radicals are considered. Without a clear and
well-tested method of extending the treatment to such a regime, we
cannot justify its inclusion. However, the inclusion of the modified
rates for hydrogen atoms alone has no significant effect on the
results of this model.

The grain surface is defined through $E_{D}$ and $E_{b}$, where
$E_{b}(i) = \frac{1}{2} E_{D}(i)$ for all species $i$. Values are
representative of an amorphous water ice surface. As in GH06, we
interpolate experimental binding energy values for certain key species
\cite[]{Collings} by simple addition or subtraction, to produce values
for all other species in our set. Therefore, to the measured CO value,
we add the H value to get HCO, and again to get H$_2$CO. Attention is
paid to species that possess an -OH functional group, as hydrogen
bonding to the water-ice surface may significantly increase their
binding energies \cite[]{Collings}. Hence, the structurally isomeric
radicals CH$_3$O and CH$_2$OH are treated differently:
\begin{eqnarray*}
E_D(\textrm{CH}_3\textrm{O})=E_D(\textrm{CO})+ 3 \times E_D(\textrm{H}) \\
E_D(\textrm{CH}_2\textrm{OH})=E_D(\textrm{CH}_3\textrm{OH})-E_D(\textrm{H})
\end{eqnarray*}
The measured value for methanol is elevated due to hydrogen bonding to the ice surface. Hence CH$_2$OH is much more
strongly bound than CH$_3$O, and so it requires greater dust temperatures to become mobile, or evaporate. This
distinction is critical, as the two species are chemically different since their radical sites are on different atoms.
Through a similar construction of binding energies, the distinction propagates through to their products. Thus, we may
investigate, for example, the difference between the formation of structural isomers methyl formate (HCOOCH$_3$) and
glycolaldehyde [CH$_2$(OH)CHO], and the difference in their own behavior due to their different binding energies.

The evaporation of water, the primary constituent of icy grain mantles,
should result in the co-desorption of any other species remaining in
the ices \cite[]{Collings}. In order to take account of this structural
aspect of mantle evaporation, we allow $E_D$ values to be no greater than
that of H$_2$O. Diffusion barriers are unaffected by this adjustment.

\subsubsection{New grain-surface and gas-phase photodissociation routes}

Excluding atomic hydrogen, which is mostly accreted from the gas
phase, the grain-surface primary radicals derive from the icy
grain mantles formed during the cold collapse phase; either
directly (CO), or through cosmic ray (CR) -induced photodissociation of
molecular ices. In addition, HCO and CH$_3$O/CH$_2$OH may be
formed {\em via} the hydrogenation of CO and H$_2$CO, respectively.

\cite{Gredel} calculated rates for the CR-induced
photodissociation of a number of astrochemically important species,
including those concerned here. However, the products remain uncertain.
We retain the treatment of Gredel et al. for H$_2$O, CH$_4$ and NH$_3$
to give H, OH, CH$_3$, NH, and NH$_2$. For H$_2$CO we include an
additional channel, producing HCO + H; see Table \ref{tab1}.

The CR-induced photodissociation of CH$_3$OH is especially
important to this model. Gredel et al. suggest two photodissociation
branches, CH$_3$ + OH and H$_2$CO + H$_2$. However, several
potential channels have been shown through both
theory and experiment to be important in methanol dissociation
mechanisms, including those producing the radicals CH$_2$OH and CH$_3$O
\cite[see][and references therein]{ChangI,ChangII}.
The adoption of the Gredel et al. channels is convenient for
purely gas-phase models, which do not typically include CH$_3$O/CH$_2$OH
radicals. However, our model includes both radicals, because of their importance
to complex surface chemistry. Given that the intramolecular rearrangement
involved in the H$_2$CO + H$_2$ channel of methanol dissociation is rather
unlikely in an ice matrix, we have replaced this channel with
CH$_3$O + H and CH$_2$OH + H. The chosen rates are slightly lower than
those for the CH$_3$ + OH channel, in agreement with the H-atom
ejection channels of other molecules. The two structural isomers are
assumed to be formed at the same rate.
These branching ratios may strongly influence structural isomerism in
the surface-formed complex molecules; we will investigate these effects
in depth in a future study.

All of the new complex species added to this network are assigned
CR-induced photodissociation channels. In the absence of experimental
data for many of these species, assumptions must be made about the
products and rates. Molecules are assumed to dissociate primarily into
their constituent functional groups, which is facilitated by our explicit
treatment of a large number of radical species. We avoid channels that
result in a great degree of structural rearrangement, which would
be unlikely in an interstellar environment. Where not otherwise available,
representative rates are selected from the existing ratefile, according to
the pair of atoms or groups whose bond is broken. This is a simplistic
approximation that ignores the specific mechanics of photodissociation in
a particular molecule, which are necessarily dependent on its UV absorption
spectrum, the spectrum of the CR-induced radiation field, the partition of
energy within the molecule, and its specific quantum state. However, this
method does take indirect account of the energy required to break the bonds,
and allows for the most accurate estimation of dissociation rates possible
given the incomplete laboratory information available. 

All photodissociation
channels are applied to gas-phase and grain-surface species with the same
rates and products, as in previous models. The accuracy of this approach is 
difficult to ascertain, as there is currently very little data on relative rates 
in the gas phase and on grain surfaces. \cite{Kroes}, and references therein, show that 
the first absorption band of crystalline water ice is centered at an energy 
$\sim$15\% higher than the gas-phase band, indicating that the dissociation rates
may be different for this molecule.

Photodissociation caused by the ISRF is also considered, using the same
approach. However, this mechanism is relatively unimportant, as visual
extinctions are typically high, except during the earliest stages of
the isothermal collapse phase, at which time the abundances of complex
molecules are negligible.

\subsubsection{New gas-phase ion--molecule reactions, dissociative recombination}

Gas-phase destruction routes involving the dominant ions He$^+$, C$^+$, H$_3^+$, HCO$^+$, and H$_3$O$^+$ have been
added for each of the new species introduced through grain-surface chemistry. Reactions of new species with C$^+$ ions is assumed typically to result in charge transfer, leaving complex molecular structure intact, whilst reaction with He$^+$ results in fragmentation. Reaction with molecular ions leads in most
cases to protonation. The complex ionic species which result from the new reactions have been added to the network,
where not otherwise present.

Rate coefficients are calculated using experimental (or otherwise, computed) dipole moments or polarizability data,
available from the NIST Computational Chemistry Comparison and Benchmark Database \citep{NIST_CCCBDB}. In the case of
linear, non-polar molecules, simple Langevin rates are used, while the method of \cite{Herbst86} is adopted for all
other molecules. Where no calculated or experimental values are available, dipole moments are estimated by comparison
to structurally similar molecules. Since the Herbst \& Leung method produces a linear dependence on dipole moment, we
should expect the calculated rates to be as accurate as our estimate of this quantity. Measured dipole moments for
complex molecules are typically in the range of 1 -- 5 debye, so ion--molecule reaction rates are unlikely to be
inaccurate by more than a factor of a few.

The introduction of new ionic species requires new dissociative recombination reactions to be introduced. The branching
ratios of such reactions have recently come under particular scrutiny. \cite{Geppert06} have shown that the CH$_3$OH +
H channel accounts for less than 5\% of protonated methanol recombinations; the strongest channels represent three-body
break-ups. For protonated complex molecules, we assume that two-fragment channels represent 5\% each, and that the
remainder is evenly split between channels with three (or more, where applicable) fragments. In analogy with the new
photodissociation reactions, we allow fragmentation primarily between functional groups within the molecule (including
the molecule--proton bond). Other channels are allowed if they result in stable (and especially, saturated) products,
with minimal structural re-arrangement of the molecule. Total rate coefficients of $k = 3 \times 10^{-7}$ cm$^{3}$
s$^{-1}$ are assigned, in keeping with other large-molecule values.

\subsection{Physical model and initial conditions}

We adopt the same two-phase physical model as GH06. In the collapse phase, the nascent hot core undergoes isothermal
collapse at 10 K, from a density of 3$\times10^3$ to 10$^7$ cm$^{-3}$. The collapse phase begins at a visual extinction
of 1 magnitude, growing to over 200. We include the H$_2$ and CO self-shielding functions of \cite{Lee}, to ensure a
more reliable treatment of hydrogen and CO in the initially diffuse physical conditions of the collapse phase. For the
purposes of self-shielding, we designate an H$_2$ abundance of 1/3 and a CO abundance of $10^{-5}$ to the outer
envelope, and assume a total hydrogen column of $1.6 \times 10^{21}$ cm$^{-2}$. As the central visual extinction grows
throughout the collapse, we allow the H$_2$ and CO column densities to increase according to
\begin{eqnarray*}
N(\textrm{i}) = N_{\textrm{init}}(\textrm{i}) + X(\textrm{i}) \times (1.6 \times 10^{21}) \times (A_{V}-1)
\end{eqnarray*}
\noindent where $X($i$)$ is the fractional abundance with respect
to total hydrogen as computed in the code. Initial abundances are
the so-called low-metal abundances of \cite{Graedel} except for
species He, C$^+$, N, and O, whose values are selected from the most
recent diffuse cloud values (Wakelam \& Herbst, unpublished), see
Table \ref{tab-init}. The behavior of sulfur and the other heavy
elements included in the model on grain surfaces is not well
understood.  Indeed, the form and location of sulfur in dense
regions is an unanswered question in astrochemistry; we do not
attempt to answer it in this paper, and we adopt the canonical (low)
dense cloud abundances for atoms heavier than oxygen.

In the warm-up phase, the collapse is halted, and the temperature grows from 10 to 200 K, over three timescales:
$5\times10^{4}$, $2\times10^{5}$ and $1\times10^{6}$ years, corresponding to models F (fast), M (medium), and S (slow),
respectively; see Table \ref{tab-mod}. Following Viti et al. (2004), GH06 identified these timescales with high-,
intermediate- and low-mass star formation, respectively. However, \cite{Aikawa2007} argue that the warm-up timescale is
dependent on the ratio of the size of the warm region to the infall speed, rather than on the overall speed of star
formation. This would suggest the contrary relation between mass and warm-up timescale. This point is discussed in
relation to chemical abundances in later Sections. We concentrate on the $T_2$ temperature profile adopted by
\cite{Garrod06}, where $T(t) = 10 + k \cdot t^2$, measured in Kelvin, at time $t$. Gas and dust temperatures are
assumed to be well coupled, due to the high densities and visual extinctions, so we assume $T_{K}=T_{grain}$ at all
times. 

While we use a single-point model, the resulting time-dependent data may also be interpreted as representing a range of distances from the hot-core center, with the innermost parts being the most evolved and achieving the highest temperatures. As such, we may understand the chemistry of the colder, more extended regions of a hot core using this single model.

\section{Results and Discussion}

The results of models F, M, and S are presented, corresponding to different warm-up timescales. Discussion of the
chemistry is restricted here to general trends, while specific, key aspects are discussed further in following
Sections. A more detailed discussion of the chemical behavior of individual species or classes of species is given in
the Appendix.

Figures 4 -- 6 show abundances with respect to time and temperature for models F (fast warm-up), M (medium), and S
(slow), all of which undergo the same collapse-phase evolution. Each Figure panel shows a subset of the molecular
species included in the model, beginning with simple ice-constituents (panel a). Solid lines indicate gas-phase
abundances; dotted lines of the same shade/color indicate the grain-surface abundance of that species. Included in
these plots are many of the most commonly observed species, as well as a suite of unobserved complex organics predicted
to form {\em via} the new grain-surface chemical network.

Table 3 shows the peak gas-phase abundances of species plotted in Figures 4 -- 6, for each evolution timescale, as well
as the temperatures at which those peak values are achieved.

The results demonstrate that the consideration of a surface chemistry involving a small set of radicals derived from
dominant grain-surface ice components produces a wealth of information on complex organic molecules -- many of which
have not yet been detected in interstellar space. The grain-surface production of species investigated by GH06 is
maintained and their structural isomers are also formed, along with yet more structurally-complex molecules.

The composition of the ice is crucial to the chemistry both on
grains and in the gas phase, as methanol and formaldehyde provide
most of the basic radicals that produce organic species. The
grain-surface abundances of these species also directly determine
their peak gas-phase abundances upon evaporation; surface
chemistry is not capable of strongly diminishing their surface
abundances. These species' major destruction route on the grains is
photodissociation by the CR-induced UV field, producing reactive radicals.

The injection of H$_2$CO into the gas phase strongly influences
the chemistry at intermediate temperatures. It provides the material
for the formation of other species, but also acts as the dominant reaction partner
for gas-phase ions, facilitating the survival of the other newly-formed
species in the gas phase for significant periods. For model F, this includes
complex molecules like methyl formate and dimethyl ether.

In general, those species formed on grains during the warm-up
phase benefit from longer evolution timescales, as more time is
spent at their optimal temperatures of formation. Later, following the
main period of formation, abundances of surface species are
somewhat attenuated by CR-induced photodissociation, especially
over the long timescale of model S. Additionally, the longer periods
at high temperature seen in models M and S allow hydrogen-abstraction
reactions to destroy grain-surface aldehyde group-bearing species.

Long warm-up timescales also limit the long-term survival of molecules that evaporate in advance of the remainder of
the ices. When the major ice constituents H$_2$O, NH$_3$ and CH$_3$OH evaporate, they become the dominant reaction
partners for gas-phase ions (similar to H$_2$CO at earlier times), damping ionic abundances and thus limiting the
destruction of other species. Complex molecules that spend long periods in the gas phase before this may be strongly
diminished.

\subsection{Structural isomers}

HCOOCH$_3$, CH$_2$(OH)CHO and CH$_3$COOH are structural isomers that have been detected in several star-forming regions
\cite[see][and references therein]{Snyder2002}. However, their relative abundances are puzzling, particularly in light
of the failure of gas-phase chemistry to account for methyl formate \cite[]{Horn}. Their relative abundances in the hot
core Sgr B2(N-LMH) are, respectively, 52:1:2 \cite[]{Snyder2002}. It is arguable that if all three were formed by
similar processes on grain surfaces, for example by single-atom addition reactions, then their observed abundances
should be similar, assuming comparable destruction mechanisms.

In fact, the structural differences in these species can be
explained by different combinations of the primary radicals
considered here. In this model, methyl formate and glycolaldehyde
have similar formation routes based on addition of HCO to
CH$_3$O or CH$_2$OH, whose production rates are the same. Hence, the
resultant molecules are similarly abundant. However, the lower abundance
of acetic acid relative to methyl formate is reproduced in our model,
especially with shorter warm-up timescales, because it has a very
different formation route. The secondary radical CH$_3$CO combines with
OH, but the CH$_3$CO is derived not from direct addition of CH$_3$ and
CO, but from photodissociation of acetamide, or hydrogen abstraction from
acetaldehyde, both of which are formed earlier. These routes would suggest
a correlation between the abundance of acetamide and/or acetaldehyde, and
that of acetic acid. The relative abundances of methyl formate and
glycolaldehyde may be influenced by disparities in the CH$_3$O/CH$_2$OH
branching ratios resulting from methanol photolysis and/or formaldehyde
hydrogenation. However, it is clear that this type of grain-surface
chemistry can address the observed relative abundances of structural
isomers.

This model also permits the study of the structural isomers dimethyl ether and ethanol. They are formed by addition of
the CH$_3$ radical to either CH$_3$O or CH$_2$OH, which primarily occurs when CH$_3$ becomes mobile, around 30 -- 40 K.
As above, the equal branching of CH$_3$O/CH$_2$OH production routes results in grain-surface formation of these species
in very similar quantities. However, dimethyl ether desorbs strongly around 70 K, while the -OH group of C$_2$H$_5$OH
ensures that ethanol co-desorbs with water at high temperatures. While longer warm-up timescales can result in very
large quantities of dimethyl ether forming on grains, its early evaporation results in fast gas-phase destruction (see
Section 3, above). This means that gas-phase production of this species (following methanol evaporation) is always
dominant over grain-surface formation, a result in accordance with the analysis of \cite{Peeters}. Therefore, dimethyl
ether should not be a good indicator of grain-surface chemistry {\em per se}. The substantial quantities of ethanol
observed in star-forming regions may be assumed to be such an indicator, and indeed suggests that CH$_3$O is
well-supplied on grain surfaces. Calculated abundances for dimethyl ether and ethanol are correspond with typical
observed values.

\subsection{Rotational temperatures}

One of the most striking results of this work is that the peak temperatures of many hot-core molecules (Table 3) show
agreement with observed rotational temperatures.  In particular, the low-temperature molecules identified by
\cite{Bisschop}, CH$_2$CO, CH$_3$CHO, and H$_2$CO, display excellent agreement. In the model, these species evaporate
from the grains at low temperatures, as determined by their binding energies to water ice. For long warm-up timescales,
high gas-phase abundances are sharply defined in a narrow, low-temperature range.  For short warm-up timescales, high
abundances are maintained to high temperatures.

\cite{Bisschop} also identify HCOOH as a `cold' molecule, and,
ignoring the late-time peak caused by grain-surface evaporation,
HCOOH does indeed peak at low temperatures in our model. This
peak is caused directly by reaction of evaporated formaldehyde
with OH molecules in the gas phase to form formic acid.
\cite{Nummelin} observe a low rotational temperature for HCOOH
in Sgr B2(N), at $T_{rot}$=74 K, albeit with sizeable error
margins. Such evidence may indicate that HCOOH is strongly
{\em destroyed} on grains at late times, before evaporation
takes place. It would not be sufficient to suggest that it is merely
{\em not formed} on grains, as much of the formic acid present
on grains at late times has its origin in the accretion of the
gas-phase species formed from formaldehyde at $\sim$40 K. This
may indeed suggest that hydrogen abstraction processes acting on
aldehyde group-bearing species are yet stronger than we assume here.

The model shows strong CH$_3$CHO evaporation at low temperatures,
which is qualitatively in agreement with observed rotational
temperatures \cite[]{Nummelin,Bisschop}, although these are not
well-defined. Our model also suggests a high-temperature peak for
long warm-up timescales, due to formation from evaporated hydrocarbons,
whose abundances grow slowly on grains. High-temperature
acetaldehyde detections may therefore signify long lifetimes in
hot-core sources.

CH$_3$CN, like HCOOH, achieves an early- and a late-time peak,
although the latter, due to evaporation of CH$_3$CN, is much
stronger. The former peak is produced indirectly by the
evaporation of HCN from grains. \cite{Bisschop} identify
CH$_3$CN solely as a `hot' species, but \cite{Nummelin} obtain a
low rotation temperature of 46 K in their Sgr B2 NW position,
which was chosen for its lack of star-forming activity.

H$_2$CO is also desorbed at low temperatures in this model, and is important in the formation and survival of other
species. \cite{Bisschop} find H$_2$CO rotational temperatures from $\sim70$ -- 90 K, and \cite{vanderTak2000} find
temperatures in this range toward a number of massive YSOs. This does not necessarily contradict our somewhat lower
formaldehyde evaporation temperature, as H$_2$CO abundance remains high up to temperatures of 80 K and 200 K in the
medium (M) and fast (F) models, respectively. In addition, two of van der Tak's sources yield temperatures close to 200
K. A more intricate treatment of the icy grain mantles in the model might yield a high temperature peak in formaldehyde
separate from its evaporation at $\sim$40 K, as we do not explicitly deal with trapping of weakly-bound species in the
amorphous ice matrix \citep{Collings}. Species formed early during the collapse phase may be trapped until the entire
mantle is evaporated. This could also be the case with, for example, CO, CO$_2$, CH$_4$, and perhaps HCOOH. Indeed, 
the somewhat elevated excitation temperatures of gas-phase CH$_4$ towards some protostellar sources, reported by 
\cite{Boogert04} and references therein, 
could indicate trapping of methane in the deeper water-ice layers. Explicit modeling of the individual ice layers
would shed some light on these issues. 
As in GH06, most H$_2$CO in the current model is formed late in the collapse phase, and would therefore exist in the outermost ice layers.

Unlike the results for many other species, this model does not reproduce the low rotational temperatures for
glycolaldehyde observed by \cite{Hollis2000,Hollis2001,Hollis2004} in Sgr B2(N). They suggest a warmer ($\sim$50 K)
glycolaldehyde component surrounded by a cold ($\sim$8 K) one. In our model, glycolaldehyde is formed on the grains at
around 30 -- 40 K; however, it is not desorbed strongly until temperatures of $\sim 110$ K are reached. Excluding a
gas-phase formation mechanism for glycolaldehyde, two possible scenarios exist for its presence in cold gas. The first
scenario, as suggested by Hollis et al., is that Sgr B2(N) has undergone shocks, in which case the temperature
progression of the grains is disrupted, after a period with T$_{d} >$ 30 -- 40 K at which glycolaldehyde may form. The
second scenario involves a non-thermal desorption process, regardless of the heating mechanism (ie. protostellar
switch-on, or shocks). Such may be the one suggested by Garrod et al. (2006, 2007) for the formation of gas-phase
methanol in cold clouds, whereby the energy of formation of a surface molecule may break the surface--molecule bond
with a probability on the order of 1\%. This would result in glycolaldehyde being desorbed most strongly as it forms
on the grains, at temperatures close to the observed 50 K. It could inject no more than $\sim$1\% of the total amount
formed on grains, however; approximately 10$^{-10} n_H$ in this model.

The inclusion of hydrogen-abstraction reactions for aldehydes shows that they may be destroyed on the grains at high
temperatures given sufficient time. These mechanisms could explain the apparently large spatial scale of glycolaldehyde
in the context of shock-induced desorption. In more extended regions, shocks could desorb glycolaldehyde from grains
before they have time to be processed at high temperatures, while hot-core regions which gradually achieve high
temperatures could be depleted of their glycolaldehyde before evaporation becomes efficient. Hence, glycolaldehyde
would be a tracer of sudden desorption or temperature increase.

\subsection{Complex chemistry in low-mass star-forming regions}

The low-mass protostellar analogs of hot cores, the so-called `hot corinos', also show strong signatures of complex
molecules \citep{Cazaux,Bottinelli,Bottinelli2}. Observationally, there appear to be differences between the low-mass
and high-mass scenarios. \cite{Bottinelli3} report that the ratios of HCOOCH$_3$, CH$_3$OCH$_3$ and HCOOH to their
putative parent molecules, CH$_3$OH and H$_2$CO, appear greater for hot corinos than hot cores, with the HCOOH ratios
being more strongly affected. These observations may be explained in terms of our model: shorter periods (i.e. model F)
between the evaporation of HCOOCH$_3$ and CH$_3$OCH$_3$ at $\sim$70 -- 80 K and the evaporation of water and other ice
components at $\sim$110 K improves the survival of these more complex species (see Section 3). Also, grain-surface
destruction of complex species, especially HCOOH and other aldehyde group-bearing species, is stronger in models with longer warm-up
timescales. This may suggest that the observational differences are due to the faster transit of the gas through
temperatures of $\sim$50 -- 100 K in the hot-corino case. However, the short-timescale model (F) produces lower
abundances of complex molecules, due to the shorter periods at $\sim$10 -- 50 K at which many complex molecules are
efficiently formed. This indicates that much longer periods in this temperature range than we assume may be appropriate
for hot corinos. The recent chemical/hydrodynamic model of prestellar to low-mass protostellar evolution by
\cite{Aikawa2007} does indeed indicate longer times at low temperatures than assumed here, and much shorter times at
high temperatures.

The implications for methyl formate and dimethyl ether are predicated on their having significantly lower binding
energies than H$_2$O. An interesting test of these results would be whether the ratios with H$_2$CO and CH$_3$OH hold
for species with much greater binding energies.

Shorter high-temperature timescales would also tend to reduce abundances of the more exotic species formed on grains
{\em via} the addition of two tightly-bound radicals at high temperature. Examples of this type of molecule include
urea, (NH$_2$)$_2$CO, and ethylene glycol, (CH$_2$OH)$_2$. The model indicates that the very complex species of this
sort may be much less prevalent in low-mass star-forming regions.

\subsection{Large complex molecules}

This model predicts a number of large complex molecules to be
formed with significant abundances; see Figures 4 -- 6, panels
e -- j.  Many of these species have not been detected in the
interstellar medium. We are aware through personal communication
that directed observational searches are ongoing for some of
these more complex species. However, laboratory spectra for many
of these molecules do not exist. We hope that this model will
provide impetus for such spectra to be obtained.

The newly-predicted molecules that are most abundant and robust
to changes in evolution timescale --  and therefore the best
candidates for observational investigation --  are the OH-bearing
species CH$_3$OOH, CH$_2$(OH)$_2$ and HOCOOH; the amide-group
species NH$_2$COOH and (NH$2$)$_2$CO; and the amine-group species
CH$_3$ONH$_2$, CH$_2$(OH)NH$_2$, and NH$_2$OH. Each of these
species and the many other complex species predicted to form by
this model are discussed in detail in the Appendix.

The most abundant complex molecules seem to be those that are
derived from the reaction of primary radicals alone, and not
secondaries. Secondary radicals form at the same time as primaries
(at intermediate temperatures), but their reactive nature and relative
immobility at these temperatures mean they are most likely to be
hydrogenated to their fully-saturated forms. The secondary radicals
typically become mobile much later, at a time when they are no
longer being directly formed from primary-primary addition, but
rather by CR-induced photodissociation of larger species, or by
hydrogen abstraction in the case of species bearing the 
-CHO functional group. Longer timescales
allow more time to produce secondaries at appropriate temperatures
for them to be independently mobile.

\section{Ice composition}

The results discussed above are produced by a two-stage model of
cold collapse and warm-up. However, the initial conditions of the
collapse phase, and the physical parameters themselves, are not
well constrained. In particular, it appears that the (cold) dust
temperature is important to the ultimate composition of the icy
grain mantles \cite[]{Ruffle2}; variations of a few Kelvin can
strongly affect the efficiency of hydrogenation, which is the
dominant mechanism.

Table \ref{tab-ice} shows observationally determined ice
compositions towards two embedded YSOs, and towards the galactic
center infrared source, Sgr A* (Gibb et al. 2000). Also shown are
model values for the end of the collapse phase. The observed ice
abundances appear mostly similar in composition to those achieved
with this model, however it is unclear precisely how much processing
these ices have already undergone, or indeed whether they are a fair
representation of the state of the ice mantles near the centers
of hot cores.

In comparison to observed values, methane seems to be overproduced in
our model, whilst surface CO$_2$ is significantly underproduced. The
latter is a long-standing problem in gas-grain modeling \cite[]{Ruffle2}.
Higher initial temperatures may improve agreement in both of these
cases. The lack of grain-surface formic acid in our models may be
due to the lack of CO$_2$, which could presumably be hydrogenated to
HCOOH. Alternatively, it may be related to grain processing associated
with heating and the onset of star formation.

The gas-phase methanol abundance seen in our models during the warm-up
phase may be somewhat high at $\sim$$10^{-5}n_H$; however, as it is almost entirely derived
from grain-surface evaporation, this implies that it is overproduced
on grains during the cold collapse phase. A similar grain-surface
overabundance may also be true for formaldehyde.

Since these two molecules, along with methane, are important sources of radicals on the grains, we have run further
warm-up models with grain-surface abundances of these species reduced by a factor of 10; see Table \ref{tab-ice}. We
have investigated reductions in each of these species individually, but they may easily be examined together. We do not
include changes in the CO$_2$ abundance in this discussion, as it is not a major source of radicals. Figures 7 -- 9
show time-dependent abundances for each of the three models: F(ice), M(ice) and S(ice), corresponding to fast, medium
and slow warm-up, respectively (Table \ref{tab-mod}).

The effect of CH$_4$ reduction is in fact small, if considered in
isolation. This is due to the possibility of CH$_3$ production
from photodissociated methanol. In fact, methanol is the dominant
producer in the 30 -- 40 K temperature range. The reduction of
methanol is required for a strong effect on CH$_3$-related
species, e.g. CH$_3$NH$_2$.

The impact of reduction in H$_2$CO on grains is strongest for the shorter timescales; longer periods at low temperature
allow the hydrogenation of CO to H$_2$CO, such that for model S the H$_2$CO abundance is little altered. However, a
factor of a few lower grain-surface abundance has a strong effect on its survival in the gas phase. Formaldehyde is the
most important reaction partner for ions at intermediate temperatures, damping abundances of destructive ions, so the
long-term survival of many other species in the gas phase is also diminished. This strongly affects methyl formate,
whose release from the grains no longer coincides with large gas-phase formaldehyde abundances, even in model F(ice).

Lower formaldehyde abundance also lessens the abundance of HCO.
However, this radical may also be formed by hydrogenation of CO.
For short warm-up timescales, the reduction in both CH$_3$ and HCO
radicals (formed from H$_2$CO) greatly reduces acetaldehyde
abundances.

The reduction in methanol abundance has a strong effect on complex-molecule abundances both in the gas phase and on
grain surfaces, but its effects are largely predictable. Species formed from the radicals CH$_3$O/CH$_2$OH suffer
reductions of around 10 times, or more if they are formed also from CH$_3$ or HCO. This clearly creates additional
problems for methyl formate, as, even for model S(ice), it is only formed on grains to an abundance of a few $\times
10^{-9} n_H$. The effects are yet stronger for those species which are doubly dependent on methanol-derived radicals,
like ethylene glycol, and gas phase-produced dimethyl ether.

Clearly these reductions greatly influence the complex
chemistry in the model, and the most important of the three
ice-abundance reductions is that of methanol. These reductions also
place abundances of some of the more exotic species produced in
this model firmly beneath the level of detectability, especially
for shorter warm-up timescales.

The failure to produce sufficient quantities of methyl formate
in particular, and, to some degree, dimethyl ether, may suggest either
that methanol CR-induced photodissociation is faster than
assumed in our model, or that the hydrogenation of H$_2$CO to
CH$_3$O/CH$_2$OH, which has an activation energy barrier, takes
place more efficiently to make up for a lower contribution from
methanol. Other possibilities include some alternate route for
the reduction of methanol, perhaps a barrier-mediated
hydrogen-abstraction mechanism similar to those assumed for
aldehydes, albeit with a greater activation energy. In previous
models it has been assumed that the hydrogen abstraction
reaction OH + H$_2$CO $\rightarrow$ HCO + H$_2$O has no
activation energy barrier, whereas we assume a value of 2500 K.
GH06 found this to be the most important HCO-production mechanism,
and so the higher value (used in this model for consistency with
other similar reactions) may have a large influence on
complex-molecule formation.

Recent {\em Spitzer} surveys of low-mass YSOs \cite[]{Boogert08} 
now indicate that methanol may indeed be as abundant as our `standard'
ice values in some sources, at 1 -- 30 \% with respect to H$_2$O, whilst ammonia may be
somewhat less abundant than we assume, at 1 -- 5 \%. The influence of these new data on the
chemical models should be explored in future work.

\subsection{Comparison with observations}

The galactic-center hot-core source Sgr B2(N) is well-known for its
rich chemistry, and some of the more complex species we model here
have only been detected along that line of sight, making it an apt
subject for comparison with the model. It is almost certainly not
the case that the dynamics or morphology of Sgr B2(N)
are as simple as assumed in the model; however, it is generally
accepted that at least one hot-core source resides in the very
central region, regardless of shock dynamics in more extended
regions. Here we compare abundances and rotational temperatures observed 
toward Sgr B2(N) with values from our reduced ice-abundance models. This 
provides gas-phase and grain-surface methanol and methane abundances more 
in keeping with observations, as well more acceptable gas-phase formaldehyde values.

Table \ref{tabx2} shows calculated peak abundances for a range of molecules, along with the gas/dust temperature at
that peak, for each of the warm-up timescales of the reduced ice-composition run. For some species, both early and late
peaks are shown, especially where there is observational evidence of more than one rotational temperature component.
Beside the modeled values we list observed abundances, calculated from beam-averaged column densities assuming
$X$(H${_2}$)=0.5 and $N$(H$_2$)=$3 \times 10^{24}$ cm$^{-2}$ \citep{Nummelin}. Also shown are the associated rotational
temperatures, and the FWHM beam size. The majority of these observational values are derived from the survey by
\cite{Nummelin}. Where only abundances of isotopomers are available, we assume isotopic ratios of: C$^{12}$/C$^{13}$=70
and O$^{16}$/O$^{17}$=2044 (see Redman et al. 2002, and references therein), and N$^{14}$/N$^{15}$=100. Where two
temperature components have been detected, we list them next to the early- or late-time (cold or hot) peak model
abundance, as appropriate. 

In a simple analysis, if the emitting molecular species in the hot core is more compact than the projected beam of the telescope, the beam-averaged emission will indicate a molecular column density that is lower than the true value -- by a factor of the squared ratio of the angular source size to the beam size.
In Table \ref{tabx2}, in cases where the model indicates that species are released from grains at high
temperature, we list observational abundances adjusted for beam dilution by assuming a compact emitting source that is
5\arcsec{} in size; we assume that the $N$(H$_2$) remains constant. Such cases are indicated in the {\em notes} column. Similarly, we compare the cold species with the unadjusted abundances for a 23\arcsec{} beam based on the assumption that these species are spatially extended, and therefore fill the beam. We set in boldface those peak-abundance temperatures that approximately agree with observational rotational temperatures. We
highlight abundances that agree with observed values within approximately one order of magnitude. Such values are only
highlighted in cases where the observed or re-calculated beam-sizes are appropriate to the temperatures.

Overall, the modeled abundances and temperatures match well with the
observations, and succeed in reproducing values for both hot and cold
components. The medium warm-up timescale, model M, appears
to produce the best match. Whilst this model represents only a single point
in space, we may consider the molecules that are better represented by low
temperatures as being more distant from the hot-core center (and therefore
more spatially extended), representing material that is less advanced in
its warm-up sequence.

Discrepancies exist, and these may originate from the more
subtle chemical and physical aspects of the model that are discussed
above, including branching ratios, activation energies, and binding
energies -- especially for species such as
methyl formate. Ice structure may also complicate the determination
of gas-phase abundances at lower temperatures, prior to the total
evaporation of grain mantles. Regarding physical conditions,
shocks or radiative pumping may explain the differences between
observed rotational temperatures and model peak temperatures. Other
discrepancies could likely arise from uncertain observational parameters.
For example, molecular partition functions often exclude vibrational
contributions because only the pure rotational lines from the ground
vibrational state have been measured experimentally.

Additionally, the lack of spatial-scale information for most species
makes the beam-dilution correction quite uncertain.  Indeed, recent
CARMA observations indicate that the spatial distributions of some of
the most common hot-core molecules may in fact be much more complicated
than previously thought \cite[]{CARMA_Friedel}.

\section{Conclusions}

Presented here is the first hot-core chemical model to employ an extensive
network of grain-surface reactions for organic molecules, in addition to the
standard gas-phase ion--molecule formation routes. We use a two-stage physical
model incorporating the collapse and warm-up phases of a hot core to
explore the chemistry in \linebreak low- to high-mass star-forming regions.
The photodissociation of common ice-mantle constituents by the cosmic
ray-induced UV field (typically removing a hydrogen atom) produces
grain-surface radicals. These `primary radicals' become mobile and reactive on
the grain surfaces as dust temperatures increase during the warm-up phase. In
this way, basic component structures may attach to one another, building up
larger molecules. Further photodissociation (or hydrogen abstraction) of the
resultant complex molecules provides the main source of `secondary radicals',
leading to an even greater degree of chemical complexity. This type of chemistry,
based on a small set of radicals, can explain the formation of many complex
organics that have been detected, but which previously had no plausible gas-phase
formation routes. Many other complex organic molecules are also formed in this
chemical network, and predictions for a number of previously uninvestigated
molecules are now available.

The differing surface-binding characteristics of the grain-surface species
control the order in which radicals become mobile during the warm-up phase. At
temperatures around 30 -- 40 K, the reaction of the mobile primary radicals
HCO and CH$_3$ with more strongly-bound primary radicals CH$_3$O and CH$_2$OH results
in the formation of methyl formate, dimethyl ether, glycolaldehyde and ethanol.
These species evaporate when higher temperatures are achieved.

The production of radicals of greater complexity (secondaries) is weaker than that of primaries, due to the lower
abundances of their parent complex molecules. At intermediate temperatures ($\sim$30 -- 50 K), secondaries typically
react with the mobile primary radicals, allowing three-carbon-atom species like acetone to form. At higher temperatures
($\sim$50 -- 100 K), secondary radicals themselves may become mobile; high temperature mobility of the secondary
radical CH$_3$CO results in reaction with OH to form acetic acid. The dependence of secondary-radical formation on the
abundances of the simpler complex molecules means that primary--secondary radical additions are most effective with
long warm-up timescales; hence the most complex molecules may be detectable in the most slowly-evolving sources.

Such differences between primary- and secondary-radical processes may indeed explain the disparities in observed
abundances of structural isomers, most notably methyl formate/glycolaldehyde/acetic acid. While the precise relative
abundances calculated for these species are not a direct match to observed values in Sgr B2(N), our network provides
plausible, differing formation routes for each of the three. Agreement may be improved following further exploration of
the parameter space of this model. A more extensive treatment of the ice chemistry to include esterification reactions between carboxylic acids and alcohols may also offer more insight into the differing abundances of these structural isomers.

Complex molecule formation is not restricted to grain surfaces, but may still be
intricately linked to grain-surface processes. The evaporation of formaldehyde at
$\sim$40 K results in the gas-phase formation of other species, including HCOOH.
Other rotationally cold species like CH$_3$CHO are formed on grains and evaporate at
low temperatures. Differing grain-surface characteristics may therefore be responsible
for observed spatial and temperature displacements between different gas-phase molecules.

The initial ice composition was adjusted to agree with infrared observations of
protostellar envelopes, reducing the amount of those
ices available for primary-radical production. Comparison of this model
to molecular abundances and temperatures observed in the Sgr B2(N-LMH) hot core
results in a reasonable match in light of the uncertainties in the model,
especially for the intermediate warm-up timescale. However, the abundances of
complex molecules in the fast warm-up model are rather low in comparison to low-mass
star-forming regions, with which we identify short warm-up timescales. This may
indicate that the initial ice compositions in low- and high-mass star-forming regions
are somewhat different; or that the low- to intermediate-temperature phases of hot-corino
evolution are longer-lived than we assume in our simple model. The latter conclusion
agrees with the hydrodynamic models of \cite{Aikawa2007}.  Nevertheless, our results
support the argument that hot corinos experience shorter periods at high temperature
than hot cores.

The study presented here is only the first step in probing the many chemical and
physical influences on hot-core chemistry; a much larger parameter space exists than
has been explored here. The branching ratios for both the photodissociation of methanol
ice and the hydrogenation of formaldehyde are quite uncertain, and the activation
energies for many of the grain-surface reactions are poorly defined. All of these
quantities could strongly influence the degree of isomerism and molecular complexity
reached in grain-surface reactions; further studies are planned.

The choice of grain-surface binding energies is particularly important for methyl
formate, dimethyl ether, and acetone; the values employed here allow these species to
evaporate before most other large molecules, leaving them vulnerable to gas-phase
destruction. Their abundances are therefore more sensitive to warm-up timescales, and
to their final grain-surface abundances, prior to evaporation. The interaction of
these effects with those of branching ratios and activation energies may therefore
be complex.

This model is the first such study that can explain the chemical complexity observed in both high-mass and low-mass hot
cores. Much work remains, however, before a full understanding of these complicated environments can be achieved. Given
the variability in abundances and temperatures amongst hot cores it is unclear whether a direct comparison between the
model and observations is valid for such a small sample. We encourage systematic observational surveys that probe a
range of physical and chemical conditions in star-forming regions. Given the large number of previously uninvestigated
molecules predicted by this model, we also encourage additional laboratory studies to support astronomical searches.

The chemical network employed in this model is, of course, not comprehensive. A number
of unsaturated species, hydrocarbons in particular, have been detected in star-forming
regions, but are not explicitly treated here, due to the complexity of the
hydrogenation process, which undoubtedly involves poorly-defined activation energies.
Hassel, Herbst \& Garrod (submitted) have begun to address this issue.
A number of cyanide species have also been detected, but are omitted from this model for
the same reasons. Future models must take account of all these species, which in many
cases are very abundant; however, the degree of complexity reached in this model is
substantial.

\acknowledgments

RTG thanks the Alexander von Humboldt Foundation for a Research
Fellowship.  Support for SLWW was provided by NSF grant
AST-0540459 through the UIUC Laboratory for Astronomical Imaging,
and by the UIUC Critical Research Initiative program through the
group of Benjamin McCall. SLWW would also like to acknowledge
Geoffrey Blake for very helpful discussions regarding the chemical
network as well as support during its initial development. EH acknowledges
the support of the National Science Foundation for his research program in
astrochemistry.

\section{Appendix}

Here follow detailed discussions of the results for each species or class of species represented in Figures 4 -- 6.
Each Section corresponds to one (or more) similarly labeled panel in the Figures. While the final abundances may change
between the standard and reduced ice composition models, the overall trends discussed here are also applicable to
Figures 7 -- 9.

\subsection{H$_2$O, CO, CH$_4$, NH$_3$, and HCO$^+$ (a)}
The ice constituents CO and CH$_4$ are seen to evaporate at low
temperatures (20 -- 25 K), but NH$_3$ and H$_2$O remain on the
grains in large quantities until a temperature of $\sim$110 K is
reached.  The evaporation of H$_2$O and NH$_3$ ices strongly influences
gas-phase ion chemistry, as is illustrated by the HCO$^+$ abundance.
HCO$^+$ also significantly influences the gas-phase abundance of
more complex molecules, as discussed below.

\subsection{Commonly observed hot core species (b-c)}
As shown by GH06, dimethyl ether, CH$_3$OCH$_3$, methyl formate,
HCOOCH$_3$, and formic acid, HCOOH, are formed on the grains with
abundances as great or greater than those typically observed in the gas phase,
{\em via} the addition of heavy radicals. However, dimethyl ether and
methyl formate substantially evaporate from dust grains before the majority
of other, strongly-bound species. This is a result of revised binding energies,
which now take into account the different structures of the isomeric
radicals CH$_3$O and CH$_2$OH.

Methanol and formaldehyde are primarily formed on grains during
the collapse phase, {\em via} CO hydrogenation. Formaldehyde is also formed
early in the warm-up phase. Gas-phase methanol abundances are dictated by
its grain-surface formation and subsequent co-desorption with
water ice at $\sim$110 K. The binding energy of formaldehyde is
lower than that of methanol, resulting in its ejection from dust
grains at a temperature of $\sim$40 K. The injection of this large
quantity of formaldehyde boosts abundances of certain gas-phase
species, most notably water, methanol, and formic acid.

The injection of formaldehyde provides an abundant reaction partner
for HCO$^+$. This ion is the primary destruction partner for many
large species, including dimethyl ether and methyl formate. In
model S, the large formaldehyde abundances achieved
at $\sim$40 K do not last to the higher temperatures at which
dimethyl ether and methyl formate evaporate, hence the survival of
these species is curtailed as they are exposed to large HCO$^+$
abundances.

Following the evaporation of methanol, large amounts of dimethyl
ether are formed in the gas phase, in spite of the low formation efficiency.
The long-timescale run (model S) shows an increase in gas-phase methyl
formate abundance at around 100 -- 110 K. CH$_3$O, produced by 
CR-induced photodissociation of gas-phase methanol, 
accretes onto grains, where it may react with the
strongly-bound formamide, NH$_2$CHO, to produce methyl formate, which
then quickly evaporates. When formamide co-desorbs with H$_2$O, 
the methyl formate formation route becomes weak.

Formic acid is produced in large quantities on the grains, but
also in the gas phase following the evaporation of formaldehyde,
{\em via} OH + H$_2$CO $\rightarrow$ HCOOH + H. The formic acid
produced in the gas phase accretes onto the grains, substantially
increasing grain-surface abundances. Formic acid has a large
binding energy, and in the longer timescale models (M and S)
its surface abundance is diminished by hydrogen abstraction
by OH prior to evaporation into the gas phase. Co-desorption of
formic acid with H$_2$O results in large abundances at high temperatures.
These abundances are somewhat sustained by the gas-phase reaction of OH and
formaldehyde shown above, and by HCO$^+$ + H$_2$O $\rightarrow$
CH$_3$O$_{2}^{+}$ followed by recombination.

Ethanol, C$_2$H$_5$OH, and glycolaldehyde, CH$_2$(OH)CHO, are both 
formed on grain surfaces in this model, mainly by addition of CH$_3$ and HCO 
to CH$_2$OH; hence their grain-surface behaviour is similar to that of their 
structural isomers, dimethyl ether and methyl formate, respectively (acetic 
acid has a more complex formation route, as seen below). The CH$_3$-bearing 
species, ethanol and dimethyl ether, are each more prevalent on the grains than 
their related HCO-bearing isomers, resulting from the greater prevalence of CH$_3$-precursor 
ices. Ethanol and glycolaldehyde remain on the grains after their 
structural isomers have evaporated, co-desorbing with water. The retention of 
glycolaldehyde allows it to be destroyed by OH radicals at high
temperatures, prior to complete evaporation; this is especially true for model S. 
The resultant CH$_2$(OH)CO radical may attach to heavier 
radicals at these times. In model S, CR-induced photodissociation
of ethanol and glycolaldehyde on the grains becomes significant. 
The primary destruction processes for gas-phase ethanol and glycolaldehyde are 
ion--molecule reactions resulting in proton addition, followed by dissociative
recombination.

Acetaldehyde, CH$_3$CHO, is formed on grains by the addition of
CH$_3$ and HCO; the mobility of CH$_3$ dominates grain
surface formation because the barrier to HCO diffusion is somewhat
higher. As for most other species, longer warm-up timescales provide 
longer periods over which temperatures are tuned for efficient production. 
Acetaldehyde reaches its peak
surface abundance at $\sim$30 K, and this abundance remains static
until evaporation becomes strong at $\sim$50 K.
In model S, acetaldehyde is also produced in the gas
phase at late times. The longer timescale allows more hydrocarbons
to build up on the grains \cite[see][]{Garrod07}, and once they are
released into the gas phase, acetaldehyde can form {\em via}
C$_2$H$_5$ + O $\rightarrow$ CH$_3$CHO + H.

Ketene, CH$_2$CO, is formed on grains in significant quantities
during the collapse phase, primarily by repetitive hydrogenation
of accreted C$_2$O. Later, reaction with
accreted C$_2$ molecules, followed by hydrogenation, leads to
greater surface ketene abundances prior to evaporation at $\sim$40
K. Survival of gas-phase ketene is dependent on the abundance of 
formaldehyde in the gas phase.

Although no new CN surface chemistry has been added, the network
includes the reaction CH$_3$ + CN $\rightarrow$ CH$_3$CN. This is
the strongest route for the formation of methyl cyanide, CH$_3$CN.
It is also formed at earlier times by the hydrogenation of C$_2$N on
grains. Gas-phase formation takes place following the evaporation
of HCN, which reacts with CH$_{3}^{+}$, producing modest methyl cyanide
abundances. However, the greatest gas-phase abundance occurs
upon its evaporation at $\sim$90 K.

\subsection{Nitrogen-bearing species (d)} 
Methylamine, CH$_3$NH$_2$, is formed by simple addition
of CH$_3$ to NH$_2$ on grains, and until its evaporation it is only
appreciably destroyed by CR-induced photodissociation. NH$_2$OH is
formed initially by NH + OH addition on grains, followed by
hydrogenation; however, as OH also becomes mobile, the addition
reaction OH + NH$_2$ $\rightarrow$ NH$_2$OH becomes dominant.

HNCO is only abundant in the gas phase as a result of the
destruction of larger complex species; in particular, urea,
(NH$_2$)$_2$CO, due to its high abundance in this model,
especially for longer timescales. Gas-phase production mechanisms
for HNCO which are not dependent on the destruction of larger,
grain surface-produced species are not included here.
Nevertheless, HNCO is formed on grain surfaces by hydrogenation of
accreted OCN, but is further hydrogenated to formamide, NH$_2$CHO,
such that its abundance is very small.

HCN is formed primarily at low temperatures, beginning in the collapse phase. Gas-phase produced HCN may accrete; or,
accreted gas phase-produced CN may be hydrogenated on grain surfaces. This material evaporates at $\sim$40 K, and
survives while formaldehyde is present in the gas phase to reduce ion abundances. At such temperatures, HCN and CN are
formed by CR-induced photodissociation of CH$_3$CN, and both quickly evaporate. Later, following evaporation of NH$_3$
from grains, HCN is formed in the gas phase {\em via} C + NH$_2$ $\rightarrow$ HCN + H. The behavior and abundance of
HNC (not shown in Figures) are very similar to those of HCN.

Formamide, NH$_2$CHO, is formed on grains at low temperatures by
hydrogenation of accreted OCN. It is later formed in the gas phase
by the reaction of formaldehyde and NH$_2$ after formaldehyde
evaporates at $\sim$40 K. Accretion of this material leads to increased
grain-surface abundances. As HCO becomes mobile on the grains, formamide
is also formed by addition to NH$_2$ radicals. In model S,
grain-surface formamide is destroyed by hydrogen abstraction, prior
to evaporation. At high temperatures, formamide continues to be
formed in the gas phase from NH$_2$.

Following the evaporation of formaldehyde, formamide becomes an
important contributor of HCO radicals, {\em via} its CR-induced
photodissociation.  A similar process occurs for CH$_3$ production
from methylamine. In this way, these larger species formed at
early times become repositories for important simple radicals
until late times and high temperatures.

\subsection{Complex oxygen-bearing species formed from CH$_3$CO (e)}
Many complex molecules form from secondary radicals, which are the
products of hydrogen-abstraction reactions or CR-induced photodissociation. 
One such radical is CH$_3$CO, which is the precursor to several commonly-observed
hot-core molecules.

Acetone, (CH$_3$)$_2$CO, is formed by addition of CH$_3$ to CH$_3$CO.  At the time that CH$_3$ becomes mobile, CH$_3$CO
is primarily formed by hydrogen abstraction from acetaldehyde. The interpolated value for the binding energy of acetone
used here is relatively low, and it evaporates at around 65 K. In this way, it behaves similarly to methyl formate and
dimethyl ether, requiring ion abundances to be limited by formaldehyde for its gas-phase survival. Shorter evolutionary
timescales allow its gas-phase abundance to be sustained to greater temperatures.

Acetamide, CH$_3$CONH$_2$, is formed mainly by addition of grain-surface
CH$_3$ to HNCO, followed by hydrogenation. Its production is increased
when formaldehyde evaporates; reaction of HCO$^+$ with H$_2$CO produces more
gas-phase OH and this results in faster formation of OCN. This molecule is
accreted and hydrogenated on grains, to make HNCO, which may react to produce
acetamide.

Acetic acid, CH$_3$COOH, is formed from the addition of OH to CH$_3$CO. This formation mechanism differentiates it from
its structural isomers methyl formate and glycolaldehyde, which form from the reaction of two primary radicals. Here,
the mobility of the CH$_3$CO radical dominates that of OH. However, at the time that CH$_3$CO becomes mobile, most
grain-surface acetaldehyde has evaporated, and the major source of CH$_3$CO radicals is the CR-induced
photodissociation of acetamide.

In addition to these previously observed species, two additional
molecules form from similar reactions involving CH$_3$CO.
Hydroxyacetone, CH$_2$(OH)COCH$_3$, is formed from
reaction between CH$_3$CO and CH$_2$OH, at rates determined
by CH$_3$CO mobility. Methyl acetate, CH$_3$OCOCH$_3$, is formed
by addition of CH$_3$O and CH$_3$CO; these radicals
become mobile at similar temperatures, and so they are both
capable of reacting with stationary radicals, reducing methyl 
acetate production. Methyl acetate also evaporates before the 
majority of the ices, leaving it exposed to gas-phase ion--molecule 
reactions.

\subsection{Complex oxygen-bearing species formed from CH$_3$O and CH$_2$OH (f -- g)}
Several complex species arising from CH$_3$O and CH$_2$OH form
entirely on grain surfaces. Methoxymethanol, CH$_3$OCH$_2$OH, and
dimethyl peroxide, (CH$_3$O)$_2$, are dependent on
the mobility of the CH$_3$O radical. Methoxymethanol
achieves greater abundances than dimethyl peroxide, as the
increasing mobility (and therefore, reactivity) of the CH$_3$O
radical reduces its abundance on grains.

Methoxyamine, CH$_3$ONH$_2$, and hydroxymethylamine,
CH$_2$(OH)NH$_2$, are both formed around the same time as
methoxymethanol and dimethyl peroxide. The NH$_2$ radical is still
immobile at these times, but NH has a similar binding energy to
CH$_3$O and is therefore mobile. So while methoxyamine is
predominantly formed {\em via} direct addition of CH$_3$O and
NH$_2$ radicals, hydroxymethylamine is formed by addition of NH to
CH$_2$OH, followed by hydrogenation. Later, the NH$_2$ radical becomes mobile, and
contributes to the formation of both species. Abundances of 
NH-radical--bearing species are typically $<10^{-12}n_H$, and are
therefore not shown.

Ethylene glycol, (CH$_2$OH)$_2$, is formed far later than its
structural isomers; its precursor radical, CH$_2$OH,
becomes mobile just as water and other species are beginning to
desorb. The grain-surface ethylene glycol abundance rises
dramatically at $\sim$110 K, and even as it begins to desorb
strongly and the grain-surface abundance falls, it is still being
formed in this manner. The large gas-phase quantities of ethylene
glycol at late times arise from reactions on the bare grain
surfaces.

The structural isomers dimethyl carbonate, (CH$_3$O)$_2$CO, and
methyl glycolate, CH$_3$OCOCH$_2$OH, are formed primarily by the
addition of CH$_3$O to the secondary radicals CH$_3$OCO and
CH$_2$(OH)CO, respectively. CH$_2$OH addition to CH$_3$OCO also
becomes important at later times. The production of the
precursor radicals CH$_3$OCO and CH$_2$(OH)CO is more complex. The
high gas-phase abundances of these species at late times are the
result of the protonation and dissociative recombination of larger
species following their evaporation from the grains. When
CH$_3$O becomes mobile, CH$_2$(OH)CO is mainly formed {\em via}
hydrogen-atom abstraction from glycolaldehyde by OH
radicals. However, the evaporation of methyl formate precludes the
analogous formation of CH$_3$OCO radicals. Instead,
CR-induced photodissociation of methyl ester carbamic acid,
CH$_3$OCONH$_2$, is the strongest source. The other structural
isomer of dimethyl carbonate and methyl glycolate,
dihydroxyacetone, (CH$_2$OH)$_2$CO, is formed when CH$_2$OH
becomes mobile on the grains, and therefore behaves similarly to
ethylene glycol.

\subsection{Other complex organics (h -- j)}
A variety of even more complex organic molecules form in this
reaction network.  Methyl carbamate, CH$_3$OCONH$_2$, glycolamide,
CH$_2$(OH)CONH$_2$, urea, (NH$_2$)$_2$CO, and carbamic acid,
NH$_2$COOH, are all formed as HNCO becomes mobile at around 40 --
50 K. HNCO reacts with CH$_3$O, CH$_2$OH, NH, and OH, the products
of which may be hydrogenated to form the saturated species. The
final up-turn in the abundances of these species is the result of
primary-radical addition to NH$_2$CO. This species is formed by
hydrogen abstraction from formamide; hence the abundances of the
four species strongly correlate with that of acetamide. Carbamic
acid also benefits at late times from the addition of NH$_2$ to
COOH.

The OH radical readily reacts with several primary and secondary
radicals to form other complex organics.  The methanol derivatives
react with OH to form methyl peroxide, CH$_3$OOH, and methanediol,
CH$_2$(OH)$_2$. The high mobility of both CH$_3$O and OH as
temperature increases disfavors methyl peroxide formation, as each
radical becomes scarce. Methanediol production does not suffer in
this way.

Species containing carboxylic acid groups are also formed on grain surfaces. Carbonic acid, HOCOOH, methyl carbonate,
CH$_3$OCOOH, and glycolic acid, CH$_2$(OH)COOH, all require hydrogen abstraction from the aldehyde functional groups of other complex molecules to form their
pre-cursor radicals, hence they are formed on the grains at late times.  It is possible that these molecules can
undergo further esterification reactions, though this type of chemistry has not been included in the network.

\subsection{Products of polymerization reactions (j -- k)}
Many species are formed by addition of the important primary radicals to other primary radicals of the same type.  In
the case of HCO + HCO, the resultant OHCCHO molecule can undergo hydrogen abstraction to form the COCHO radical, which
can then react with primary radicals to form a variety of complex species. Several more complex species involving
-COCHO groups are also formed from reactions involving other secondary radicals and HCO. This type of CO-polymerization
chemistry can quickly lead to a high level of molecular complexity, and so this chemistry was limited to molecules with
-COCHO groups.  Therefore, molecules with a -CO(CO)CHO backbone or larger are not included in the network.

Several highly abundant simple species are also formed from this type of chemistry.  Ethane, C$_2$H$_6$ and hydrogen
peroxide, H$_2$O$_2$, were present in previous versions of the ratefile, while hydrazine, (NH$_2$)$_2$, is a new
addition. In spite of the new formation mechanism introduced, C$_2$H$_6$ is formed predominantly throughout the warm-up
phase by hydrogenation of C$_2$H$_4$, which is accreted from the gas phase. C$_2$H$_4$ is formed {\em via}
CH$_4$ + CH $\rightarrow$ C$_2$H$_4$ + H, fed by the evaporation of CH$_4$ from the grain
surfaces. Hydrogen peroxide is formed initially by the hydrogenation of molecular oxygen, but is later strongly
enhanced in abundance when OH becomes mobile. Hydrazine is formed mainly by the addition of NH to NH$_2$ radicals,
followed by hydrogenation, and at a late stage by direct addition of NH$_2$ radicals.
The abundance of each of these three species is fairly robust to changing evolutionary timescales.

\clearpage

\begin{figure*}
\epsscale{0.8} 
\plotone{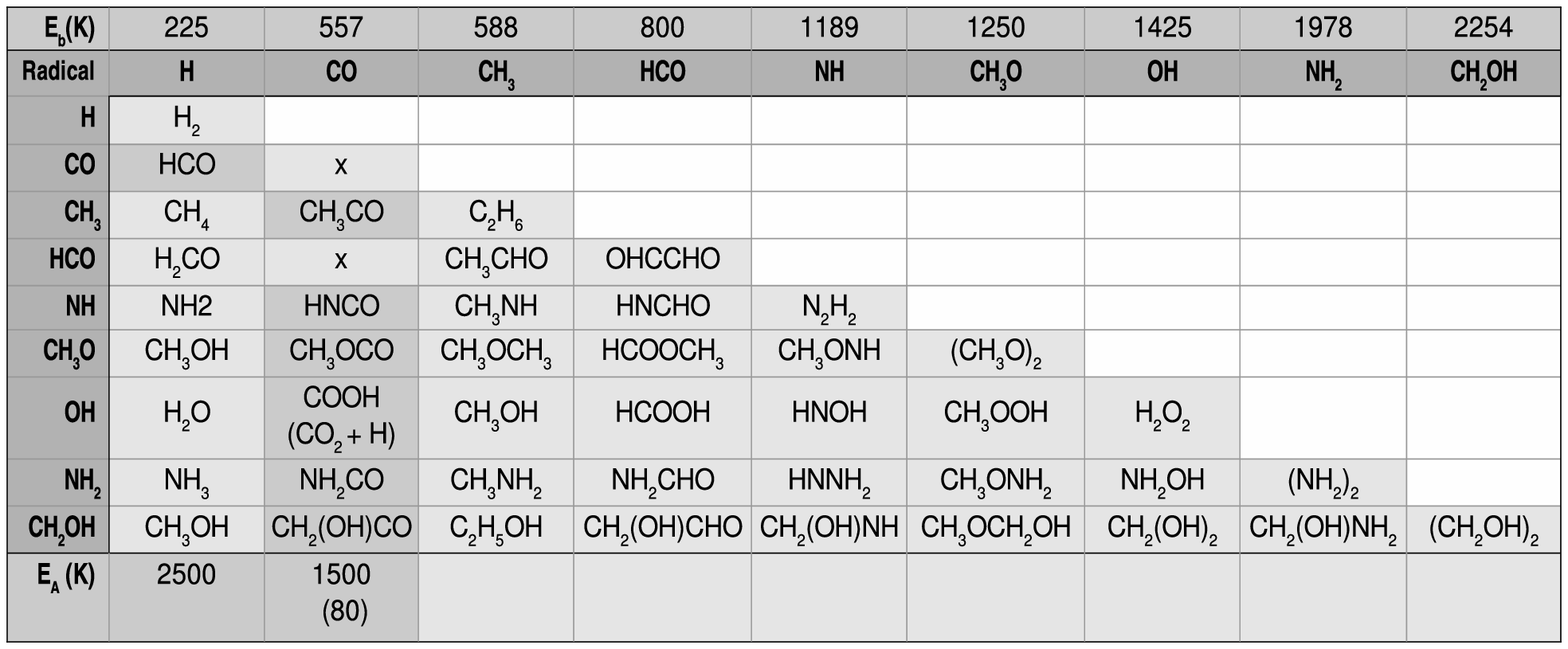} \caption{\label{fig0a} Primary radical--primary radical reactions.
Radicals are arranged in order of increasing diffusion energy barrier (shown in the top row). The product of each
reaction is shown in the box corresponding to the pair of reactants; `x' signifies reactions excluded from the reaction
set. Darker product boxes signify reactions with activation energies, and the values are indicated at the bottom of
each column.}
\end{figure*}

\begin{figure*}
\epsscale{0.95} 
\plotone{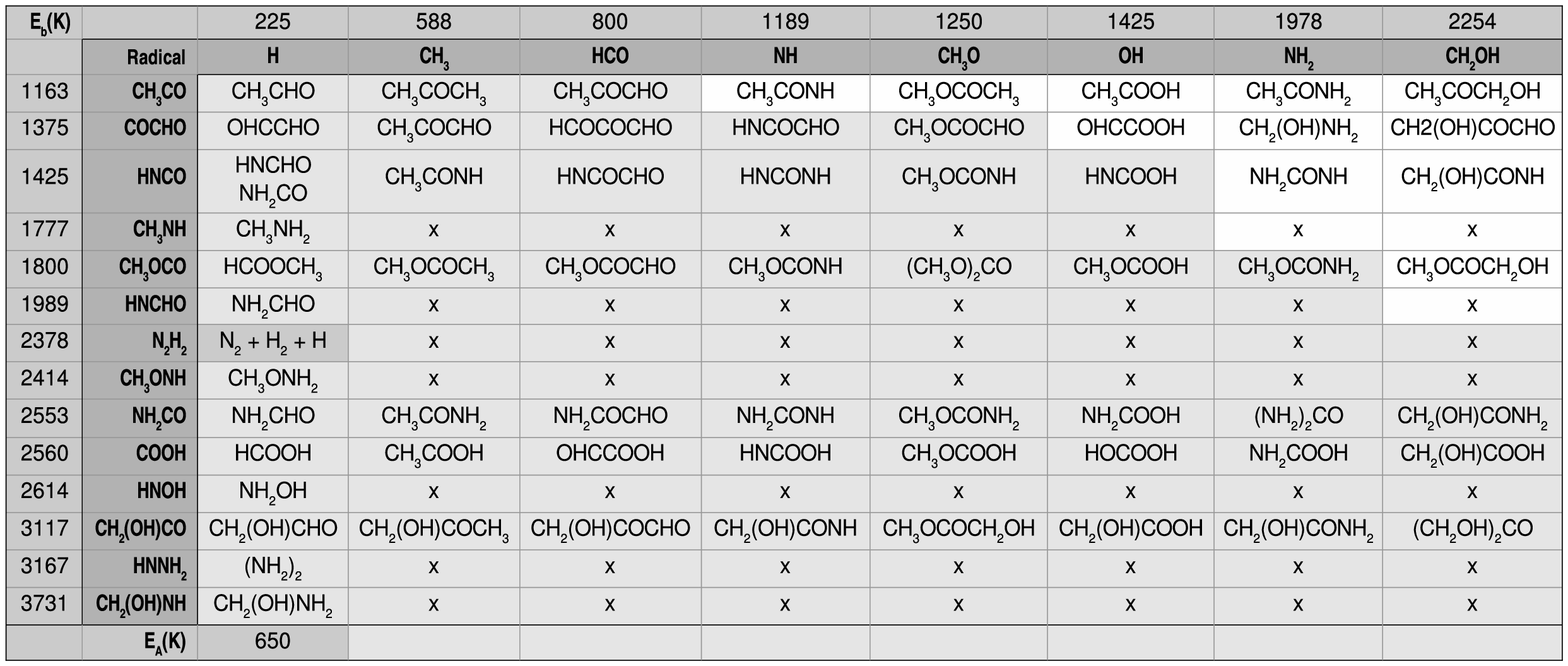} 
\caption{\label{fig0b} Primary radical--secondary radical reactions.
Primary radicals are shown along the top edge; secondary radicals on the left-hand edge. Radicals are arranged in order
of increasing diffusion energy barrier (shown in the top row and the first column). The product(s) of each addition
reaction is (are) shown in the box corresponding to the pair of reactants; `x' signifies reactions excluded from the
reaction set. Darker product boxes signify reactions with activation energies, and the values are indicated at the
bottom of each column. White boxes signify reactions whose rates are dominated by the mobility of the secondary
radical, i.e. $E_{b}($primary$)>E_{b}($secondary$)$.}
\end{figure*}

\begin{figure*}
\epsscale{1.05} 
\plotone{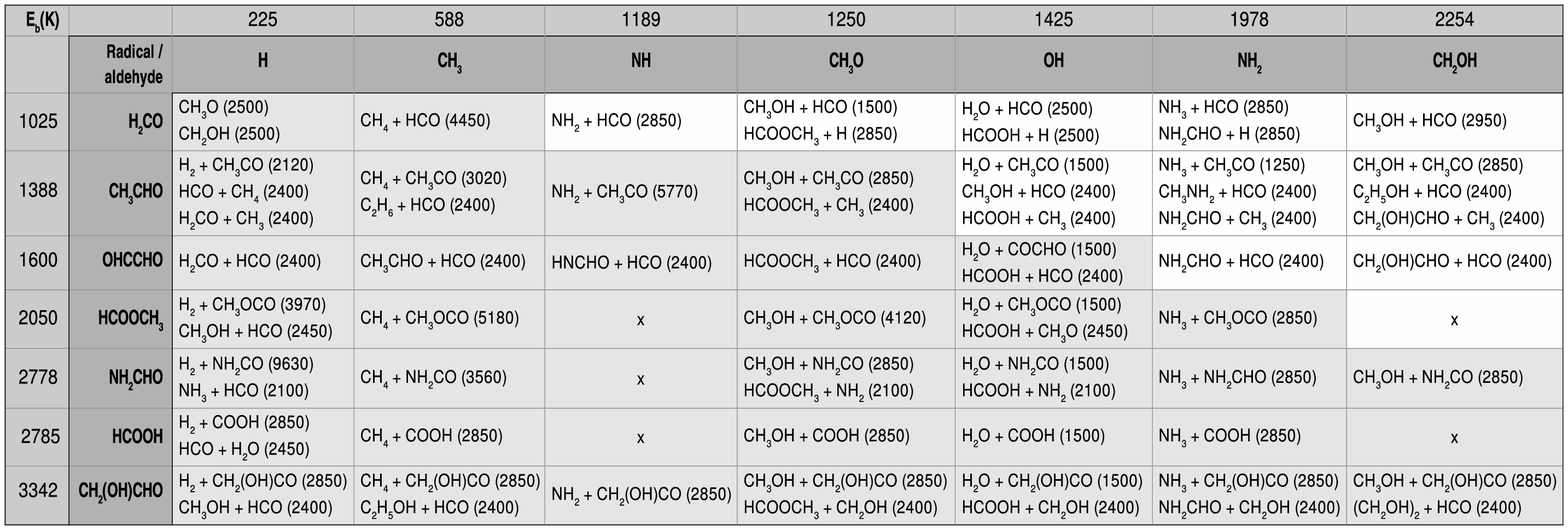}
\caption{\label{fig0c} Primary radical--aldehyde-group reactions. Primary radicals are shown along the top edge; aldehyde group-bearing species on
the left-hand edge. Species are arranged in order of increasing diffusion energy barrier (given in K; see the top row
and the first column). The products of each reaction are shown in the box corresponding to the pair of reactants; `x'
signifies reactions excluded from the reaction set. All reactions have activation energies, shown in brackets. White
boxes signify reactions whose rates are dominated by the mobility of the aldehyde group-bearing species, i.e.
$E_{b}($radical$)>E_{b}($X-CHO$)$.}
\end{figure*}

\begin{figure*}
\epsscale{0.4} 
\plotone{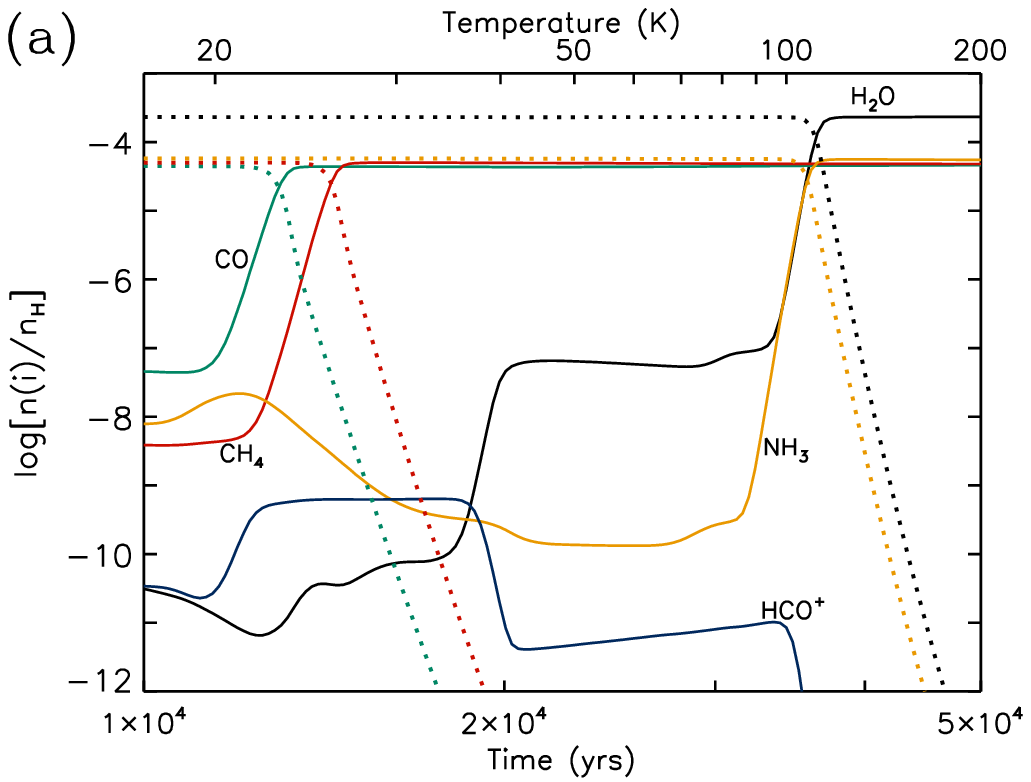} 
\plotone{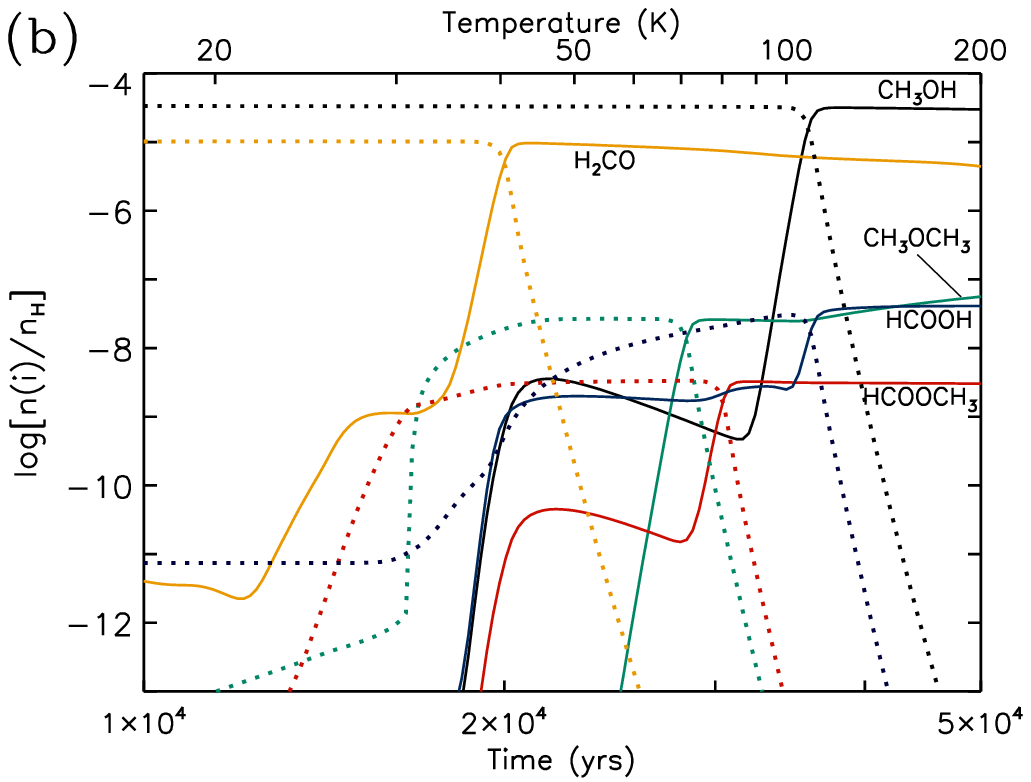}
\plotone{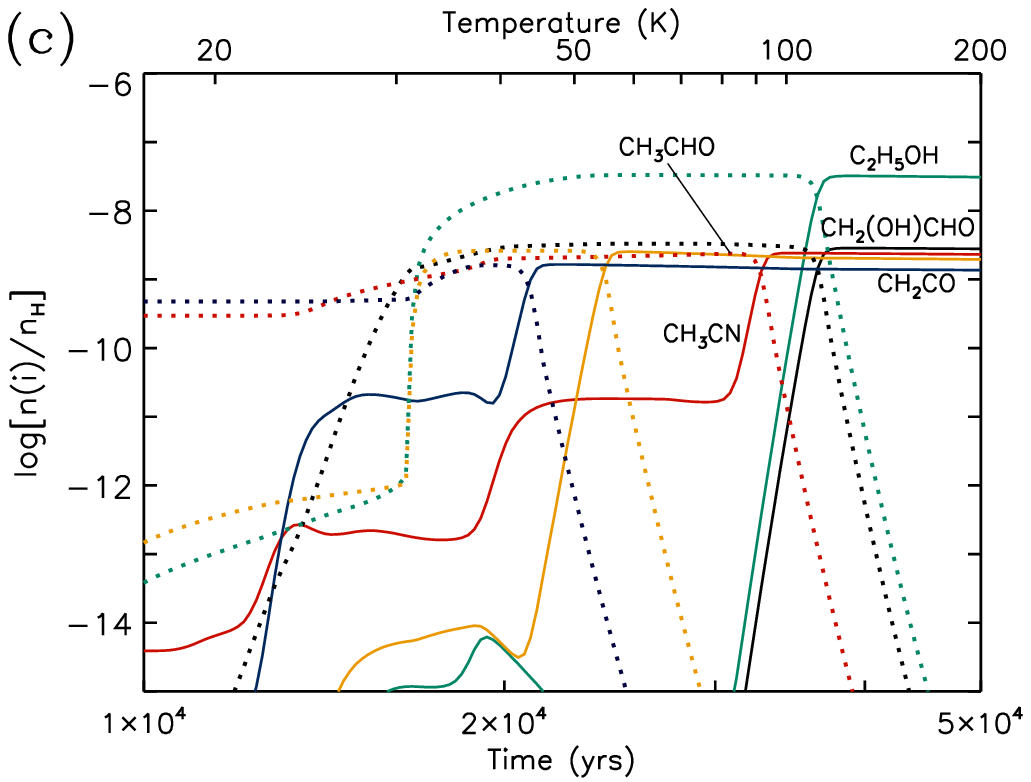} 
\caption{\label{fig1} Fractional
abundances for model F, with a warm-up timescale of $5 \times 10^4$ yr. Solid lines indicate gas-phase species; dotted lines of the same color indicate the grain-surface species. [{\em See the electronic edition of the Journal for panels d--k.}]}
\end{figure*}
\begin{figure*}
\epsscale{0.4}
\plotone{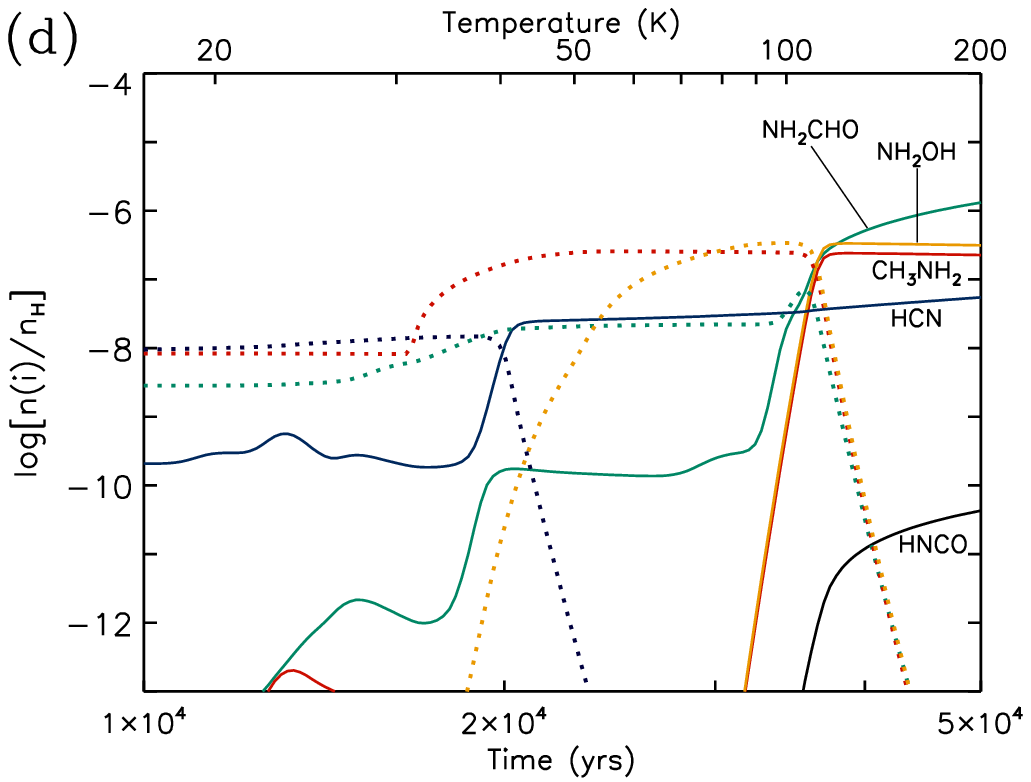}
\plotone{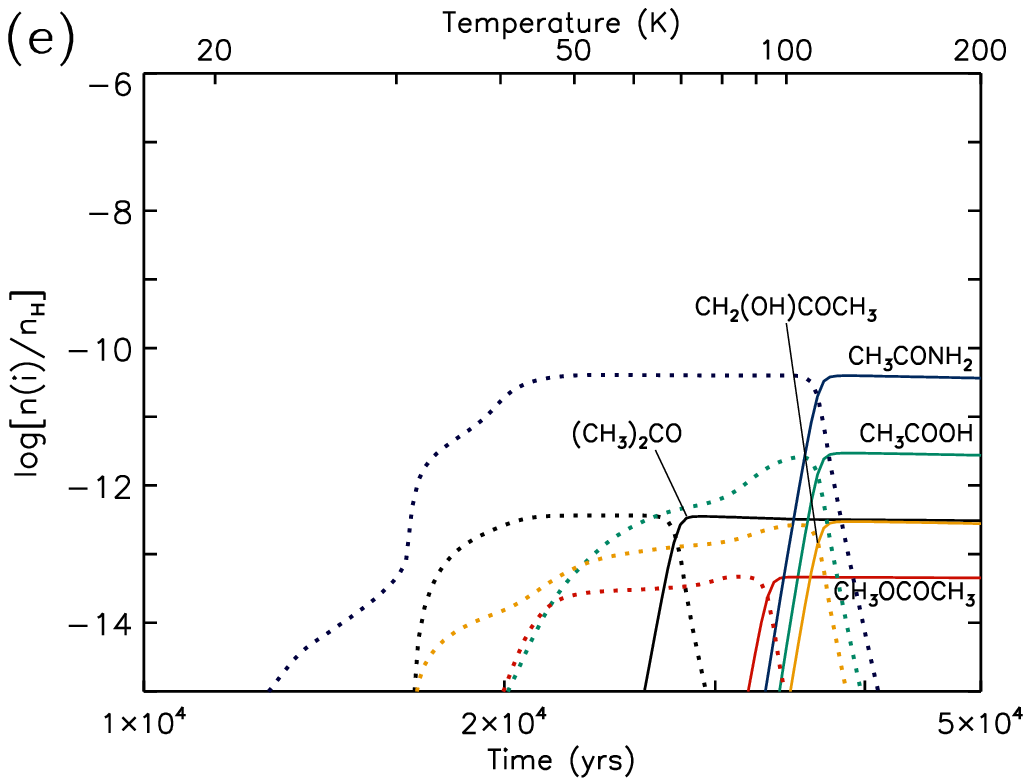} 
\plotone{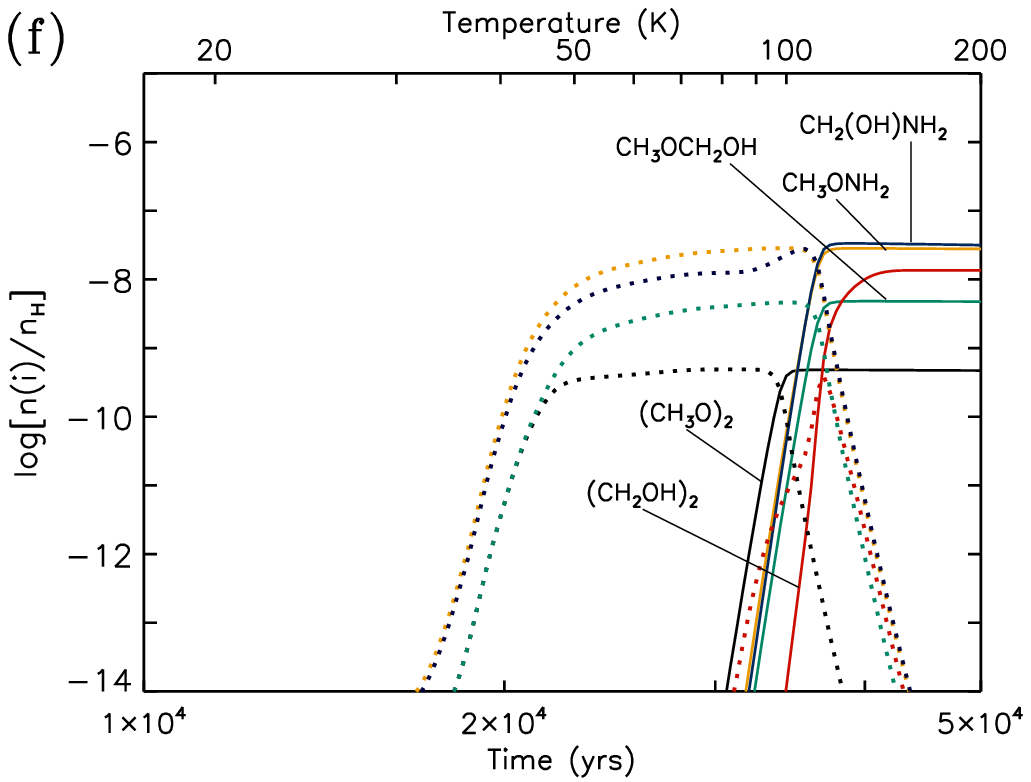} 
\plotone{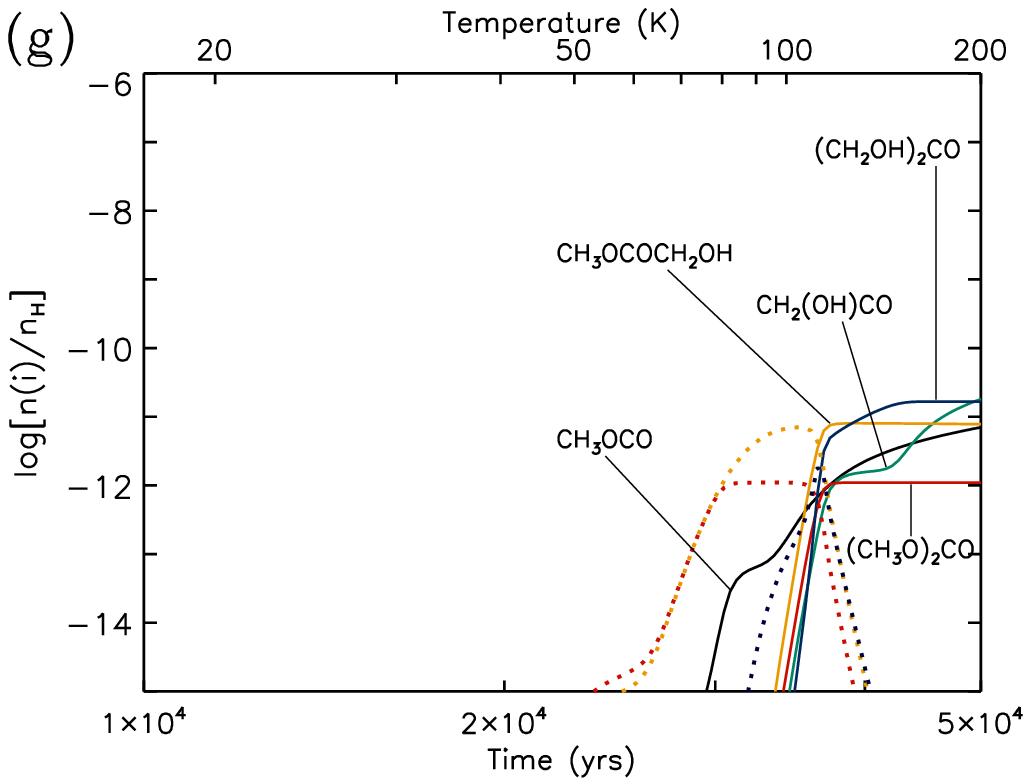}
\plotone{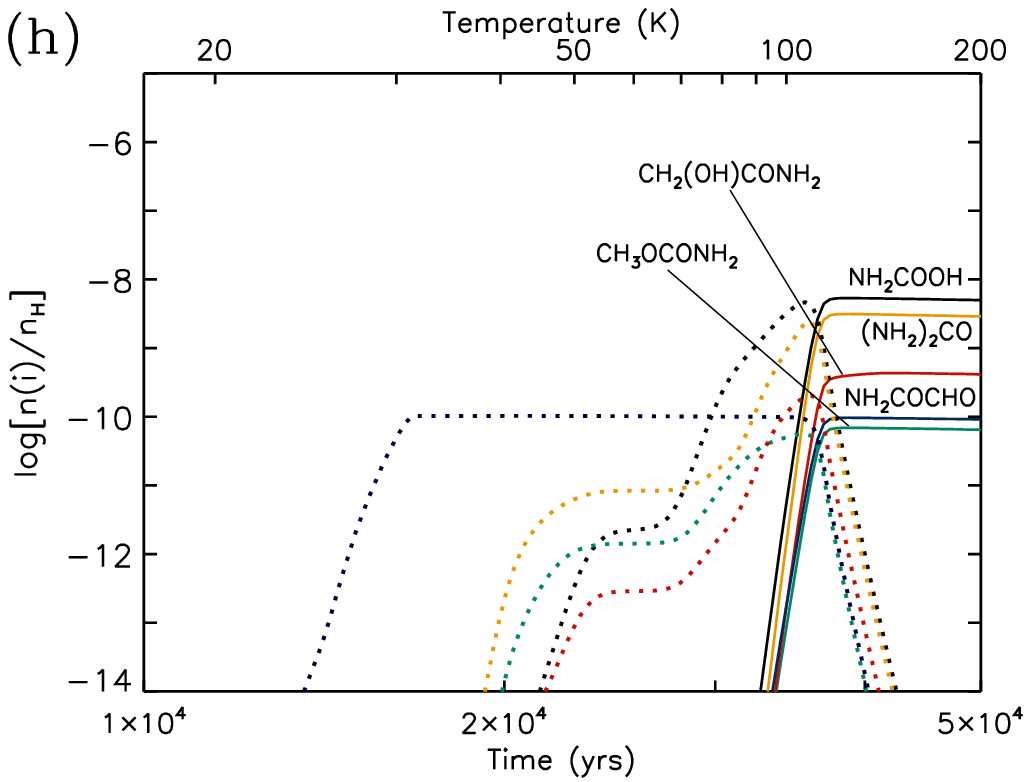}
\plotone{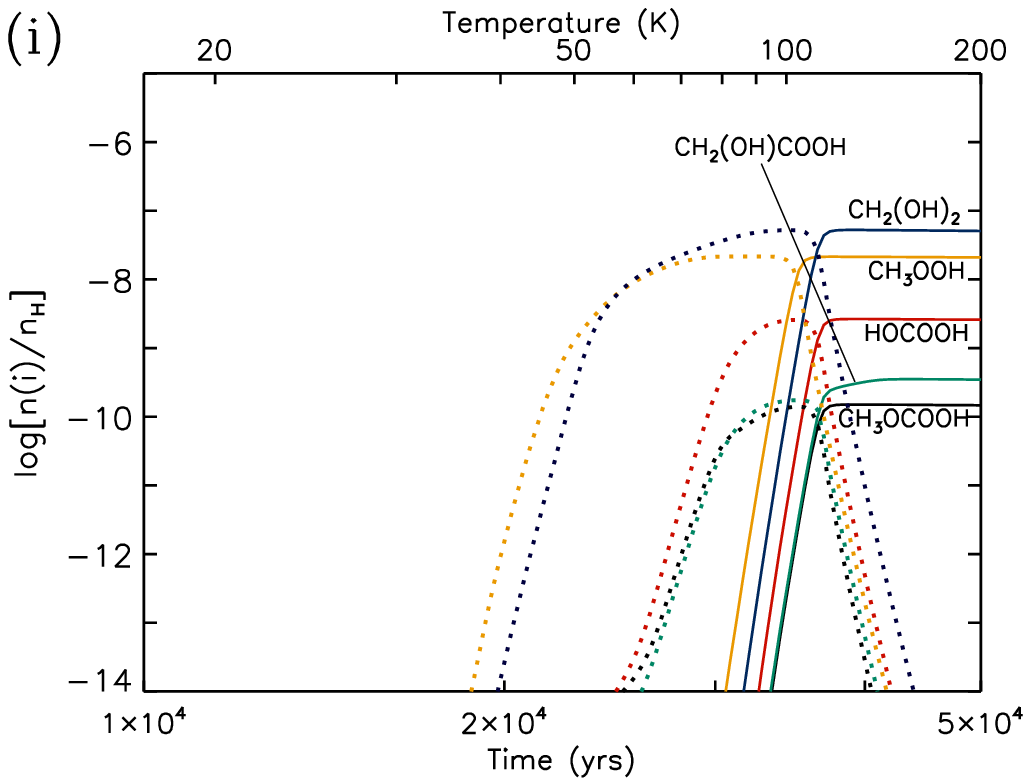}
\plotone{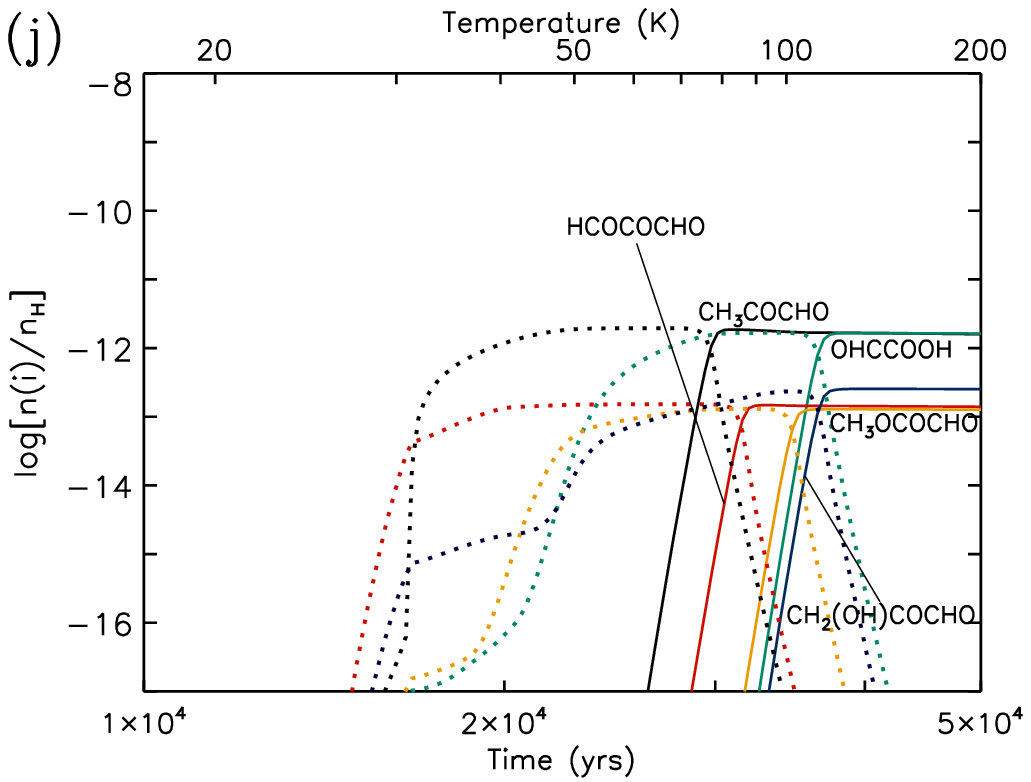}
\plotone{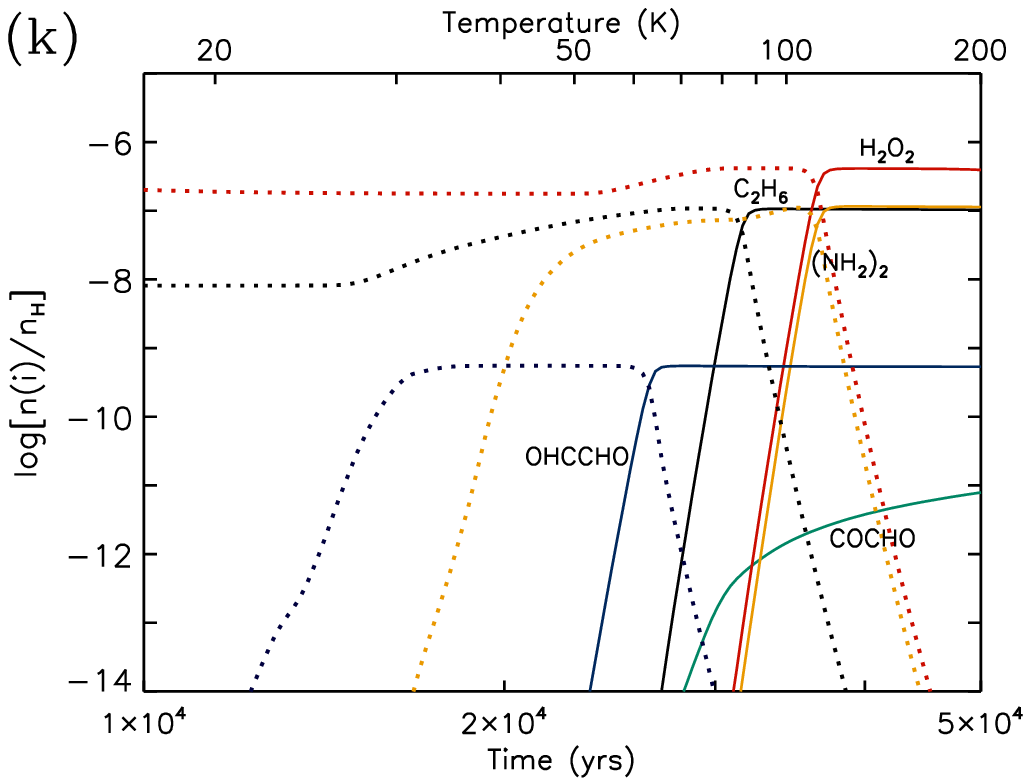}
\end{figure*}

\begin{figure*}
\epsscale{0.4} 
\plotone{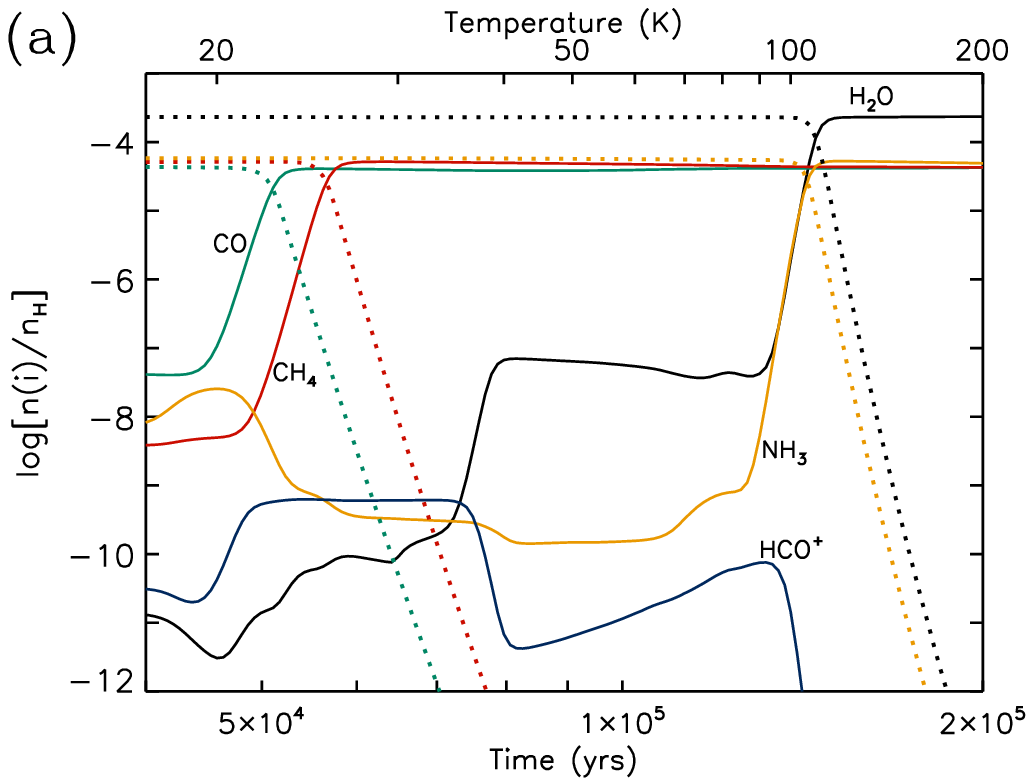} 
\plotone{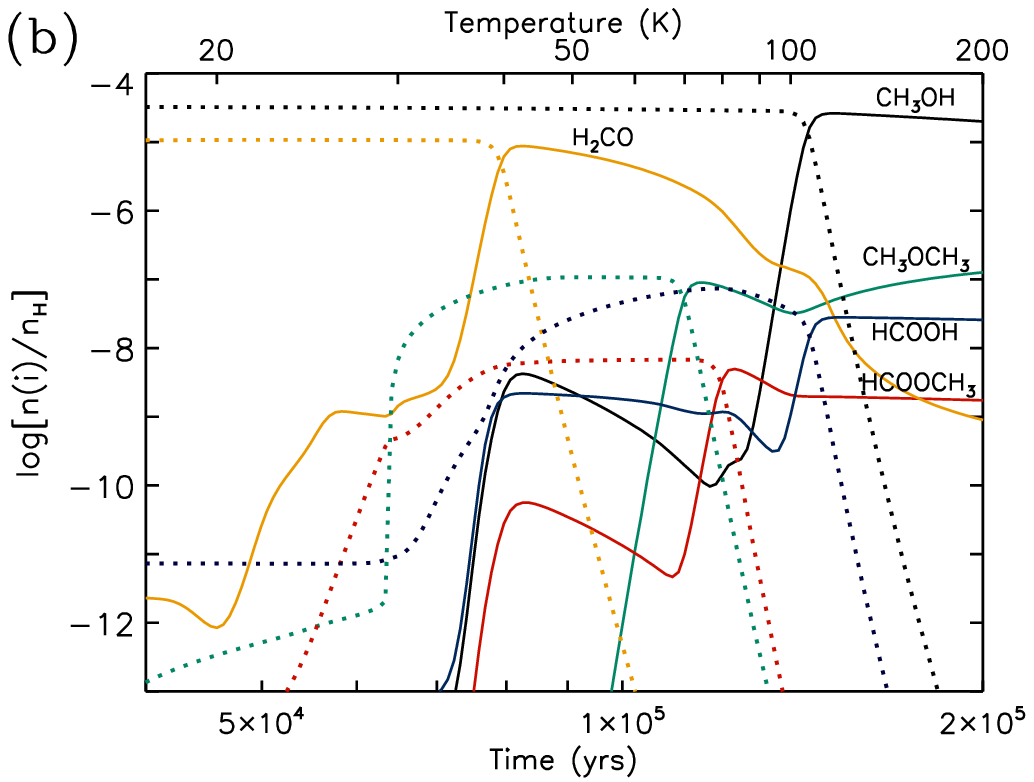}
\plotone{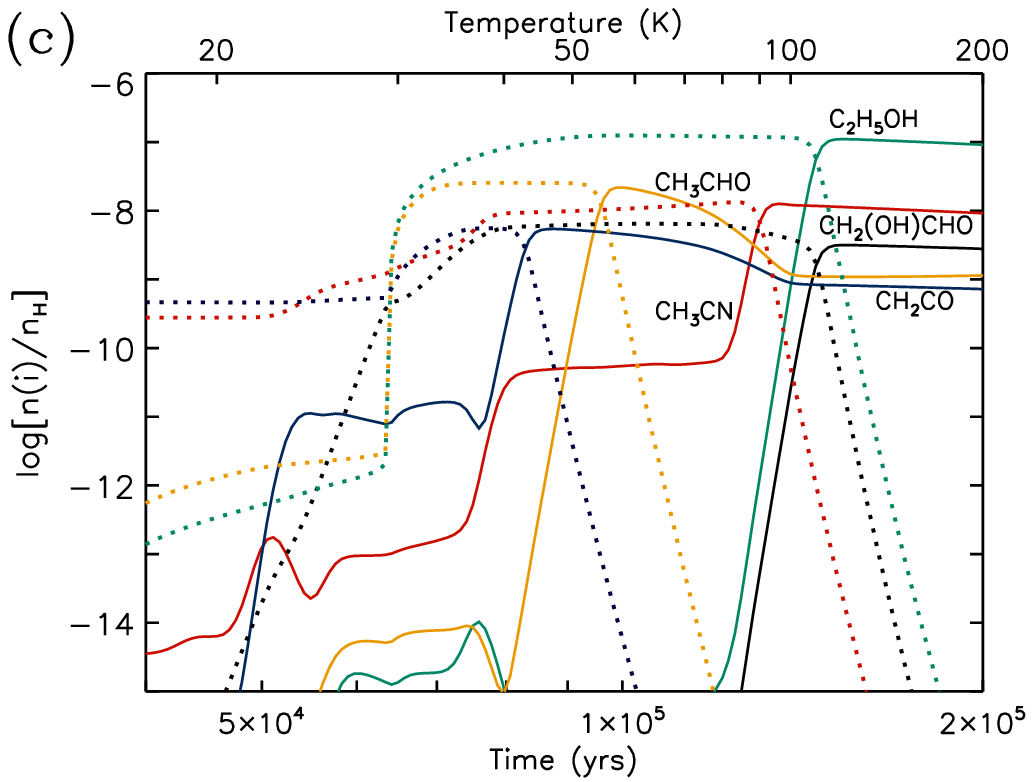} 
\plotone{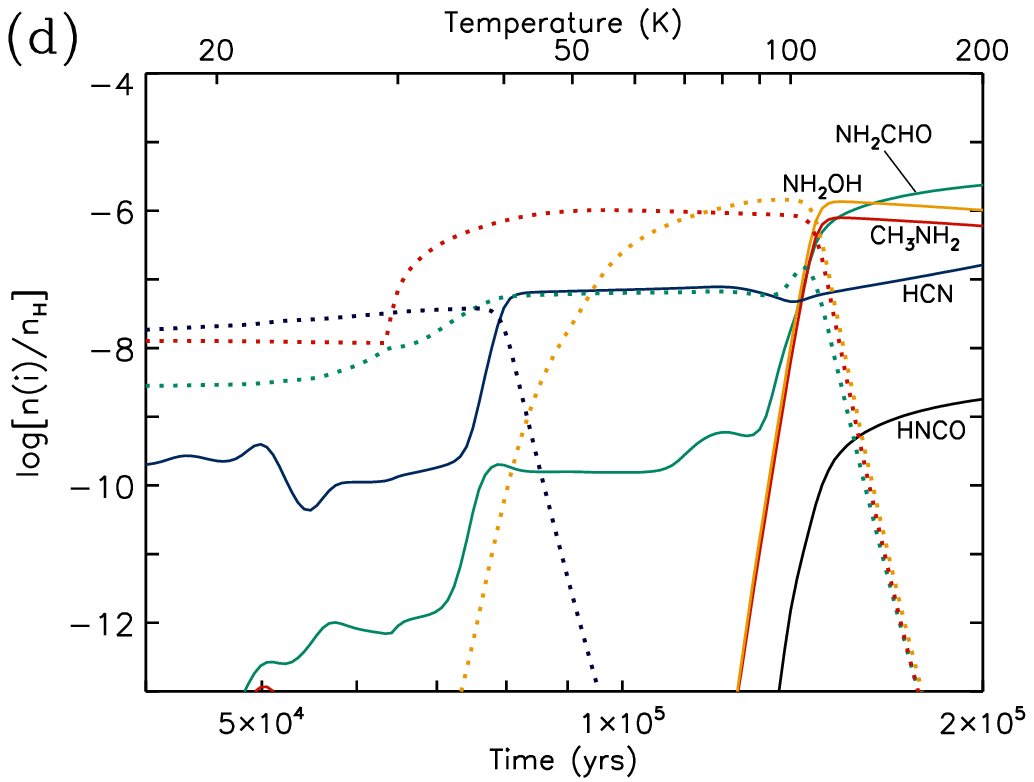}
\plotone{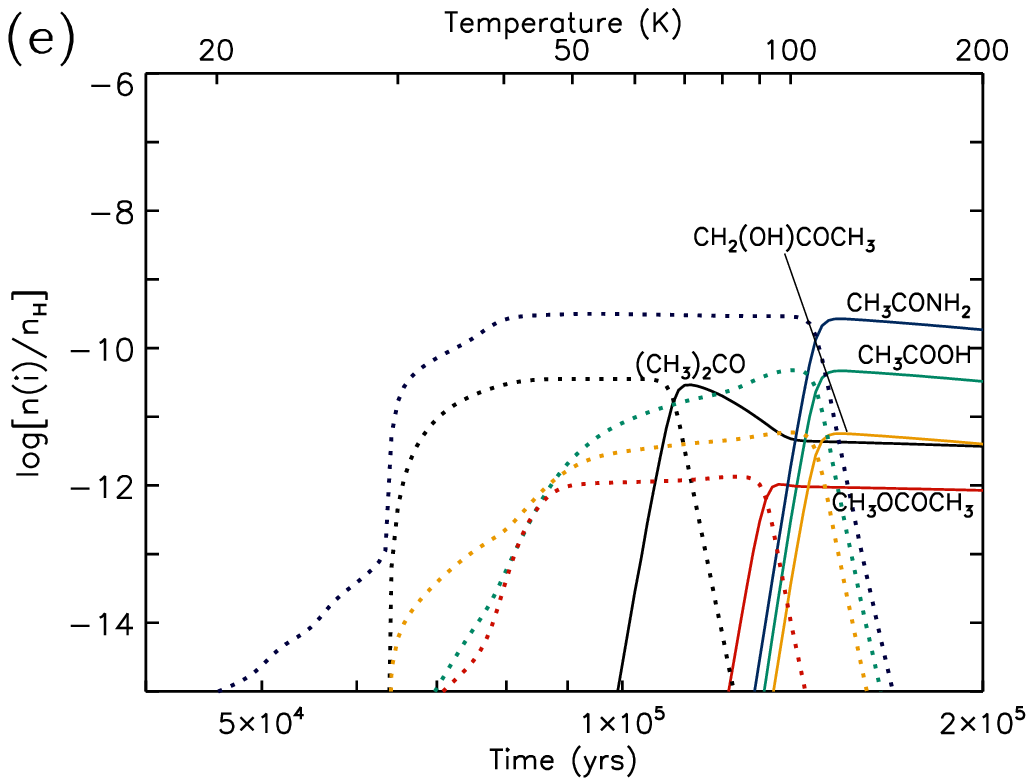} 
\plotone{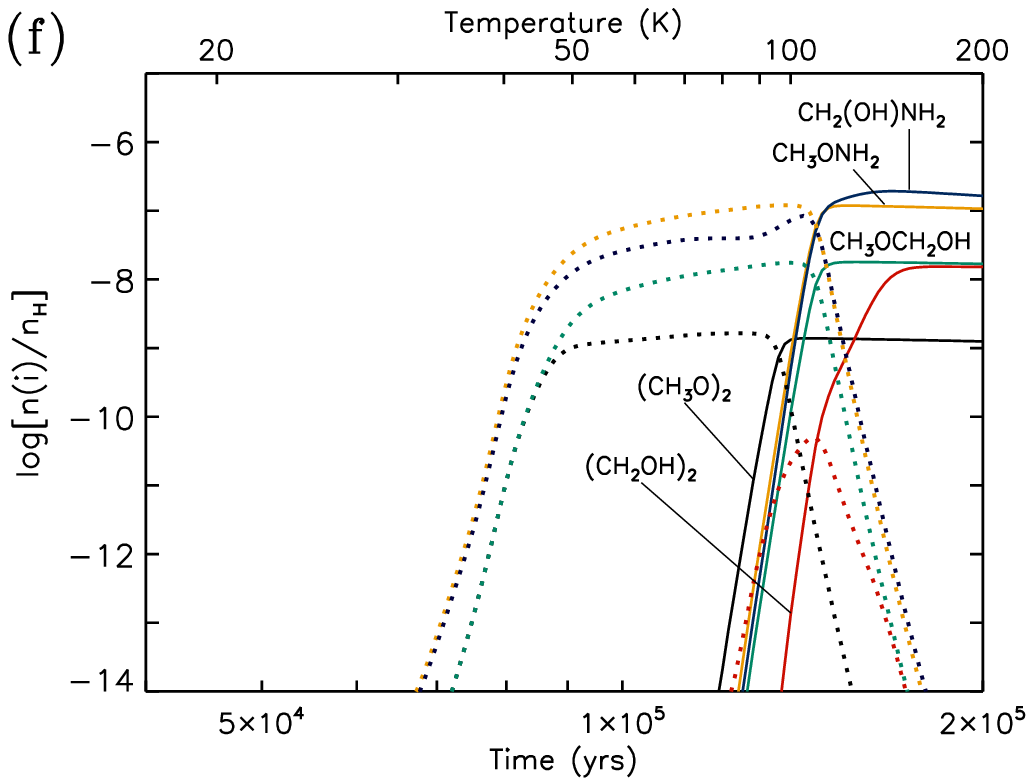} 
\caption{\label{fig2} Fractional
abundances for model M, with a warm-up timescale of $2 \times 10^5$ yr. Solid lines indicate gas-phase species; dotted lines of the same color indicate the grain-surface species. [{\em See the electronic edition of the Journal for panels g--k.}]}
\end{figure*}
\begin{figure*}
\epsscale{0.4}
\plotone{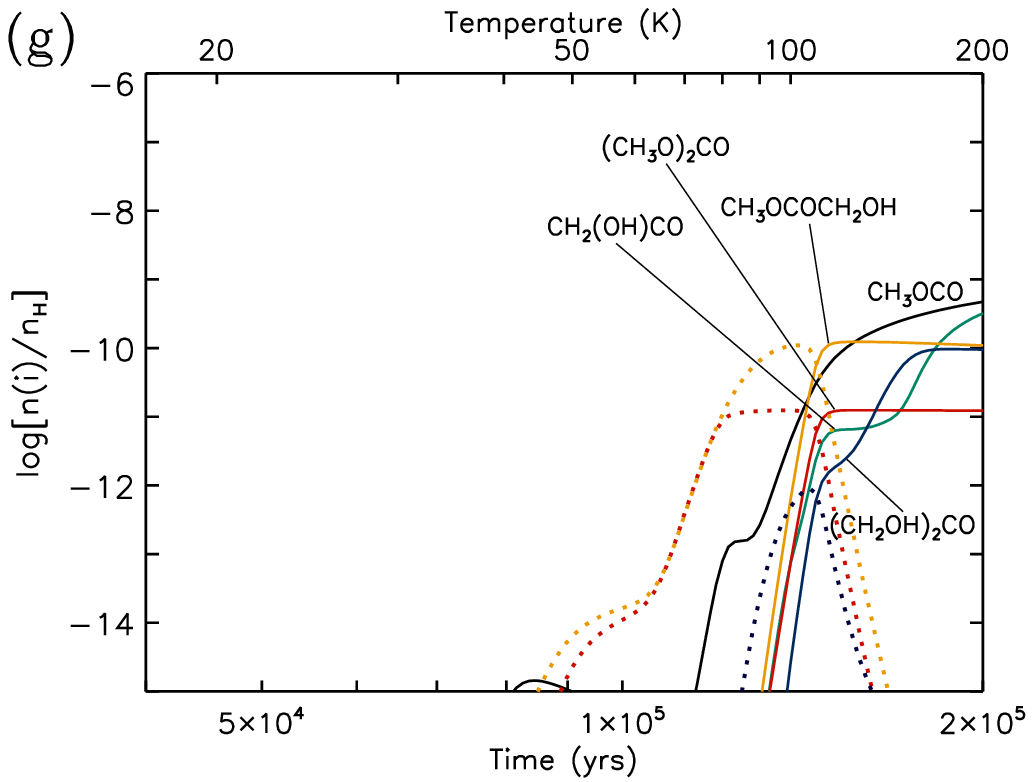}
\plotone{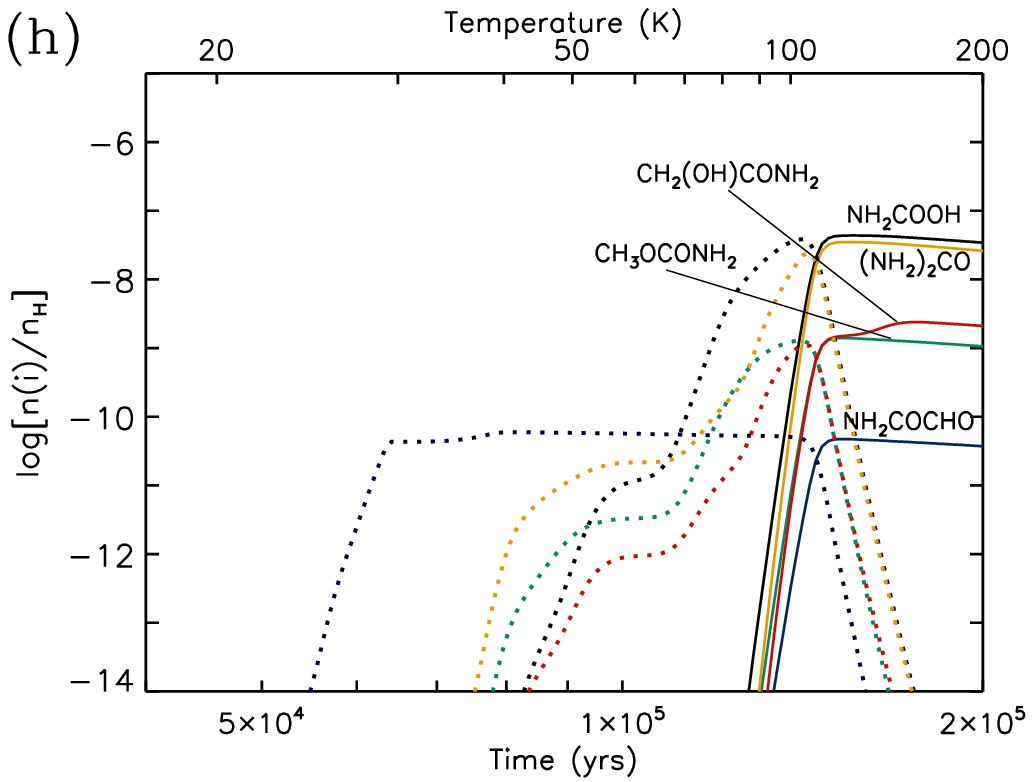}
\plotone{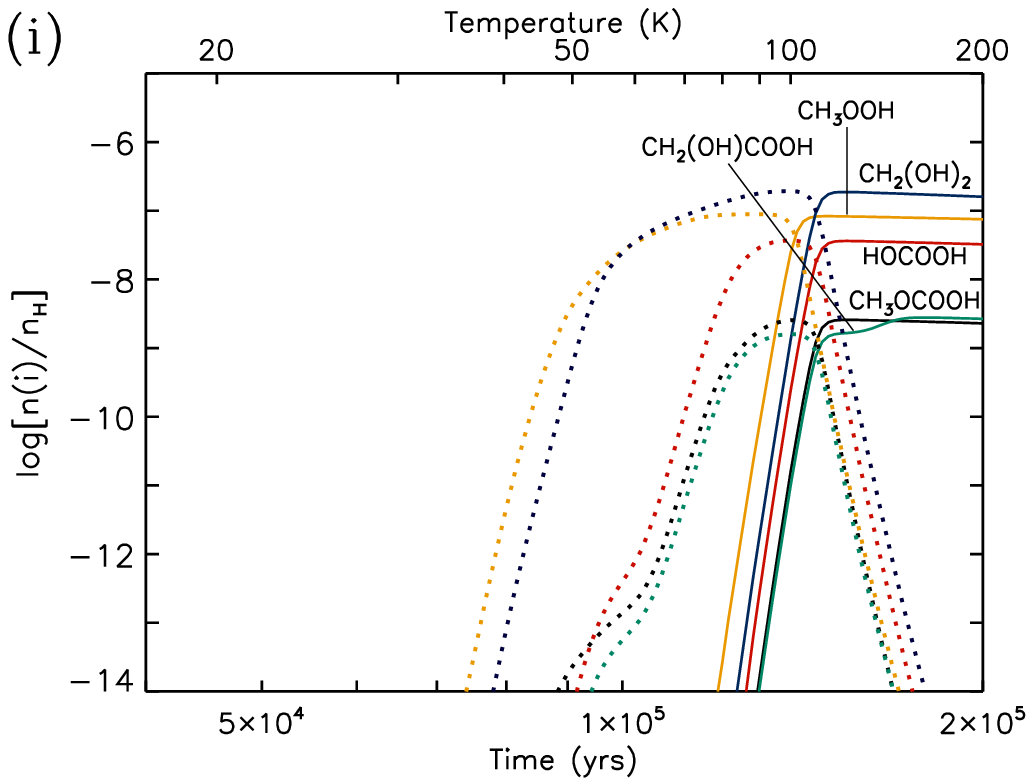}
\plotone{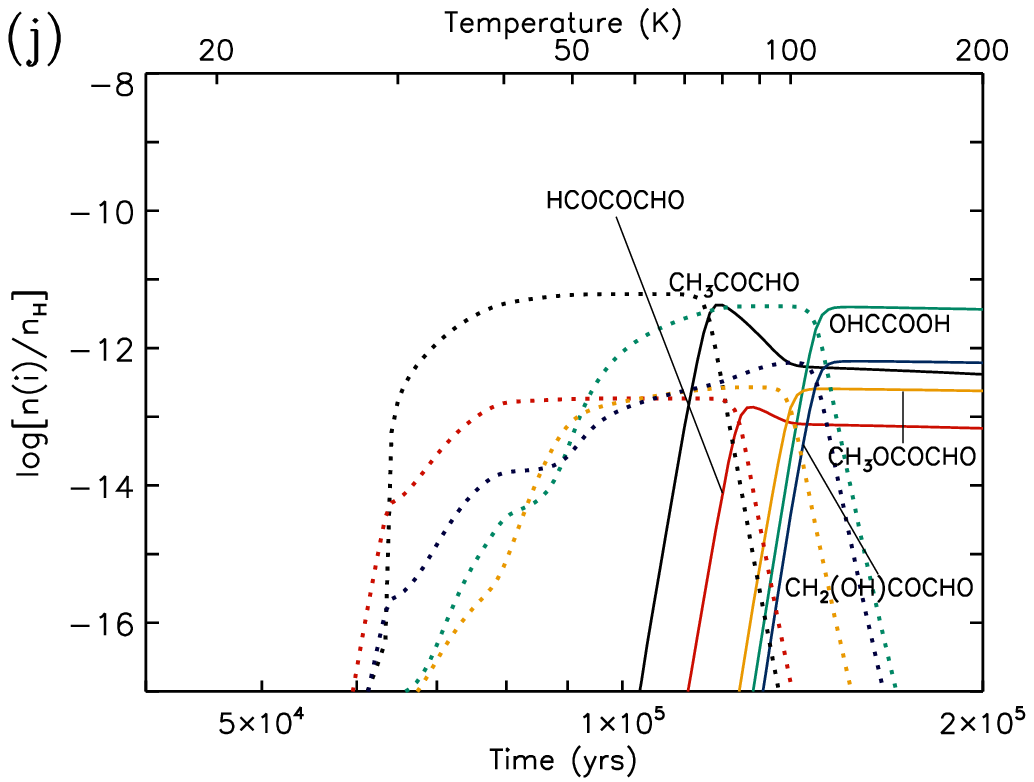}
\plotone{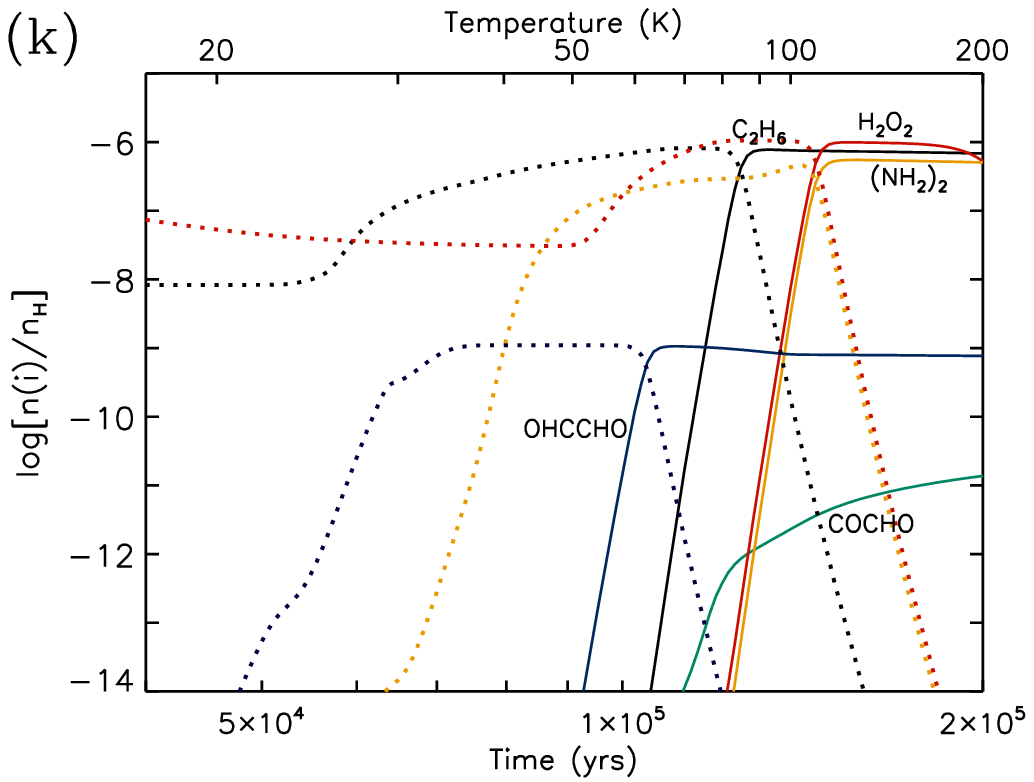}
\end{figure*}

\begin{figure*}
\epsscale{0.4} 
\plotone{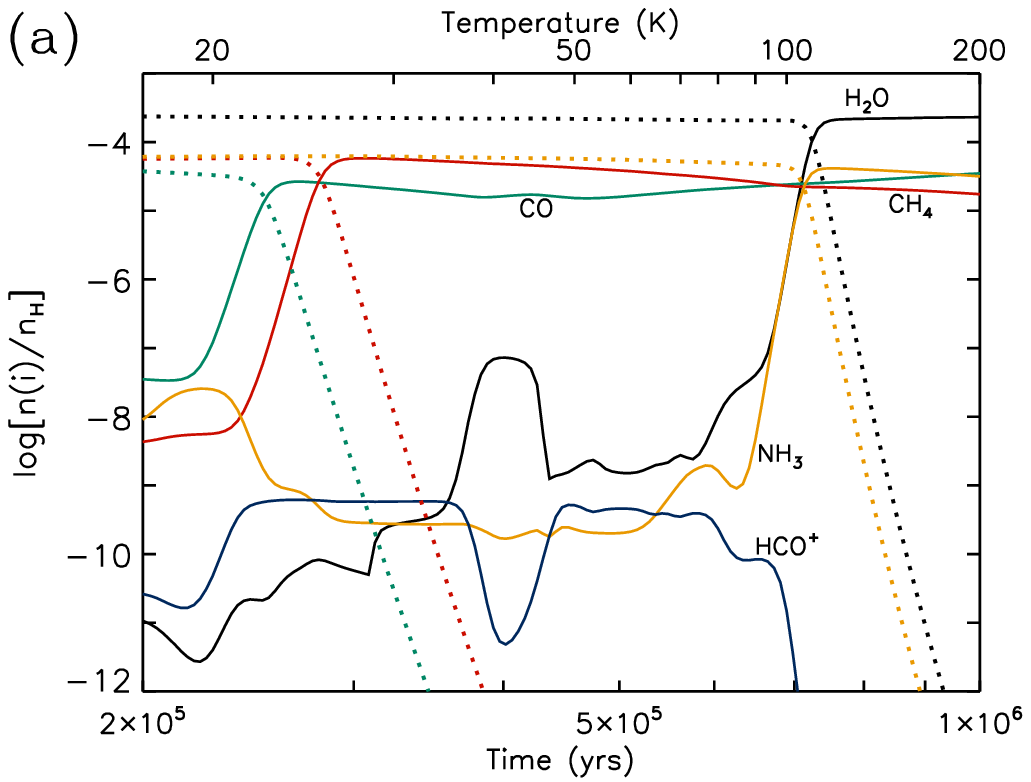} 
\plotone{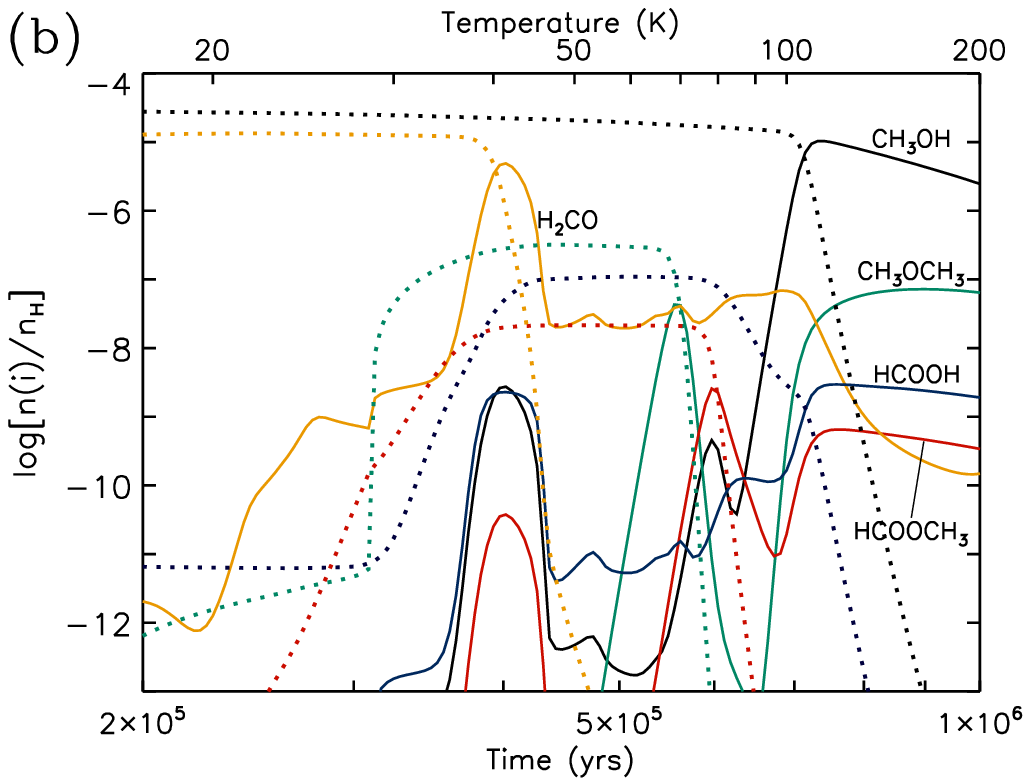}
\plotone{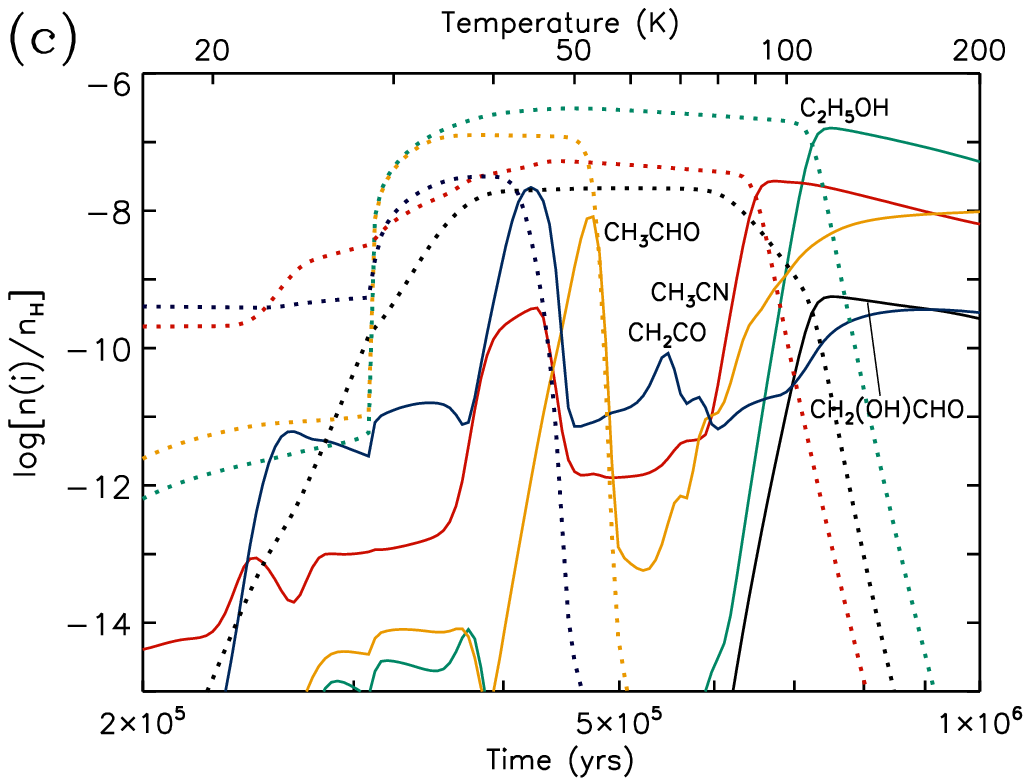} 
\caption{\label{fig3} Fractional
abundances for model S, with a warm-up timescale of $1 \times 10^6$ yr. Solid lines indicate gas-phase species; dotted lines of the same color indicate the grain-surface species. [{\em See the electronic edition of the Journal for panels d--k.}]}
\end{figure*}
\begin{figure*}
\epsscale{0.4}
\plotone{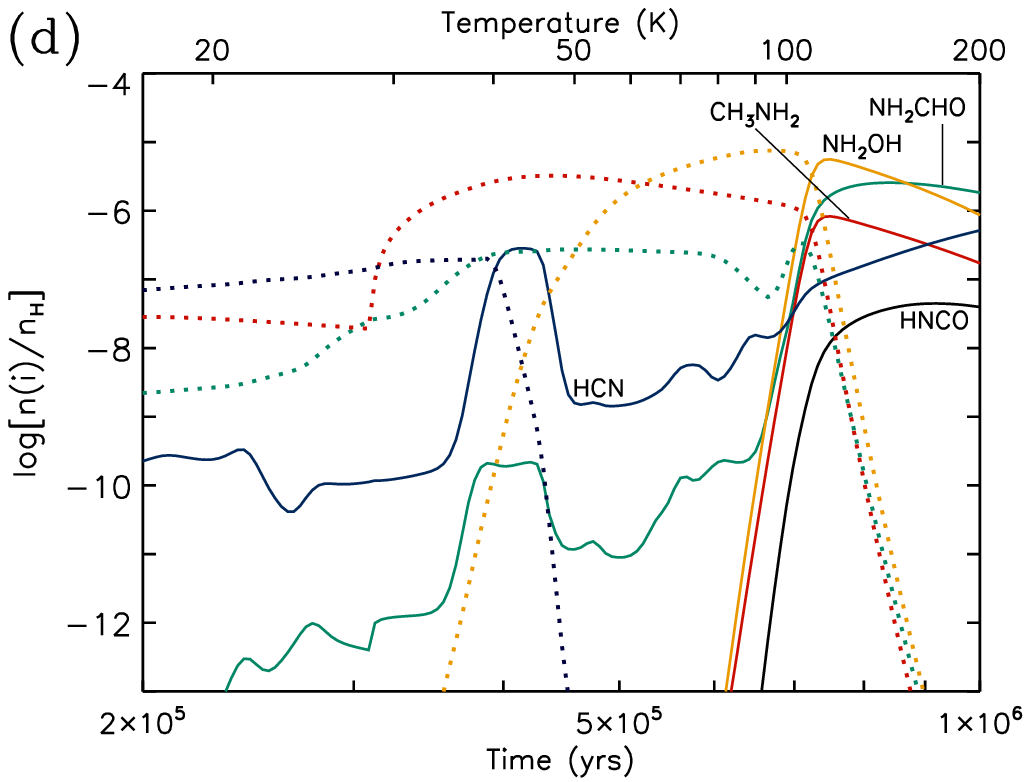}
\plotone{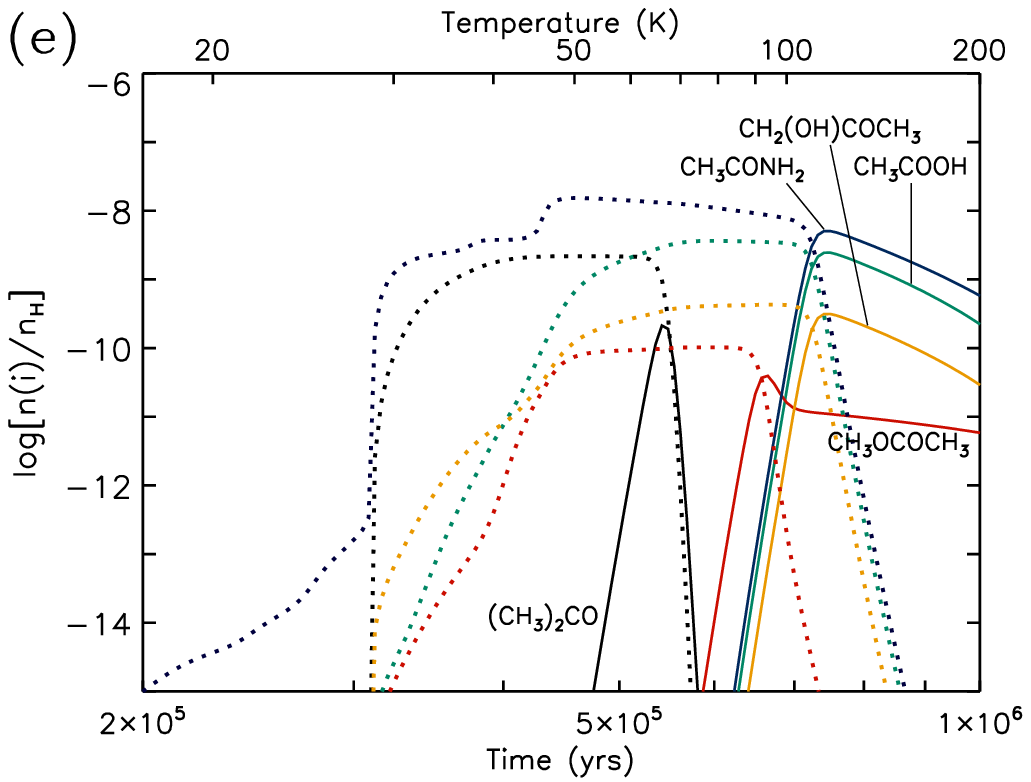} 
\plotone{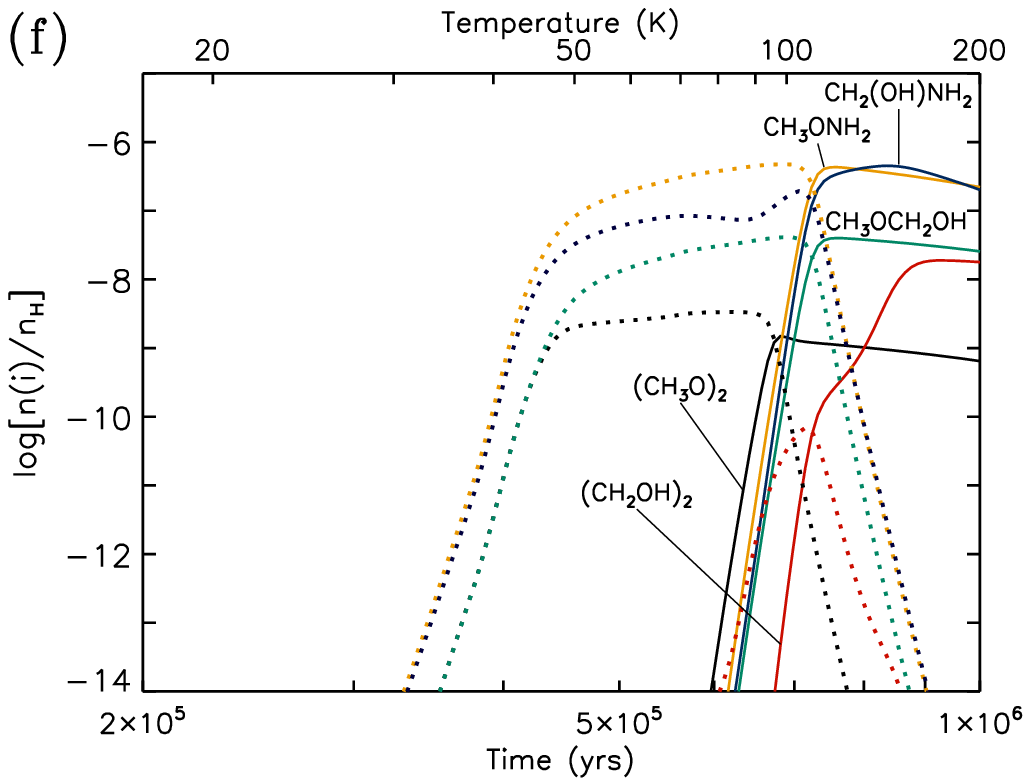} 
\plotone{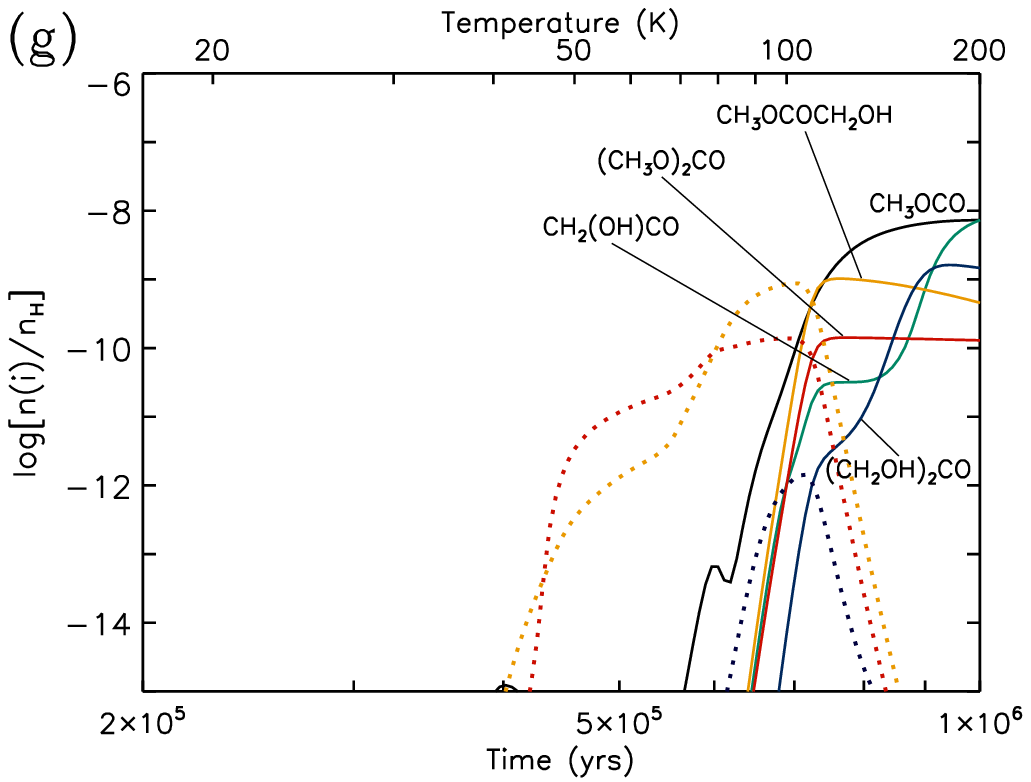}
\plotone{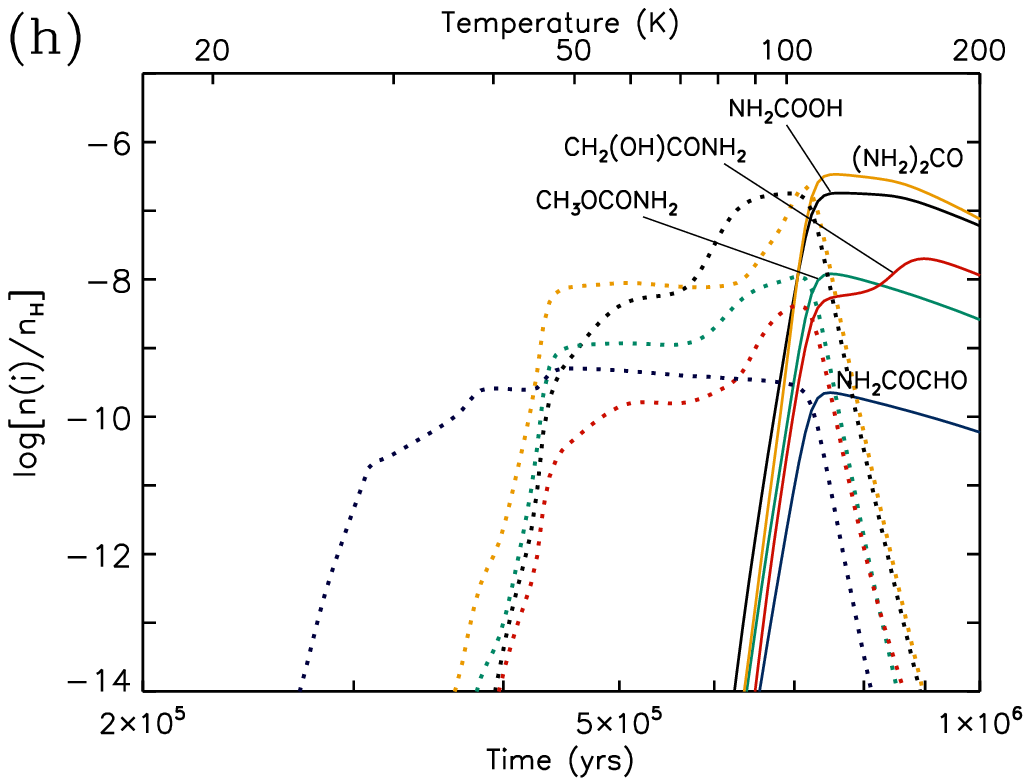}
\plotone{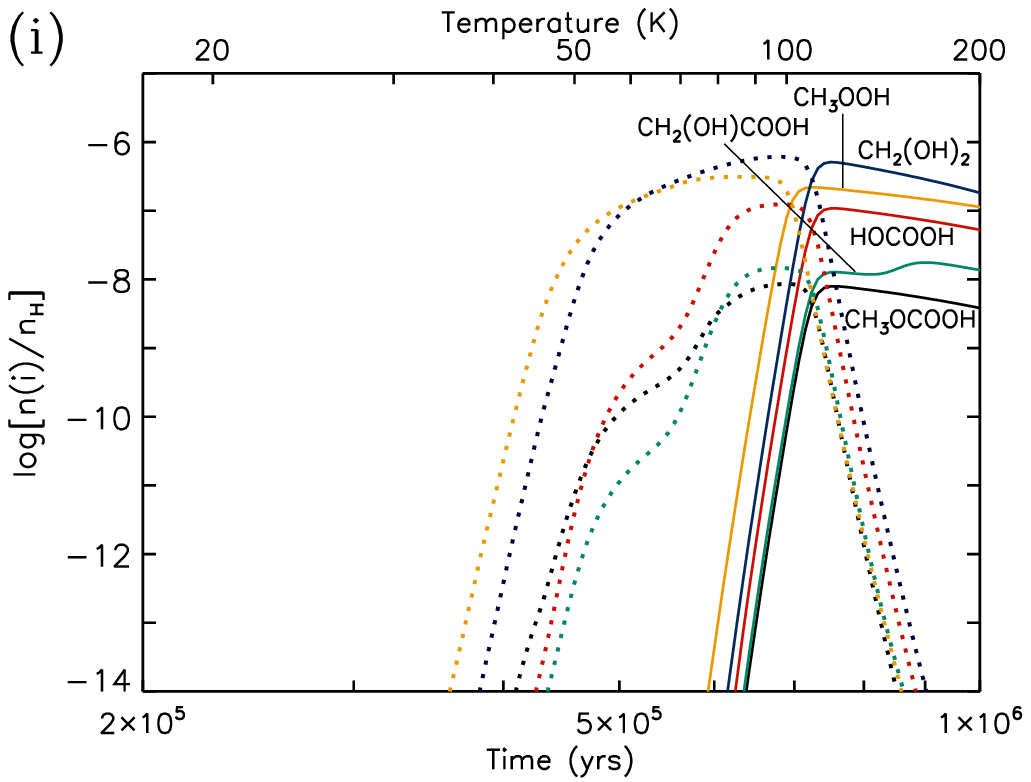}
\plotone{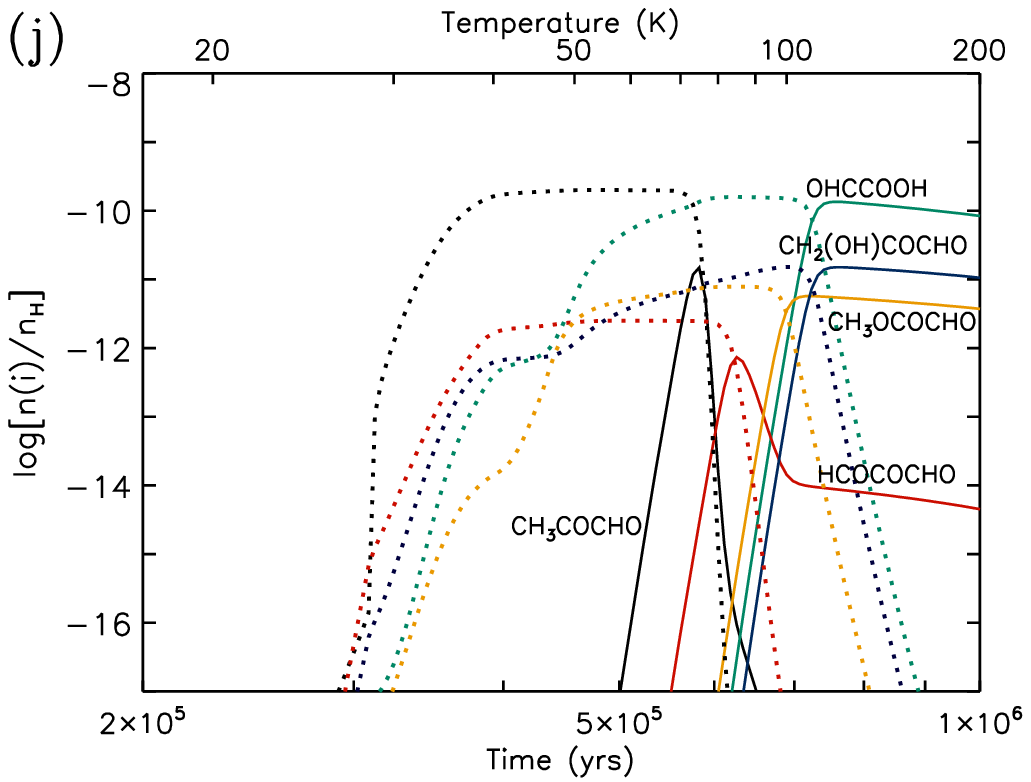}
\plotone{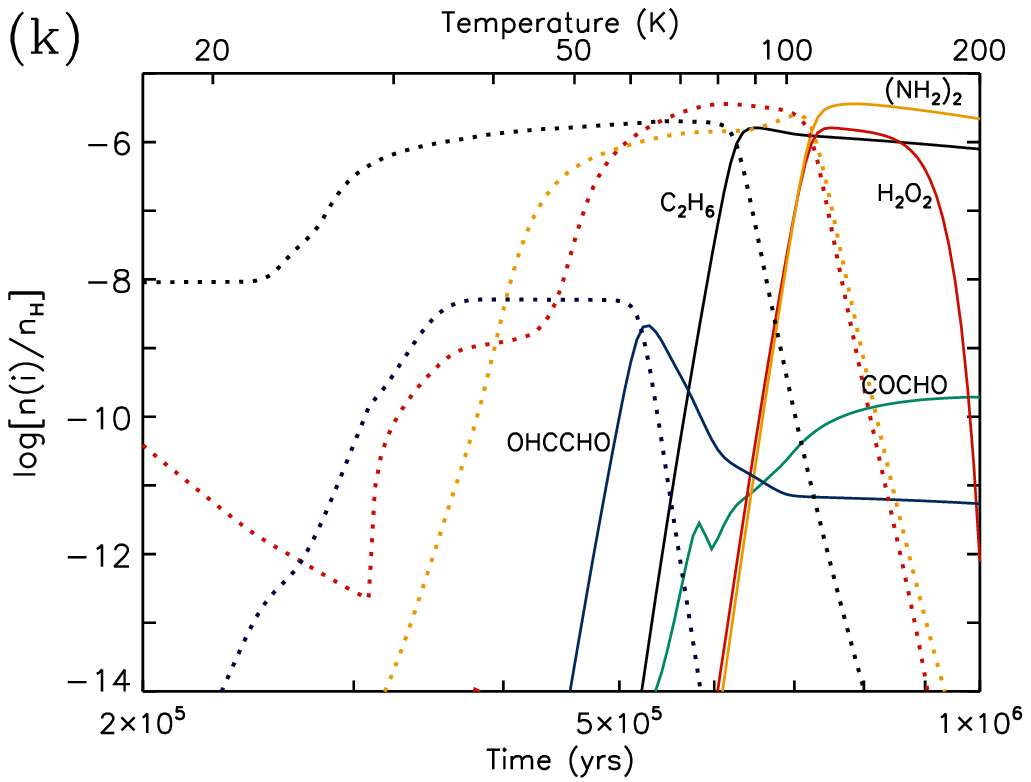}
\end{figure*}

\begin{figure*}
\epsscale{0.4} 
\plotone{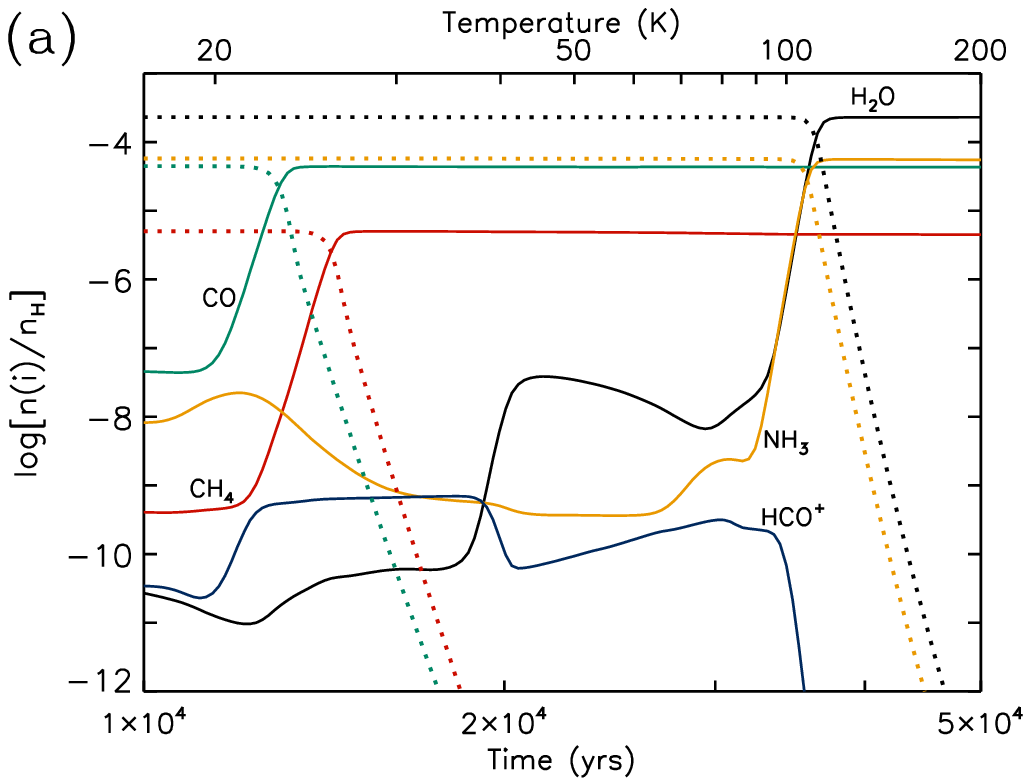} 
\plotone{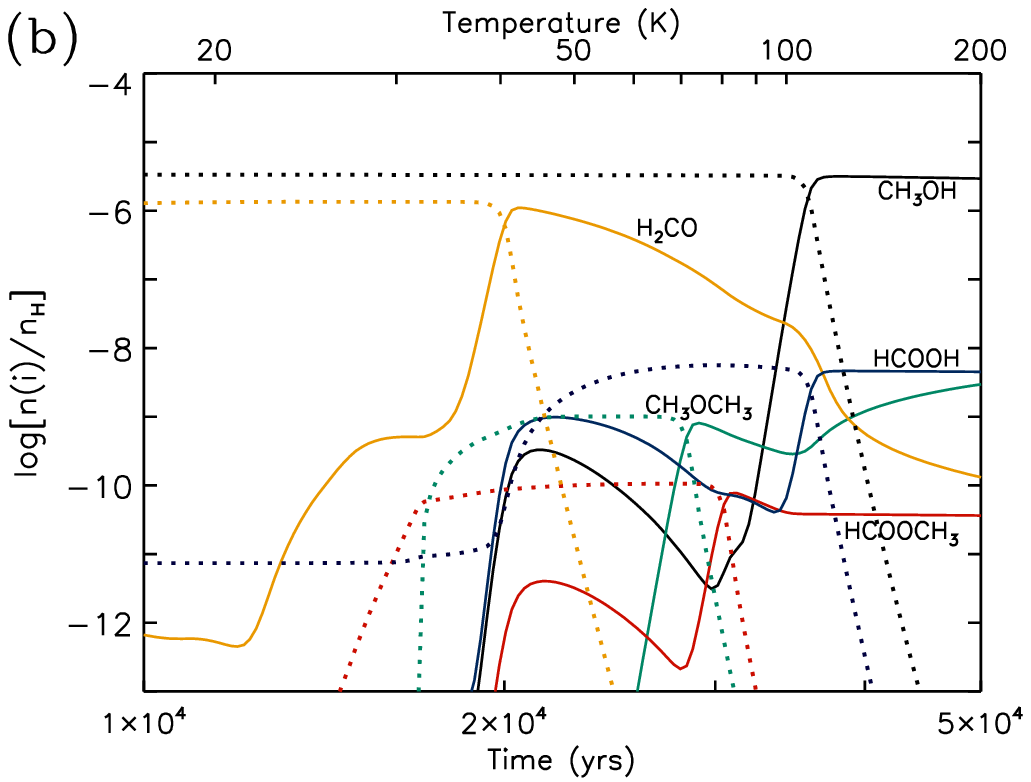}
\plotone{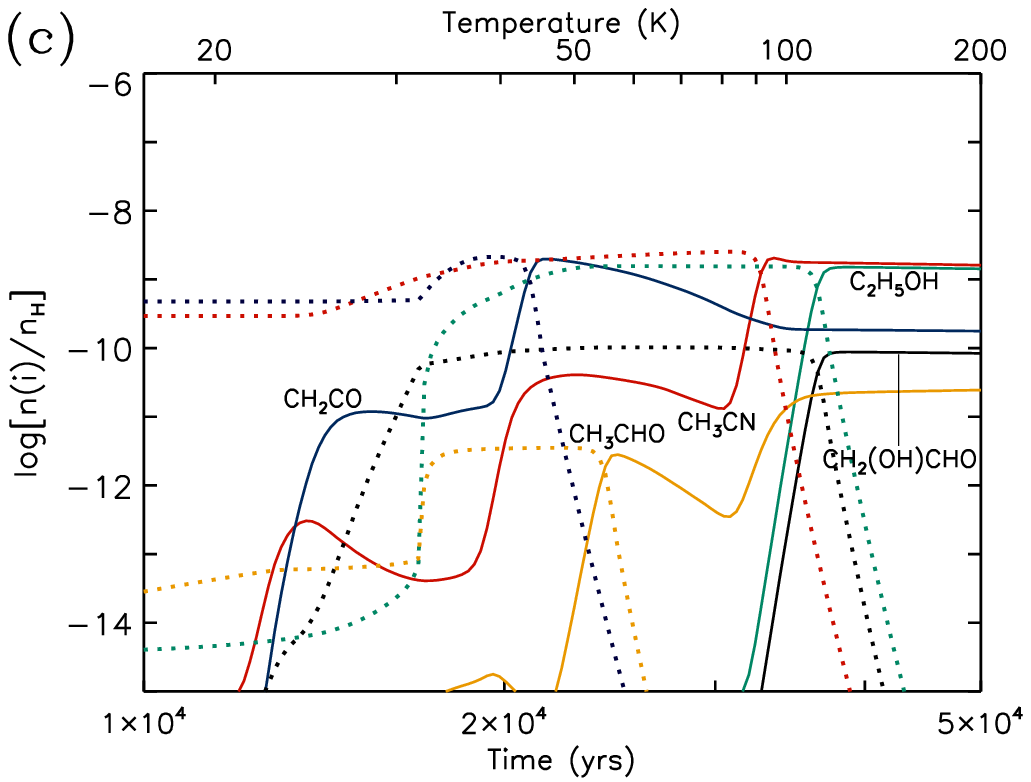} 
\caption{\label{fig4} Fractional
abundances for model F(ice), with a warm-up timescale of $5 \times 10^4$ yr. Solid lines indicate gas-phase species; dotted lines of the same color indicate the grain-surface species. [{\em See the electronic edition of the Journal for panels d--k.}]}
\end{figure*}
\begin{figure*}
\epsscale{0.4}
\plotone{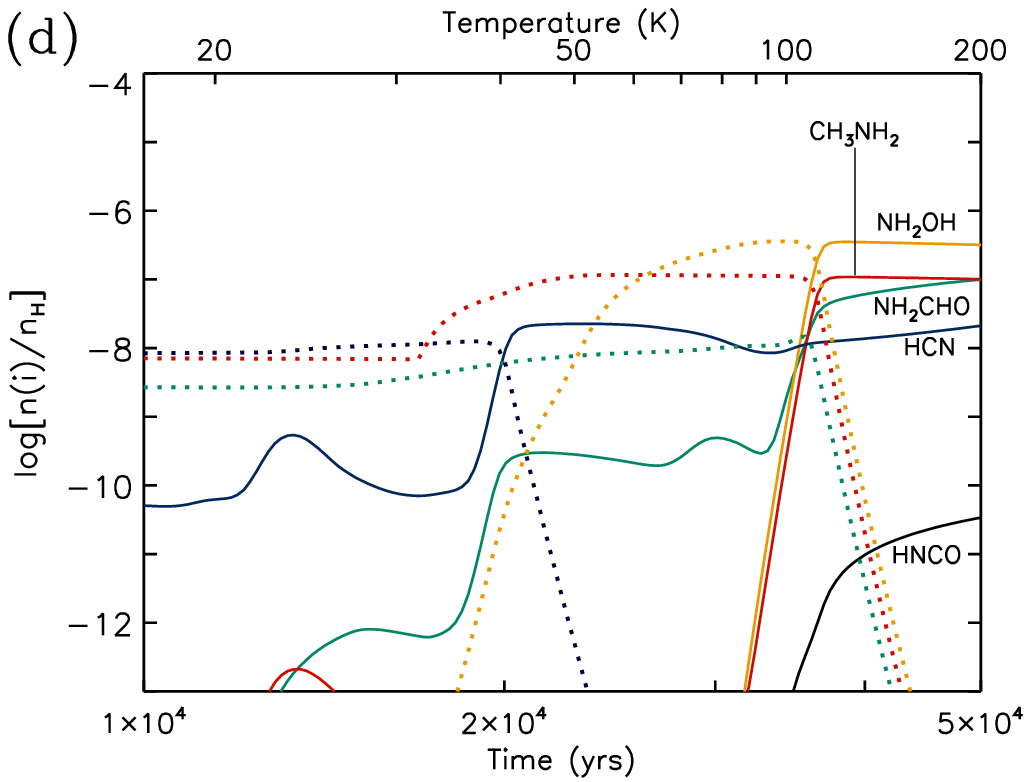}
\plotone{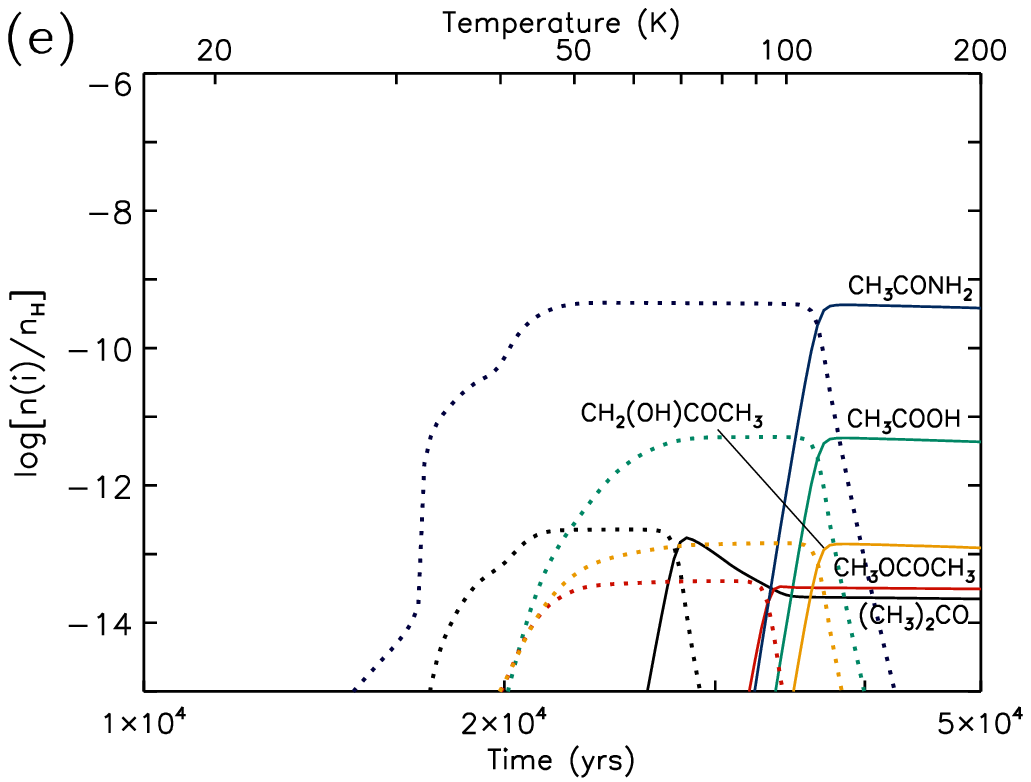} 
\plotone{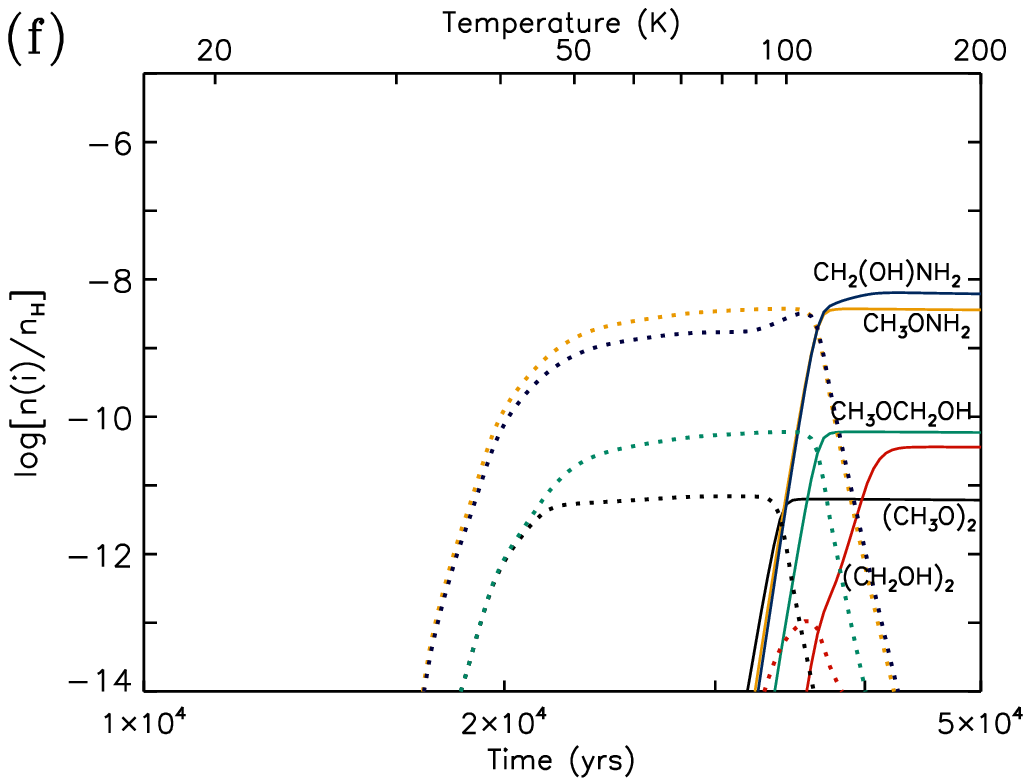} 
\plotone{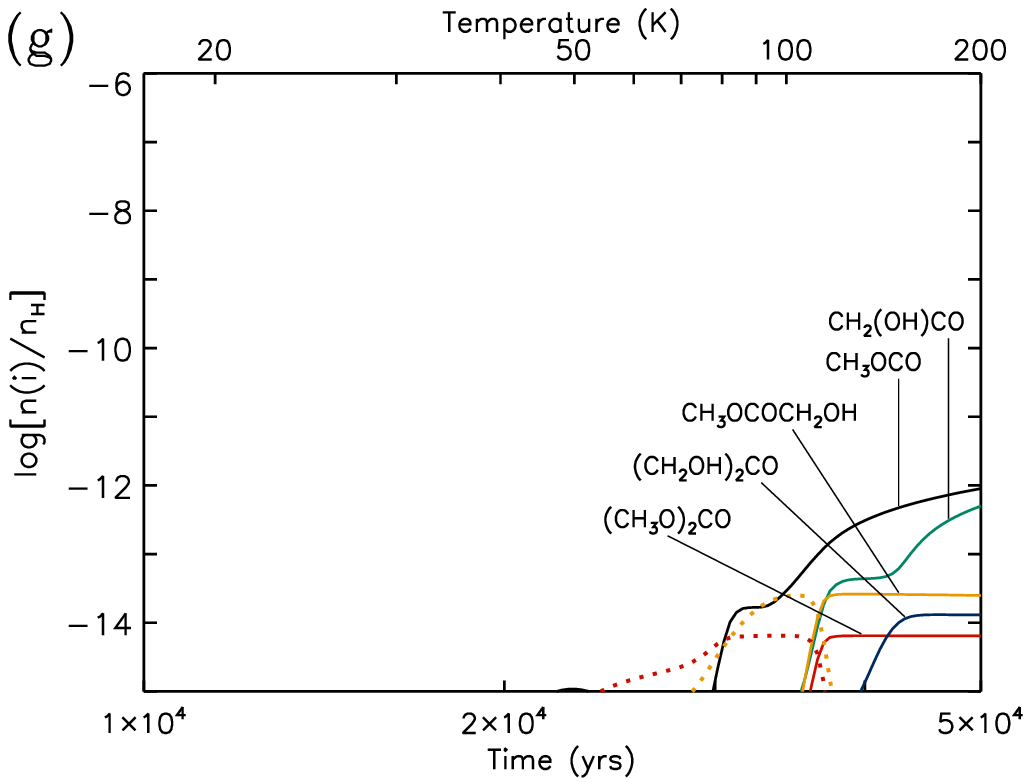}
\plotone{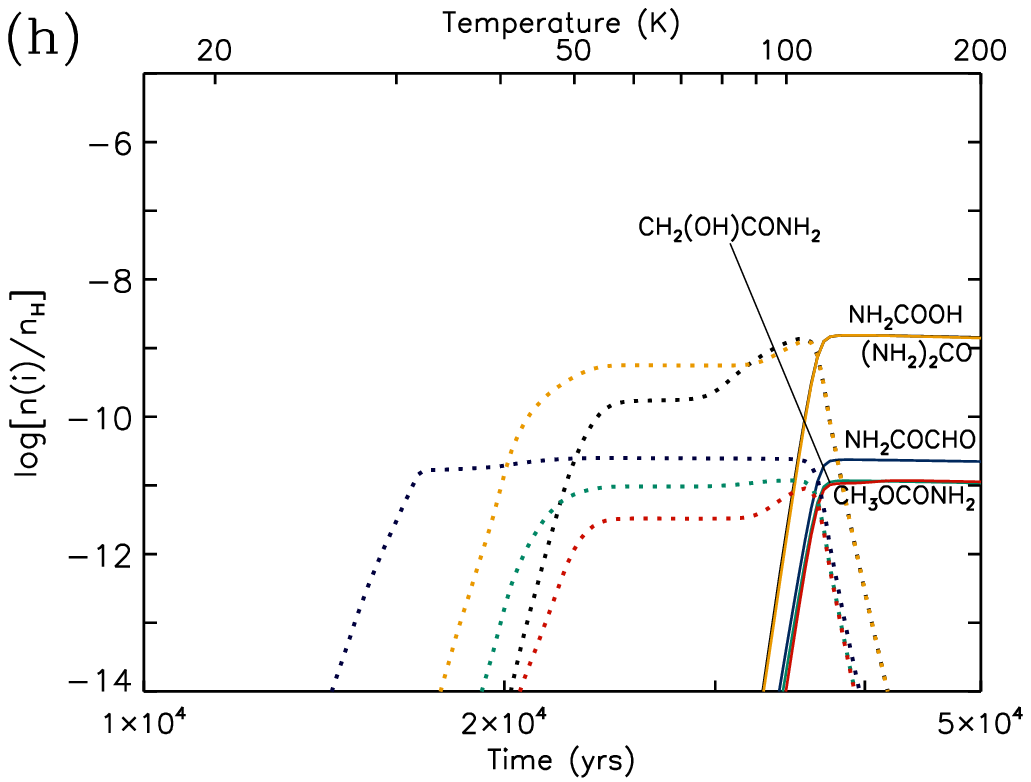}
\plotone{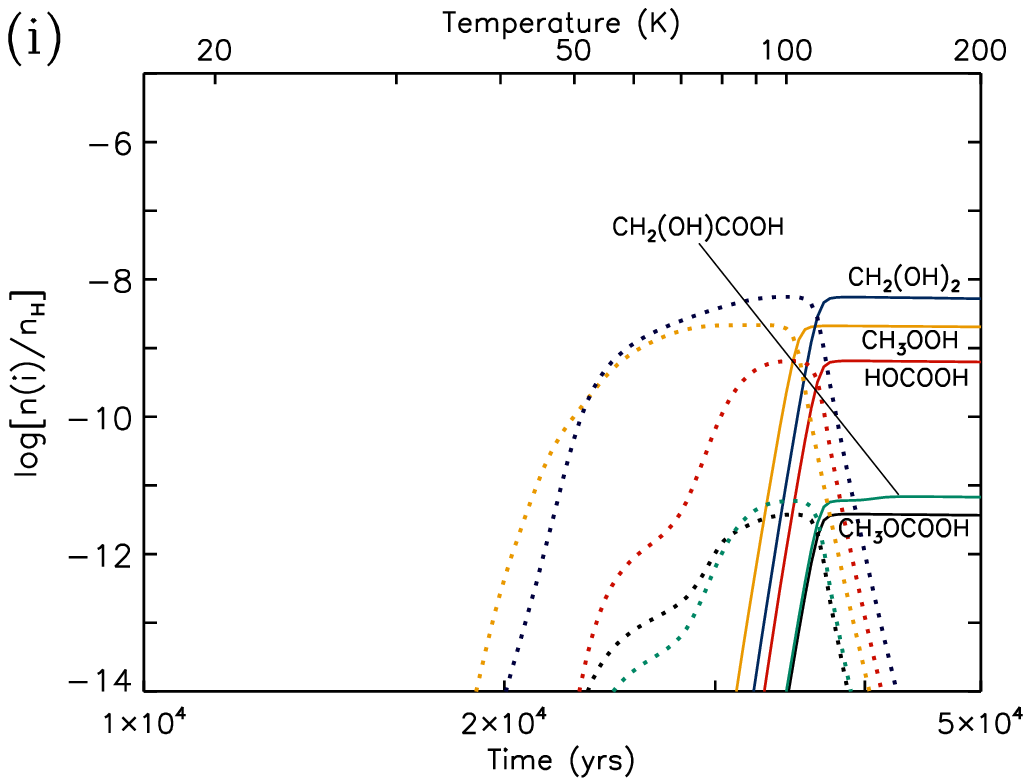}
\plotone{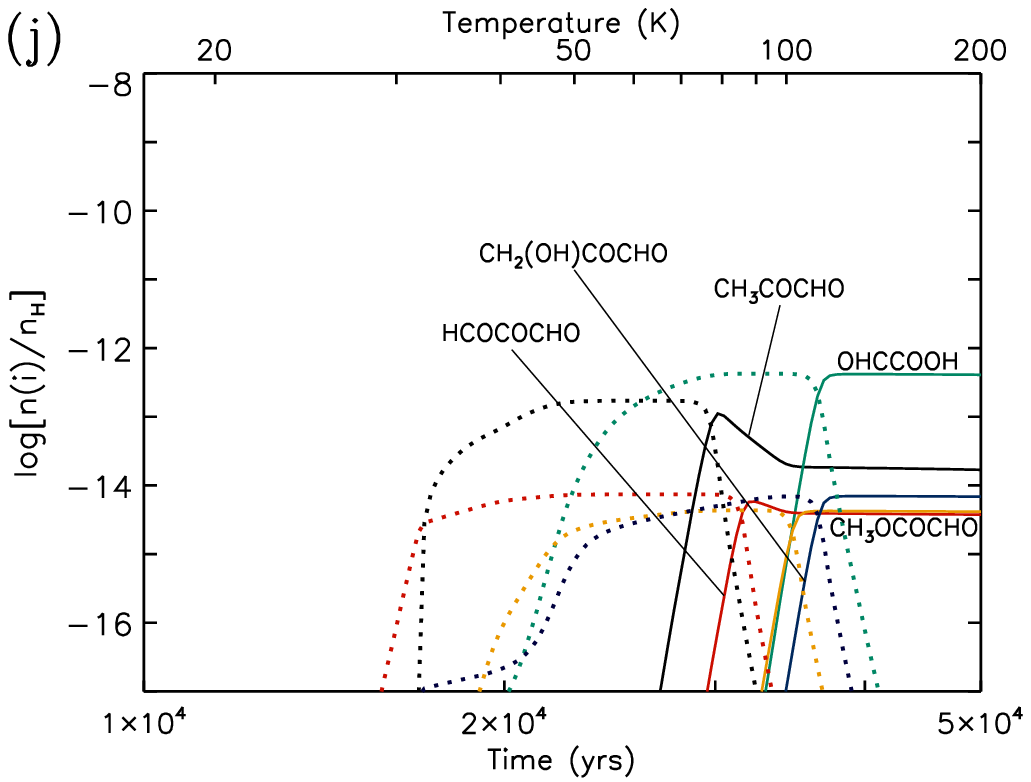}
\plotone{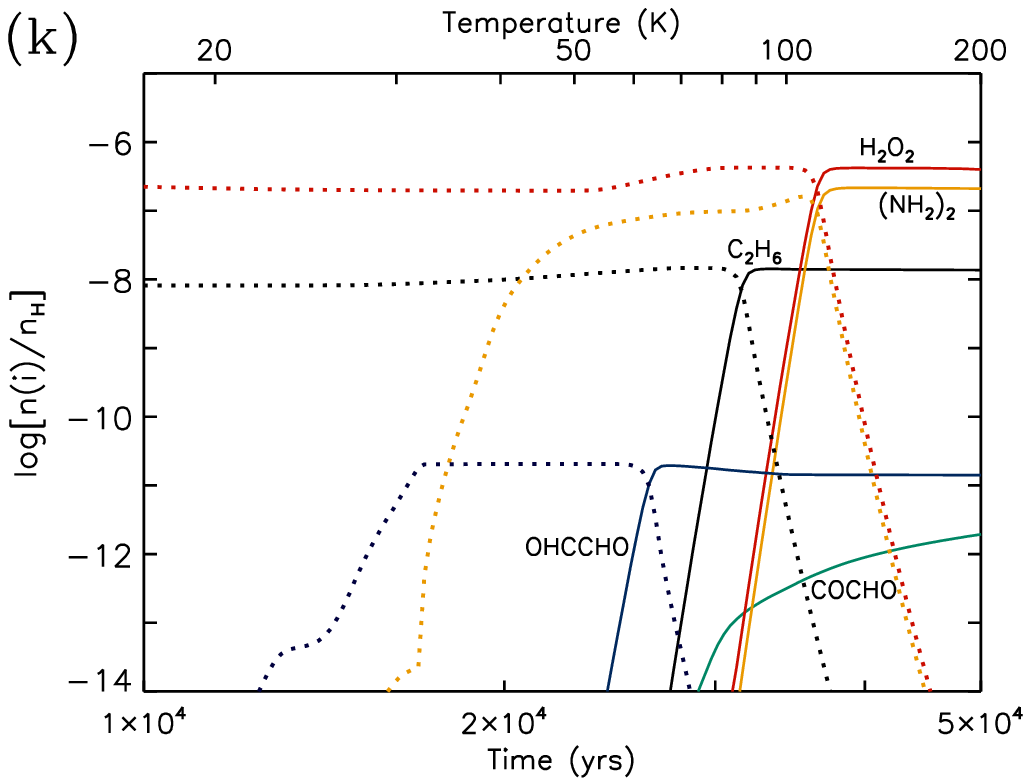}
\end{figure*}

\begin{figure*}
\epsscale{0.4} 
\plotone{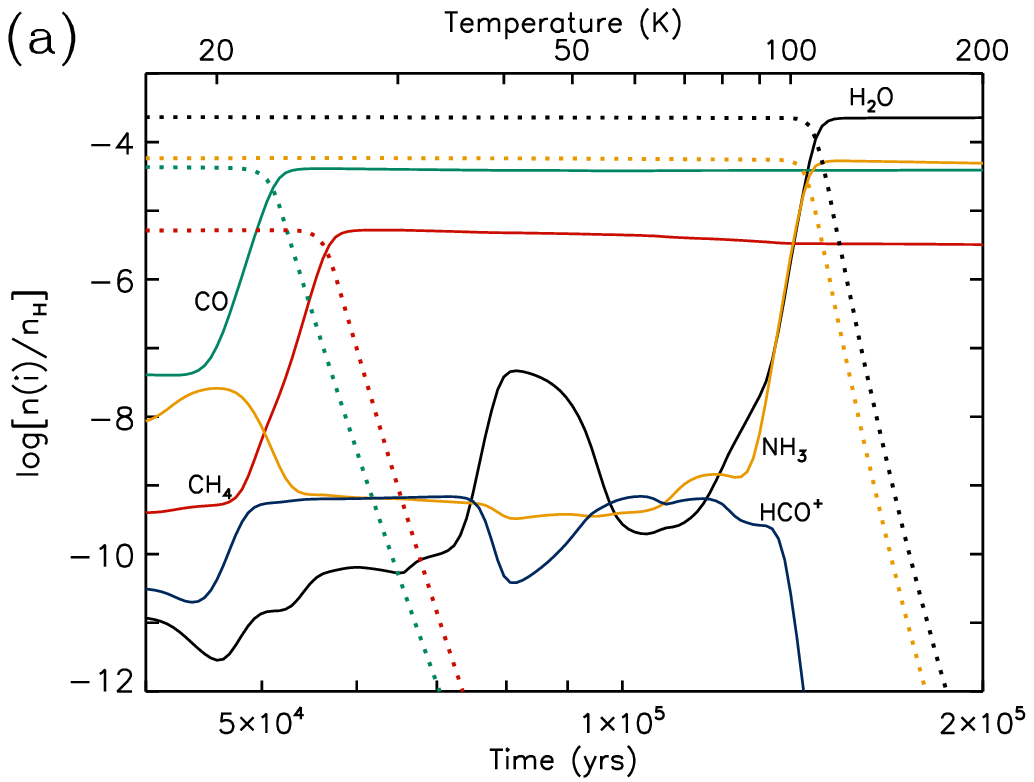} 
\plotone{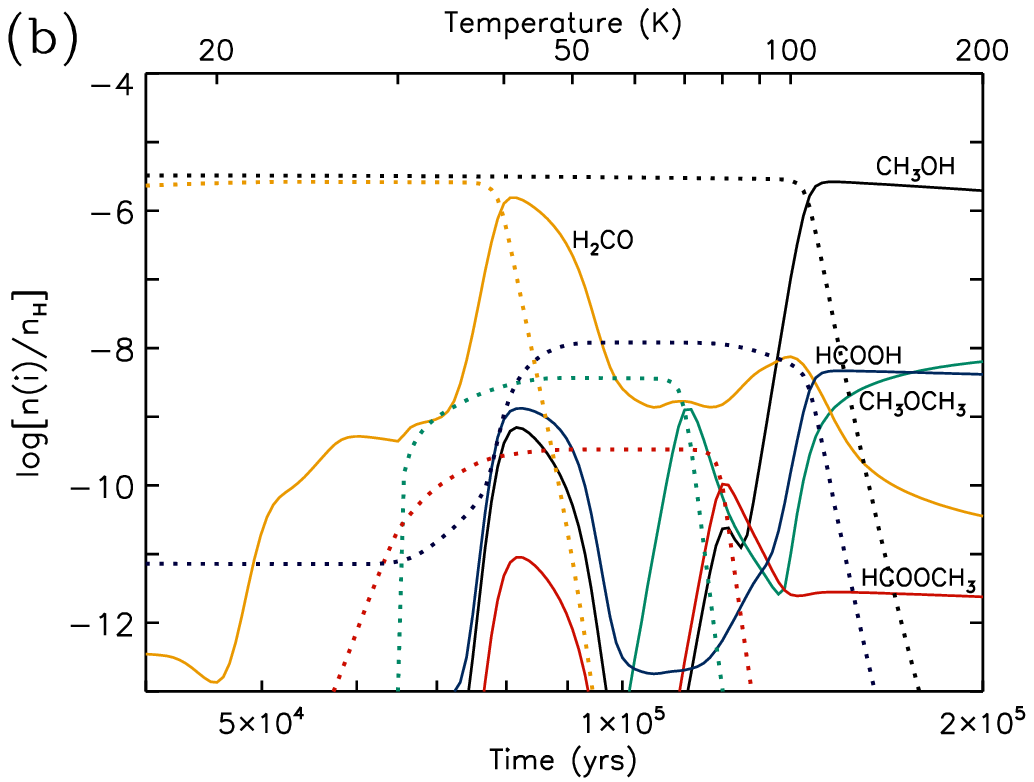}
\plotone{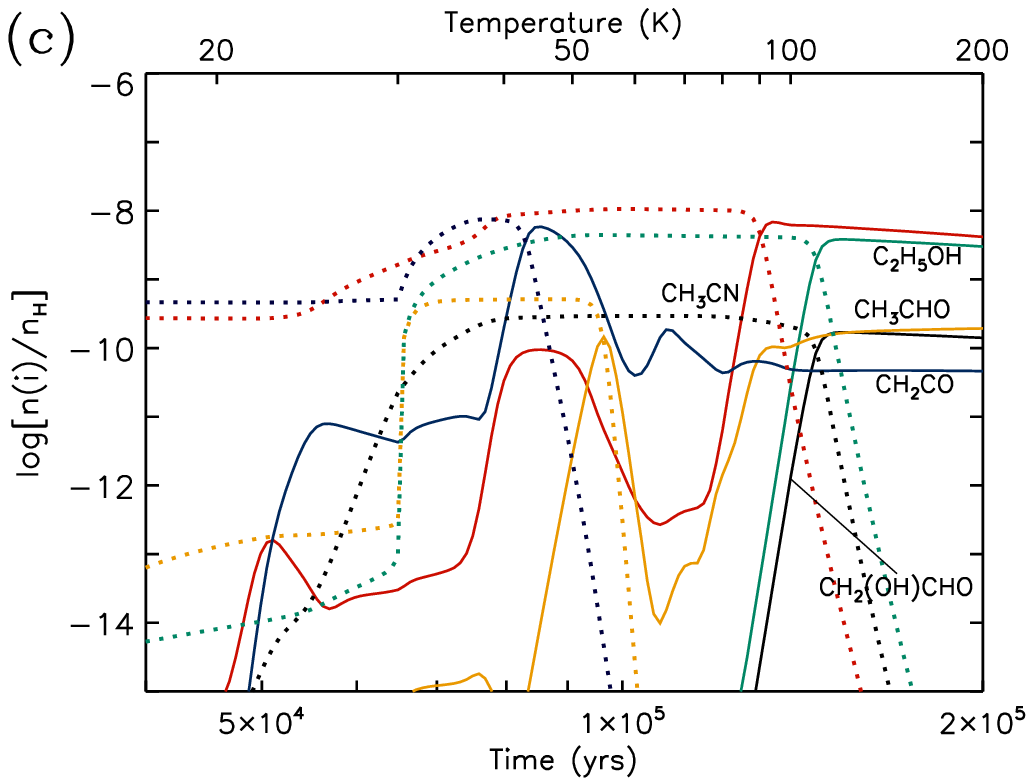} 
\plotone{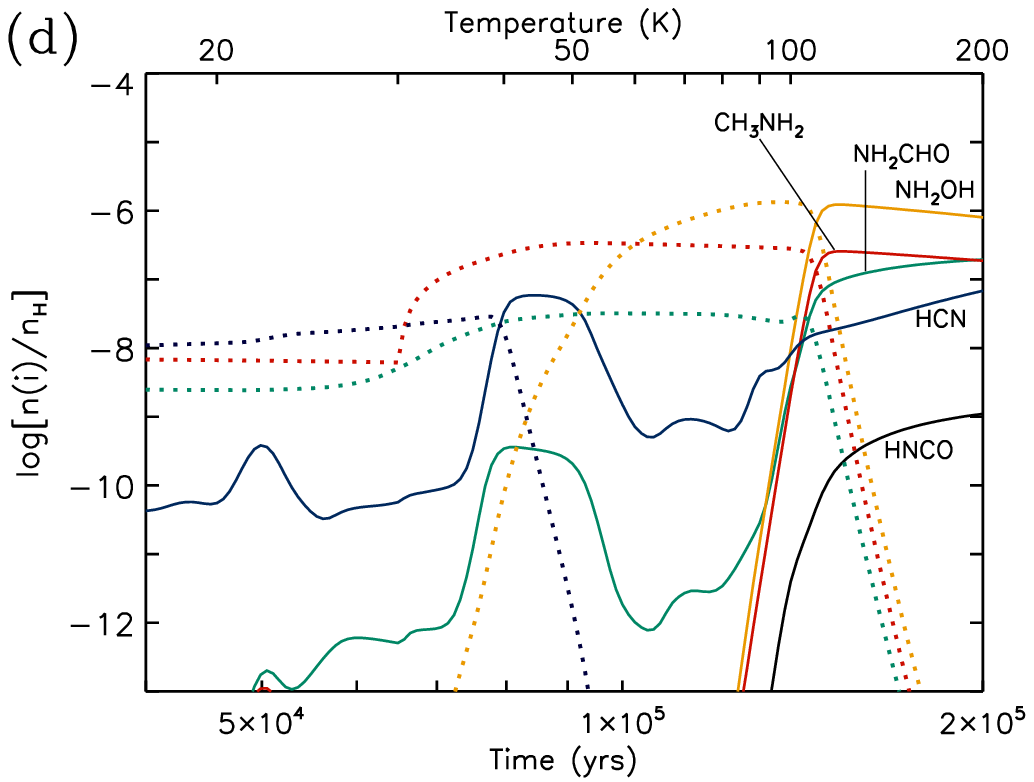}
\plotone{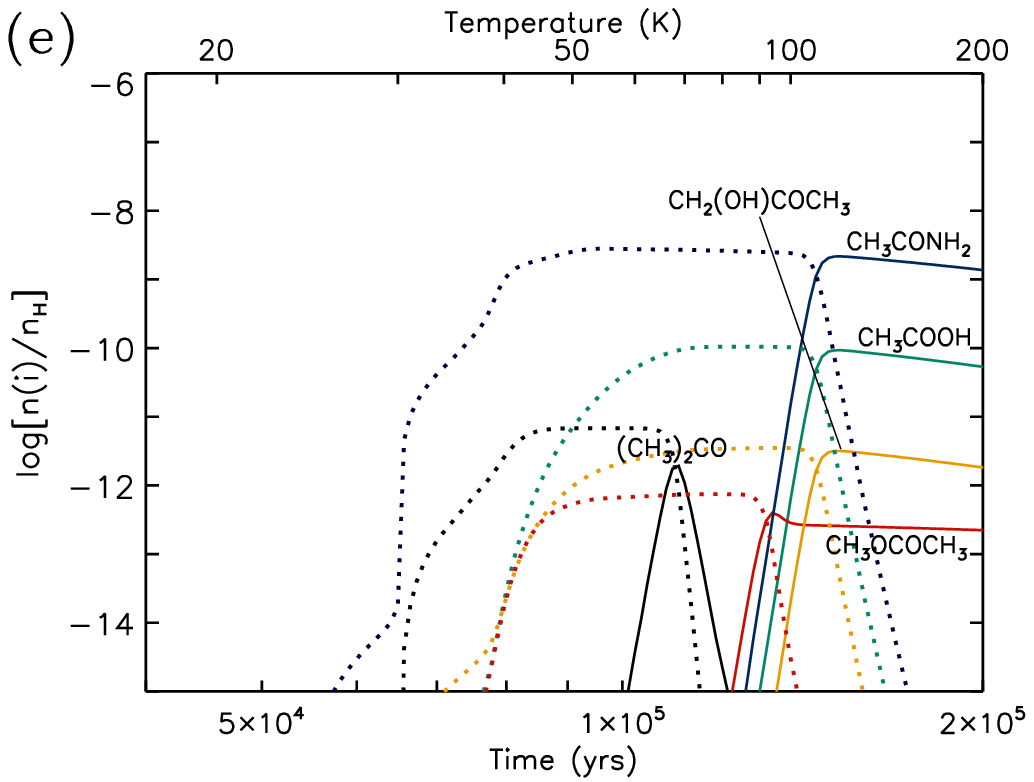} 
\plotone{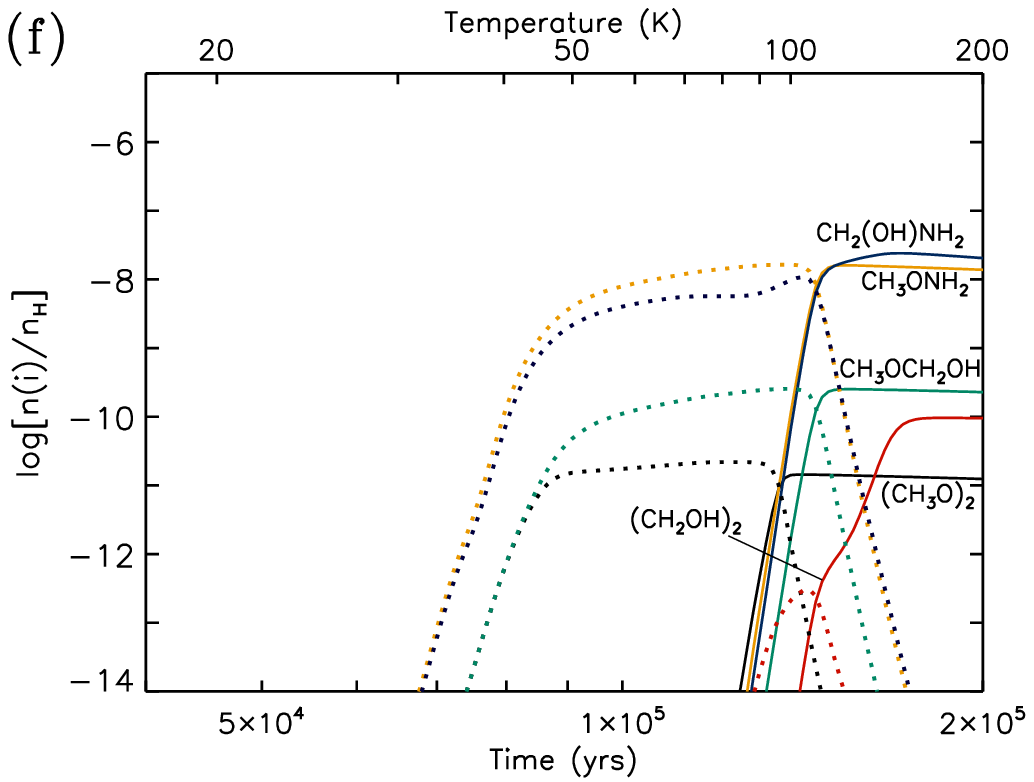} \caption{\label{fig5} Fractional
abundances for model M(ice), with a warm-up timescale of $2 \times 10^5$ yr. Solid lines indicate gas-phase species; dotted lines of the same color indicate the grain-surface species. [{\em See the electronic edition of the Journal for panels g--k.}]}
\end{figure*}
\begin{figure*}
\epsscale{0.4}
\plotone{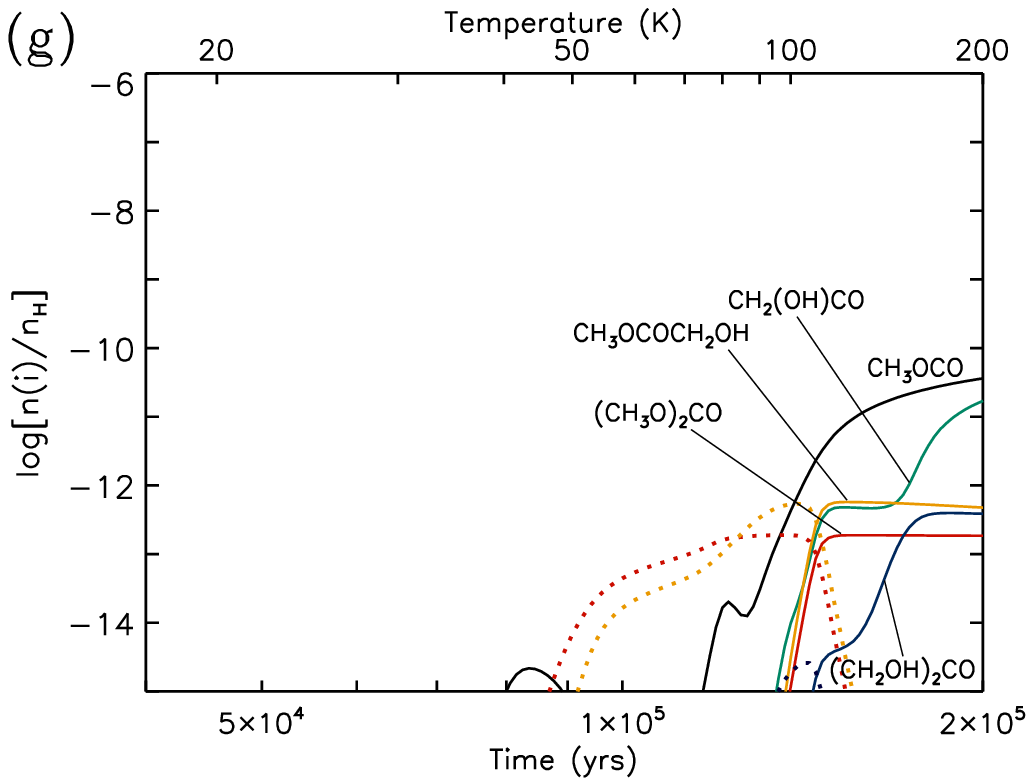}
\plotone{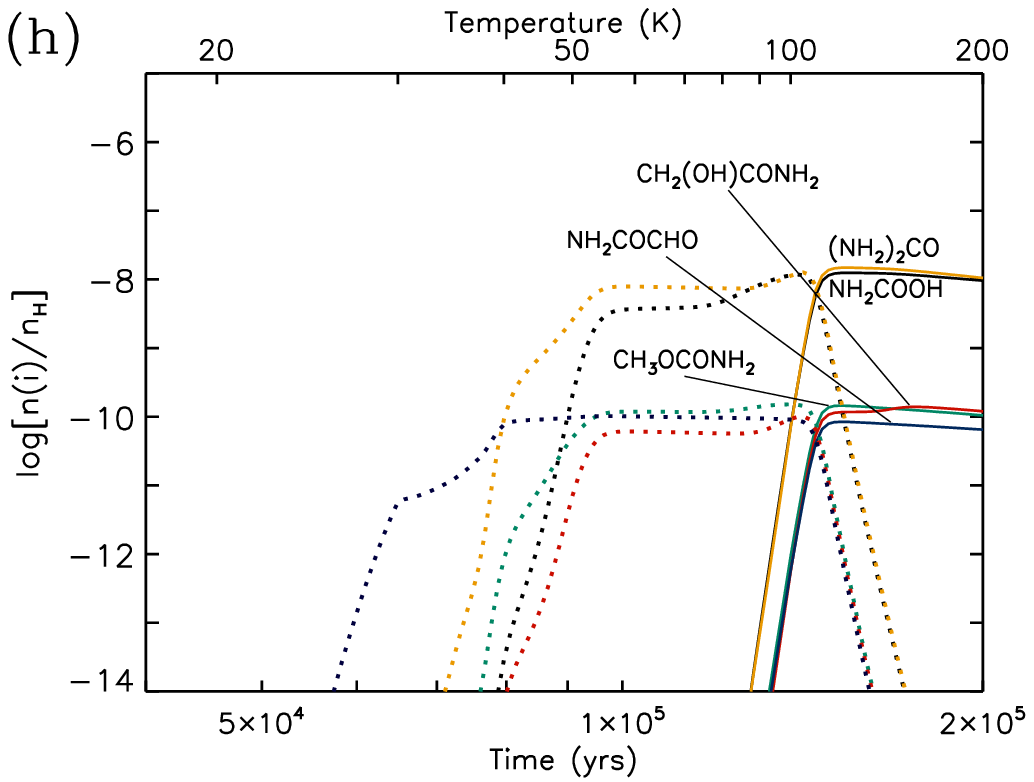}
\plotone{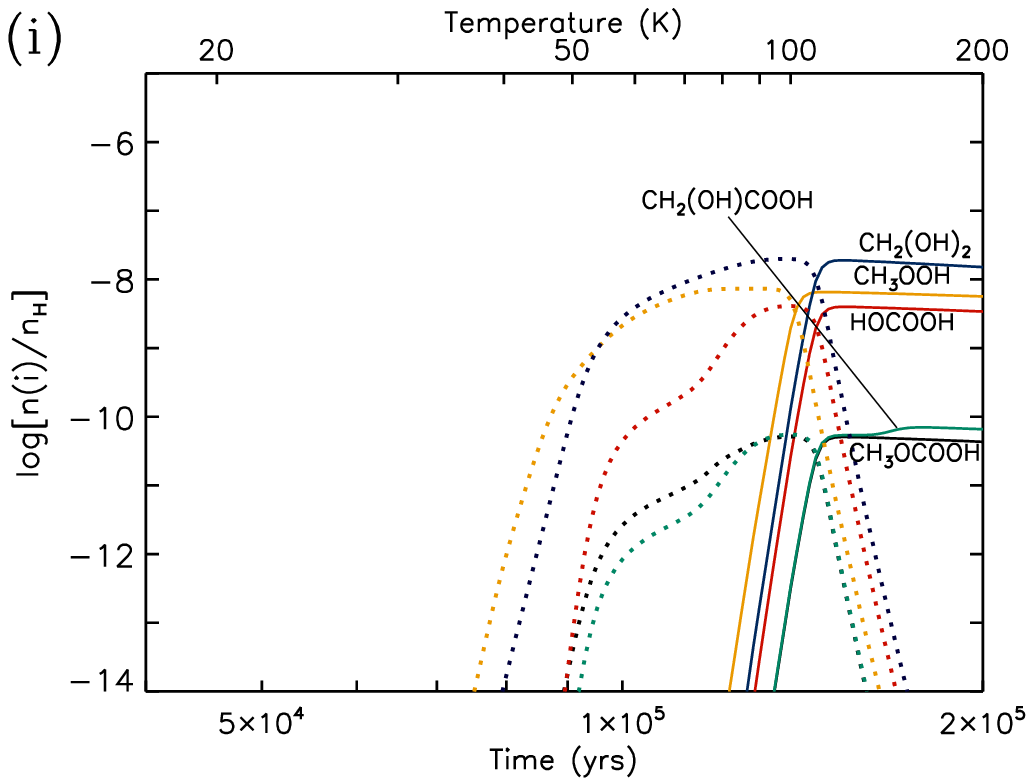}
\plotone{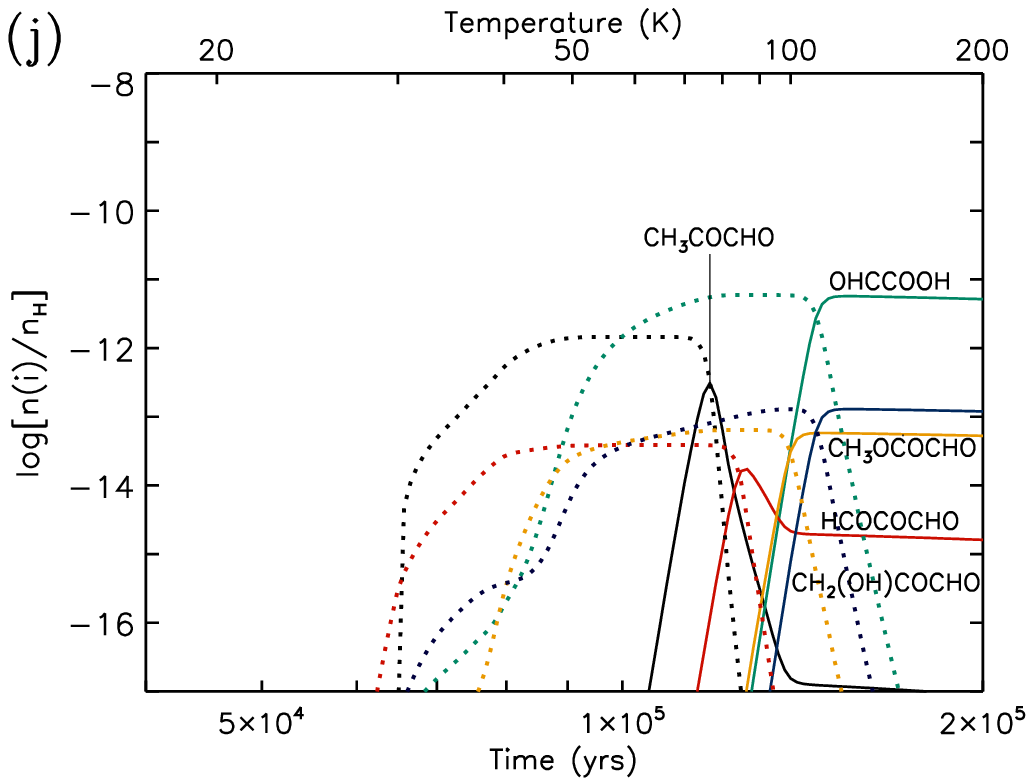}
\plotone{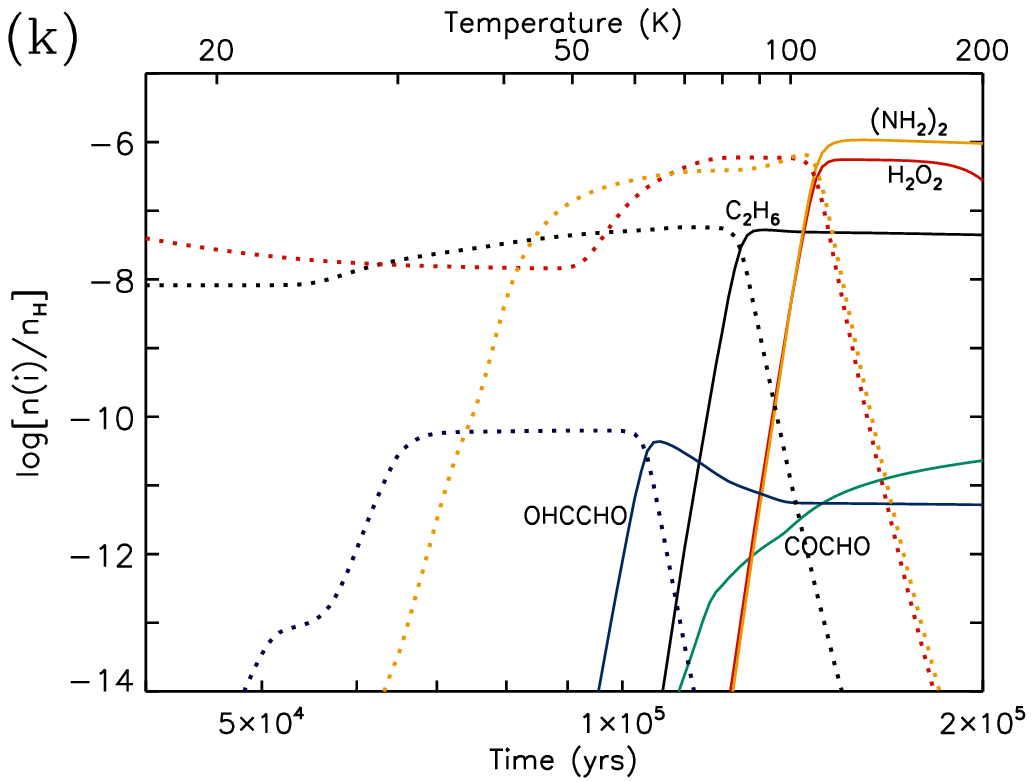}
\end{figure*}

\begin{figure*}
\epsscale{0.4} 
\plotone{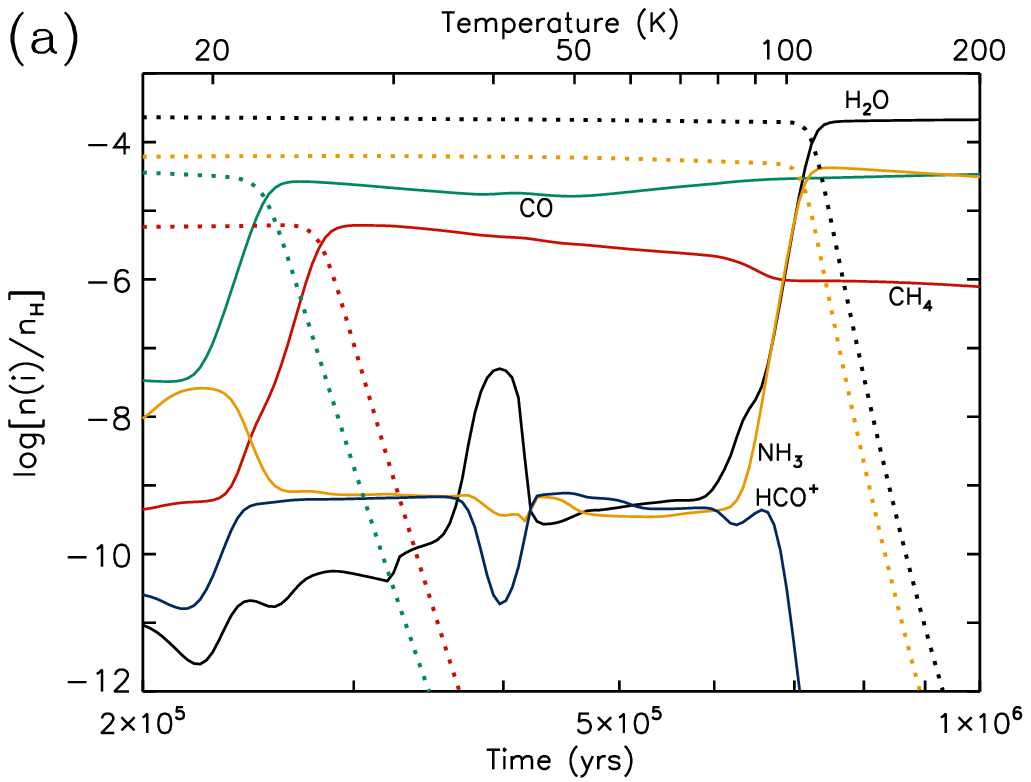} 
\plotone{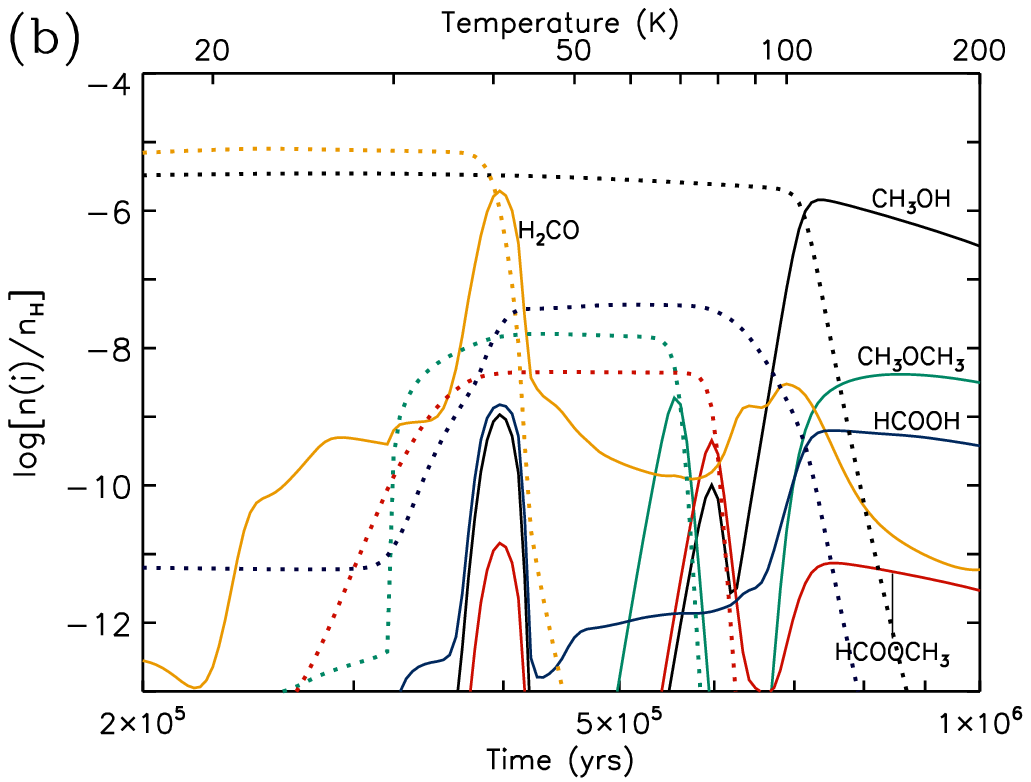}
\plotone{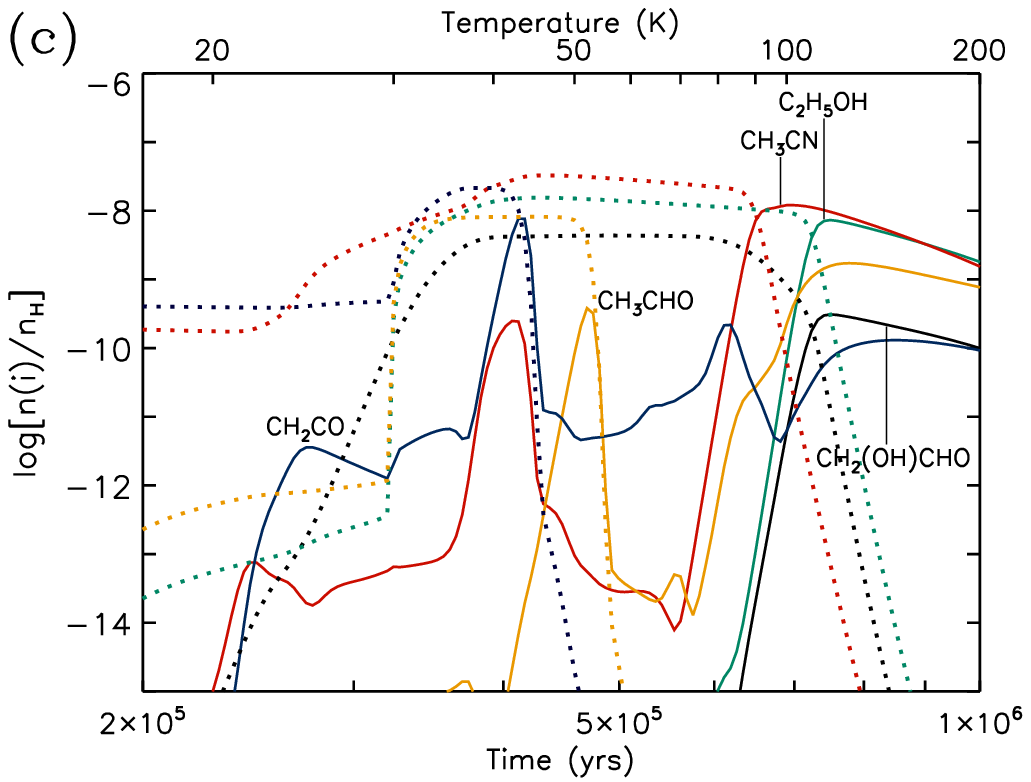} 
\caption{\label{fig6} Fractional
abundances for model S(ice), with a warm-up timescale of $1 \times 10^6$ yr. Solid lines indicate gas-phase species; dotted lines of the same color indicate the grain-surface species. [{\em See the electronic edition of the Journal for panels d--k.}]}
\end{figure*}
\begin{figure*}
\epsscale{0.4}
\plotone{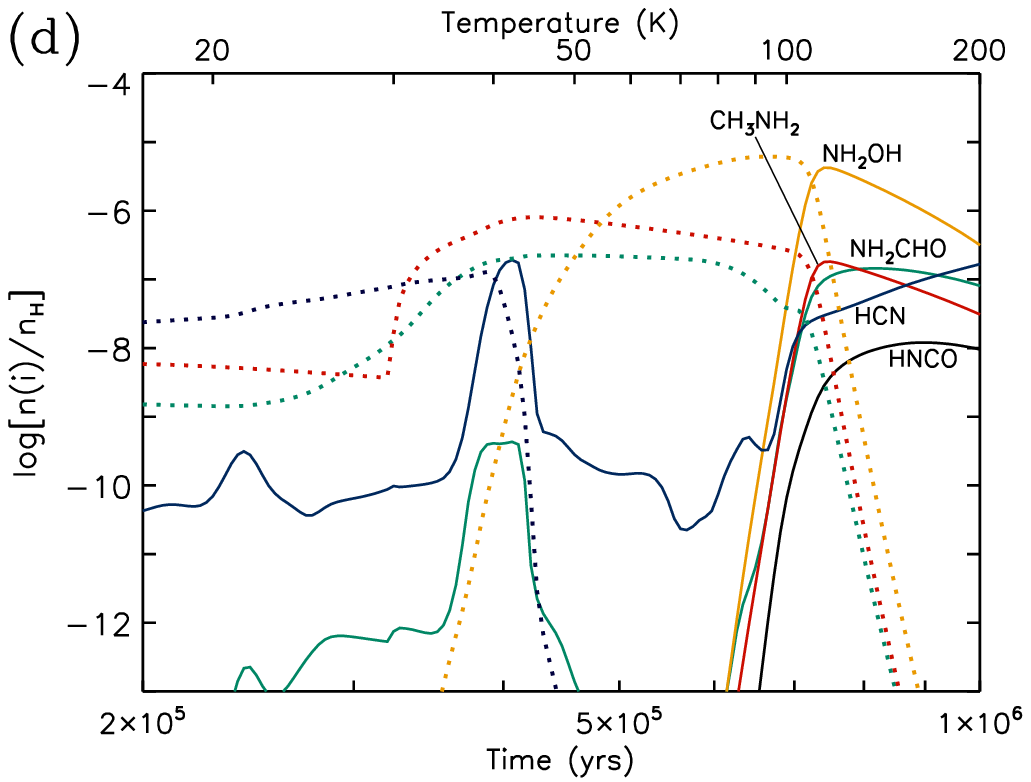}
\plotone{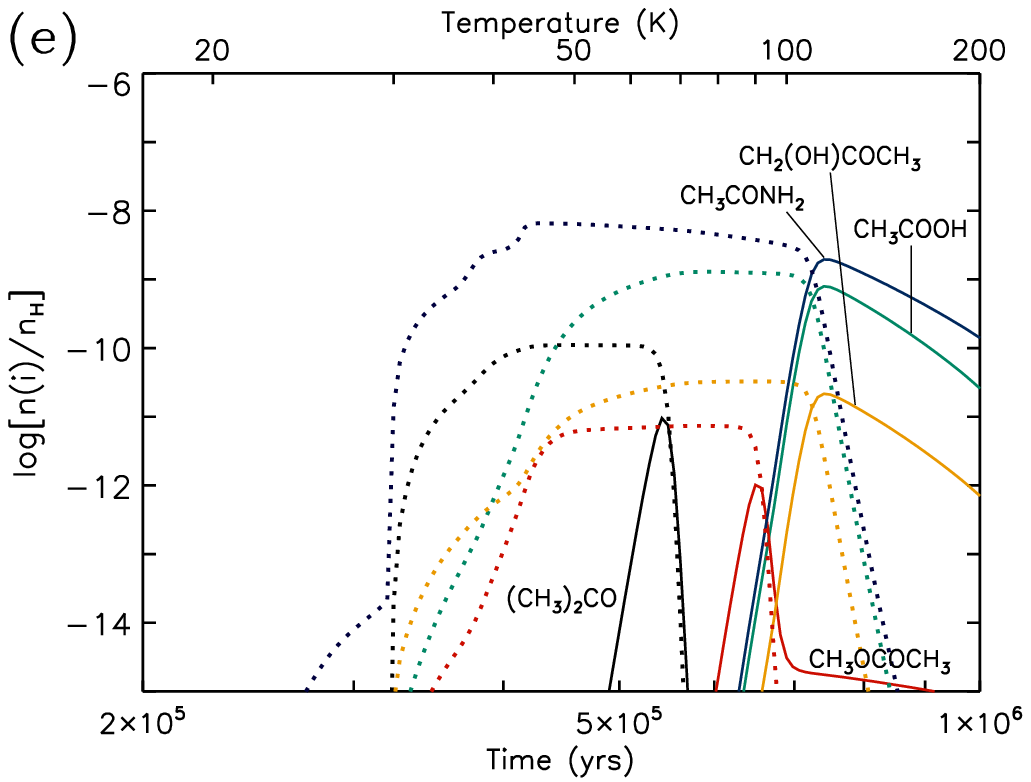} 
\plotone{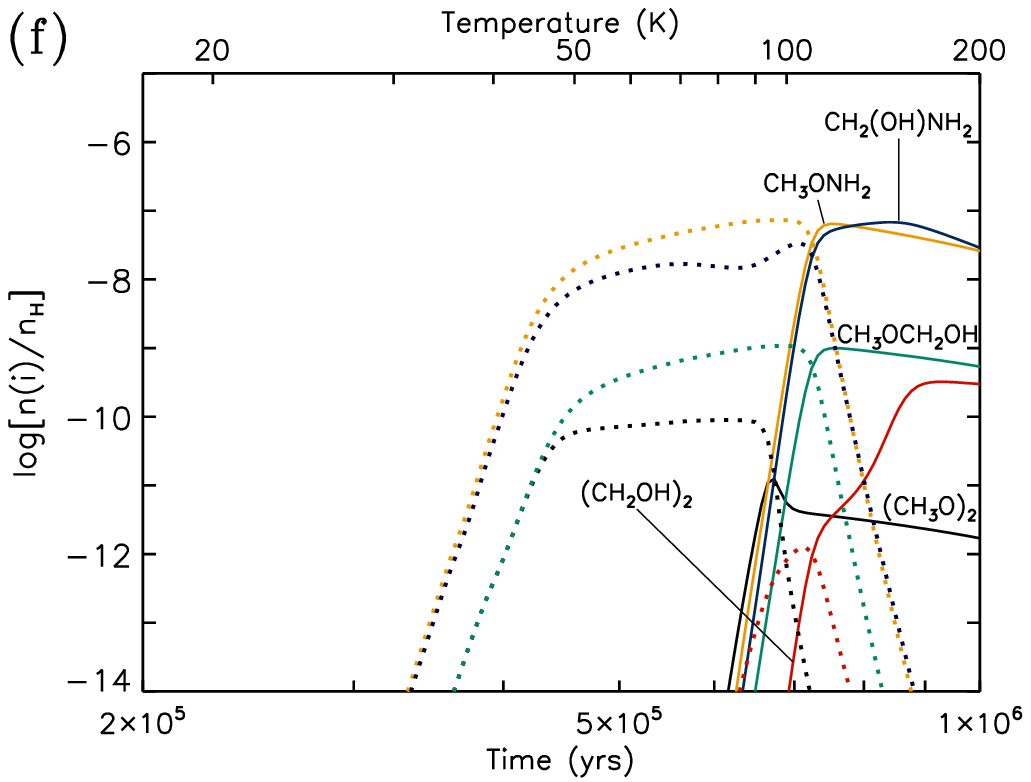} 
\plotone{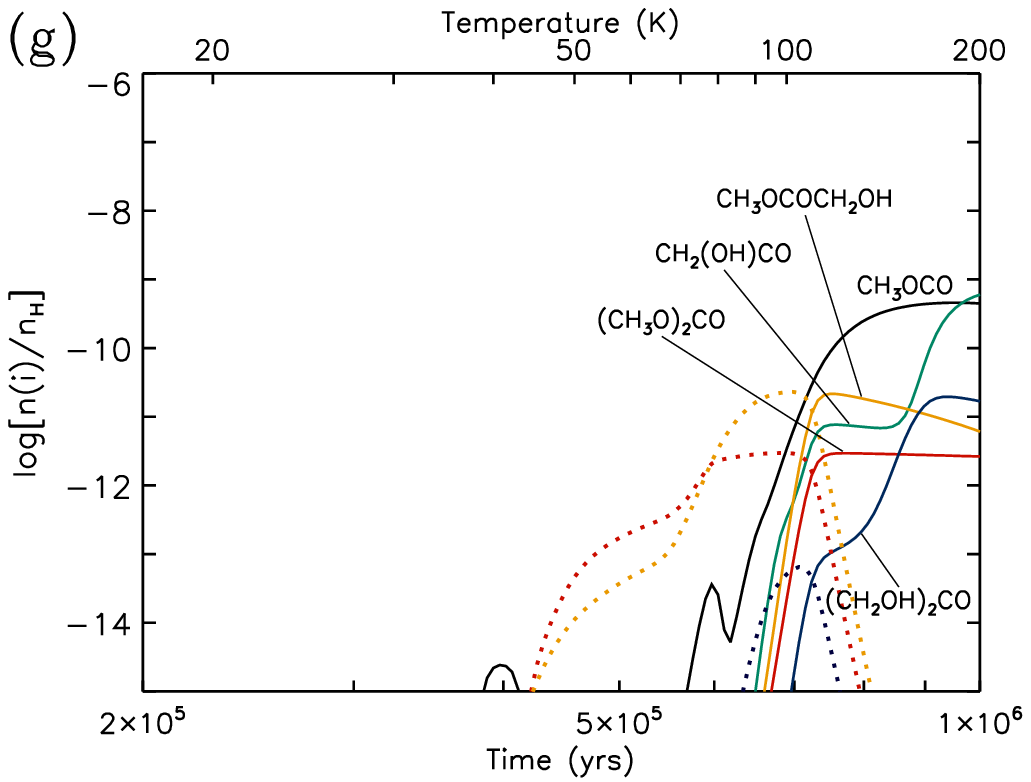}
\plotone{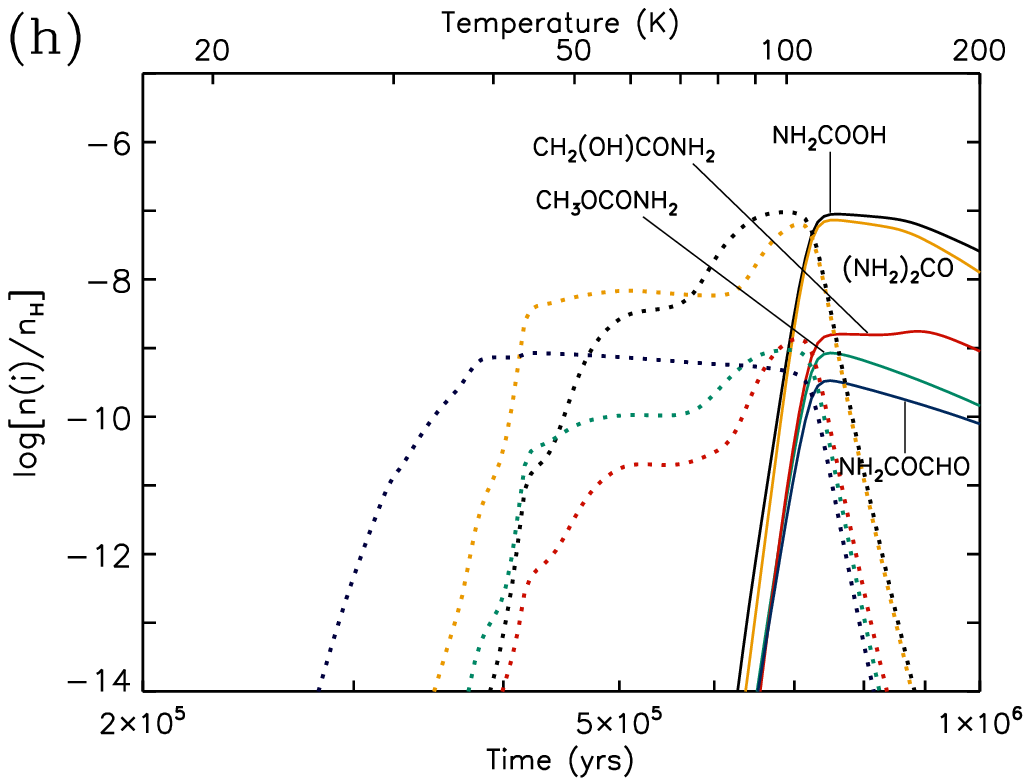}
\plotone{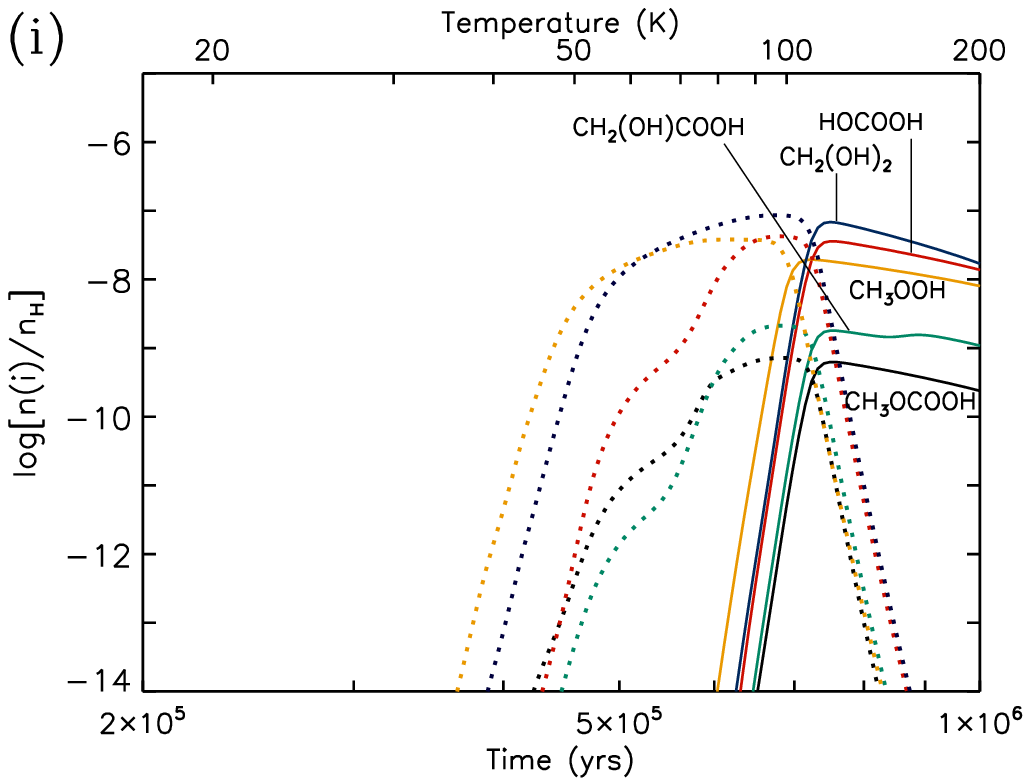}
\plotone{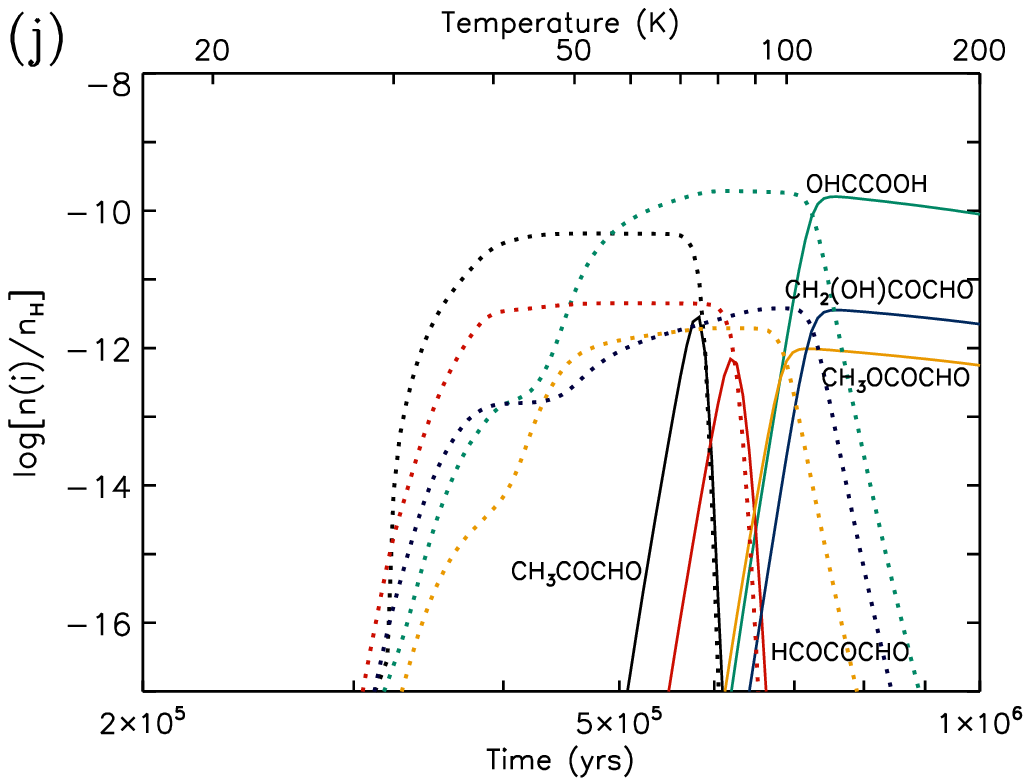}
\plotone{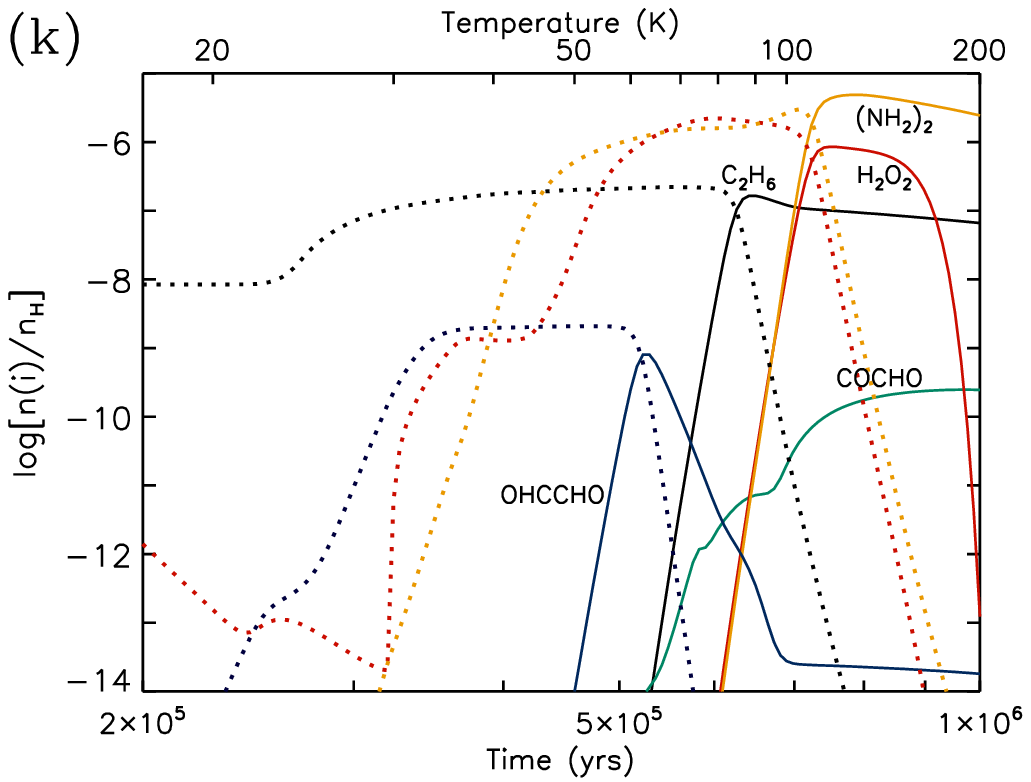}
\end{figure*}

\begin{deluxetable}{lclc}
\tablecaption{\label{tab1} Products and Rates of Cosmic-Ray Dissociation of Molecular Ice Species.}
\tablewidth{0pt}
\tablehead{
\colhead{Species} & & \colhead{Products} & \colhead{Rate multiplier, $b$ \tablenotemark{\dag}}}
\startdata
H$_2$O   & $\rightarrow$ & OH + H & 970  \\
CH$_4$   & $\rightarrow$ & CH$_2$ + H$_2$ & 2340 \\
NH$_3$   & $\rightarrow$ & NH + H$_2$ & 540 \\
NH$_3$   & $\rightarrow$ & NH$_2$ + H & 1320 \\
HCO      & $\rightarrow$ & CO + H & 421 \\
H$_2$CO  & $\rightarrow$ & {\bf HCO + H} & {\bf 1330} \\
H$_2$CO  & $\rightarrow$ & CO + H$_2$ & {\bf 1330} \\
CH$_3$OH & $\rightarrow$ & CH$_3$ + OH & 1500 \\
CH$_3$OH & $\rightarrow$ & {\bf CH$_3$O + H} & {\bf 500} \\
CH$_3$OH & $\rightarrow$ & {\bf CH$_2$OH + H} & {\bf 500} \\
\enddata
\tablenotetext{\dag}{Rate, $k=b \times \zeta$, where $\zeta$ is the cosmic-ray ionization rate.}
\tablecomments{Boldface denotes divergence from Gredel et al. (1989)
products/rates.}
\end{deluxetable}

\begin{deluxetable}{lc}
\tablecaption{\label{tab-init} Initial fractional abundances, with respect to total hydrogen, of elements and H$_2$ at
the start of the collapse phase.} \tablewidth{0pt} \tablehead{ \colhead{Species, $i$} & \colhead{$n_{i}/n_{H}$
\tablenotemark{\dag}}} \startdata
H$_2$ & $0.3\dot{3} $ \\
H & $0.3\dot{3} $ \\
He & $0.09$ \\
C & $1.4(-4)$ \\
N & $7.5(-5)$ \\
O & $3.2(-4)$ \\
S & $8.0(-8)$ \\
Na & $2.0(-8)$ \\
Mg & $7.0(-9)$ \\
Si & $8.0(-9)$ \\
P & $3.0(-9)$ \\
Cl & $4.0(-9)$ \\
Fe & $3.0(-9)$ \\
\enddata
\tablenotetext{\dag}{$a(b)=a^{b}$}
\end{deluxetable}

\clearpage

\begin{deluxetable}{lcc}
\tablecaption{\label{tab-mod} Warm-up phase model parameters}
\tablewidth{0pt}
\tablehead{
\colhead{Model} & \colhead{Warm-up timescale (yr)} & \colhead{Ice composition}}
\startdata
F      & $5 \times 10^{4}$ & Standard \\
M      & $2 \times 10^{5}$ & Standard \\
S      & $1 \times 10^{6}$ & Standard \\
F(ice) & $5 \times 10^{4}$ & Reduced \\
M(ice) & $2 \times 10^{5}$ & Reduced \\
S(ice) & $1 \times 10^{6}$ & Reduced \\
\enddata
\end{deluxetable}

\clearpage

\begin{deluxetable}{lcccccc}
\tablecaption{\label{tabx1} Peak gas-phase abundances and associated
temperatures for standard warm-up phase models F, M, and S\tablenotemark{\dag}.}
\tablewidth{0pt}
\tablehead{
\colhead{Species} & \multicolumn{2}{c}{Model F} & \multicolumn{2}{c}{Model M} & \multicolumn{2}{c}{Model S} \\
\colhead{ } &
\colhead{Peak $n[i]/n_{H}$} & \colhead{T (K)} &
\colhead{Peak $n[i]/n_{H}$} & \colhead{T (K)} &
\colhead{Peak $n[i]/n_{H}$} & \colhead{T (K)}}
\startdata
CO                   & 4.6(-5)  & 200   & 4.2(-5)  & 200   & 3.5(-5)  & 200   \\
H$_2$O               & 2.3(-4)  & 200   & 2.3(-4)  & 200   & 2.3(-4)  & 200   \\
CH$_4$               & 5.0(-5)  & 29    & 5.2(-5)  & 29    & 5.8(-5)  & 28    \\
NH$_3$               & 5.6(-5)  & 124   & 5.3(-5)  & 120   & 4.1(-5)  & 117   \\
HCO$^+$              & 6.4(-10) & 33    & 6.3(10)  & 24    & 6.2(-10) & 24    \\

CH$_3$OH             & 3.2(-5)  & 122   & 2.6(-5)  & 117   & 1.0(-5)  & 112   \\
H$_2$CO              & 9.6(-6)  & 44    & 8.7(-6)  & 42    & 4.9(-6)  & 41    \\
HCOOH                & 4.1(-8)  & 190   & 2.8(-8)  & 120   & 3.0(-9)  & 119   \\
HCOOCH$_3$           & 3.3(-9)  & 88    & 5.0(-9)  & 83    & 2.6(-9)  & 78    \\
CH$_3$OCH$_3$        & 5.6(-8)  & 200   & 1.3(-7)  & 200   & 7.2(-8)  & 163   \\

CH$_2$(OH)CHO        & 2.9(-9)  & 127   & 3.2(-9)  & 123   & 5.6(-10) & 117   \\
C$_2$H$_5$OH         & 3.2(-8)  & 124   & 1.1(-7)  & 120   & 1.6(-7)  & 117   \\
CH$_3$CHO            & 2.5(-9)  & 59    & 2.2(-8)  & 58    & 9.7(-9)  & 200   \\
CH$_2$CO             & 1.7(-9)  & 48    & 5.4(-9)  & 46    & 2.2(-8)  & 44    \\
CH$_3$CN             & 2.4(-9)  & 102   & 1.3(-8)  & 97    & 2.7(-8)  & 96    \\

HCN                  & 5.5(-8)  & 200   & 1.6(-7)  & 200   & 5.2(-7)  & 200   \\
HNCO                 & 4.3(-11) & 200   & 1.8(-9)  & 200   & 4.5(-8)  & 171   \\
NH$_2$CHO            & 1.3(-6)  & 200   & 2.4(-6)  & 200   & 2.6(-6)  & 143   \\
CH$_3$NH$_2$         & 2.4(-7)  & 124   & 7.9(-7)  & 120   & 8.3(-7)  & 117   \\
NH$_2$OH             & 3.3(-7)  & 124   & 1.4(-6)  & 120   & 5.6(-6)  & 117   \\

(CH$_3$)$_2$CO       & 3.5(-13) & 74    & 2.9(-11) & 72    & 2.1(-10) & 66    \\
CH$_3$CONH$_2$       & 4.0(-11) & 124   & 2.7(-10) & 120   & 5.0(-9)  & 117   \\
CH$_3$COOH           & 3.0(-12) & 124   & 4.7(-11) & 120   & 2.4(-9)  & 117   \\
CH$_3$OCOCH$_3$      & 4.6(-14) & 104   & 1.0(-12) & 97    & 3.9(-11) & 94    \\
CH$_2$(OH)COCH$_3$   & 3.0(-13) & 124   & 5.7(-12) & 120   & 3.1(-10) & 117   \\

(CH$_3$O)$_2$        & 4.8(-10) & 111   & 1.4(-9)  & 108   & 1.5(-9)  & 98    \\
(CH$_2$OH)$_2$       & 1.4(-8)  & 159   & 1.5(-8)  & 171   & 1.9(-8)  & 175   \\
CH$_3$OCH$_2$OH      & 4.8(-9)  & 133   & 1.8(-8)  & 125   & 4.0(-8)  & 119   \\
CH$_3$ONH$_2$        & 2.8(-8)  & 127   & 1.2(-7)  & 123   & 4.3(-7)  & 119   \\
CH$_2$(OH)NH$_2$     & 3.3(-8)  & 124   & 1.9(-7)  & 143   & 4.5(-7)  & 143   \\

CH$_3$OCO            & 7.0(-12) & 200   & 4.7(-10) & 200   & 7.3(-9)  & 196   \\
CH$_2$(OH)CO         & 1.8(-11) & 200   & 3.2(-10) & 200   & 7.3(-9)  & 200   \\
(CH$_3$O)$_2$CO      & 1.1(-12) & 130   & 1.2(-11) & 125   & 1.4(-10) & 122   \\
(CH$_2$OH)$_2$CO     & 1.7(-11) & 166   & 9.7(-11) & 175   & 1.6(-9)  & 179   \\
CH$_3$OCOCH$_2$OH    & 8.0(-12) & 130   & 1.2(-10) & 128   & 1.0(-9)  & 119   \\

CH$_3$OCONH$_2$      & 6.9(-11) & 124   & 1.4(-9)  & 120   & 1.2(-8)  & 117   \\
CH$_2$(OH)CONH$_2$   & 4.3(-10) & 145   & 2.4(-9)  & 157   & 2.0(-8)  & 163   \\
(NH$_2$)$_2$CO       & 3.1(-9)  & 124   & 3.5(-8)  & 123   & 3.4(-7)  & 119   \\
NH$_2$COCHO          & 9.7(-11) & 124   & 4.7(-11) & 120   & 2.2(-10) & 117   \\
NH$_2$COOH           & 5.3(-9)  & 124   & 4.4(-8)  & 125   & 1.8(-7)  & 122   \\

CH$_3$OOH            & 2.1(-8)  & 119   & 8.3(-8)  & 115   & 2.2(-7)  & 109   \\
CH$_2$(OH)$_2$       & 5.2(-8)  & 124   & 1.9(-7)  & 123   & 5.1(-7)  & 117   \\
HOCOOH               & 2.6(-9)  & 127   & 3.6(-8)  & 123   & 1.1(-7)  & 119   \\
CH$_3$OCOOH          & 1.5(-10) & 127   & 2.6(-9)  & 123   & 7.9(-9)  & 119   \\
CH$_2$(OH)COOH       & 3.5(-10) & 152   & 2.8(-9)  & 160   & 1.8(-8)  & 163   \\

HCOCOCHO             & 1.5(-13) & 94    & 1.4(-13) & 89    & 7.4(-13) & 85    \\
CH$_3$COCHO          & 1.9(-12) & 82    & 4.2(-12) & 78    & 1.5(-11) & 75    \\
OHCCOOH              & 1.7(-12) & 127   & 4.0(-12) & 123   & 1.3(-10) & 119   \\
CH$_3$OCOCHO         & 1.3(-13) & 116   & 2.5(-13) & 112   & 5.7(-12) & 109   \\
CH$_2$(OH)COCHO      & 2.5(-13) & 133   & 6.4(-13) & 128   & 1.5(-11) & 119   \\

C$_2$H$_6$           & 1.1(-7)  & 96    & 7.7(-7)  & 92    & 1.6(-6)  & 90    \\
OHCCHO               & 5.5(-10) & 71    & 1.1(-9)  & 68    & 2.1(-9)  & 63    \\
H$_2$O$_2$           & 4.1(-7)  & 127   & 9.9(-7)  & 123   & 1.6(-6)  & 117   \\
(NH$_2$)$_2$         & 1.2(-7)  & 127   & 5.5(-7)  & 125   & 3.6(-6)  & 128   \\
COCHO                & 7.9(-12) & 200   & 1.4(-11) & 200   & 1.9(-10) & 200   \\
\enddata
\tablenotetext{\dag}{ $a(b)=a \times 10^{b}$}
\end{deluxetable}

\clearpage

\begin{deluxetable}{lccccc}
\tablecaption{\label{tab-ice} Selection of ice compositions determined or
collated by Gibb et al. 2000, and post-collapse model values \tablenotemark{a}}
\tablewidth{0pt}
\tablehead{
\colhead{Species} &
\colhead{W33A} &
\colhead{NGC 7538 IRS 9 \tablenotemark{b}} &
\colhead{Sgr A* \tablenotemark{b}} &
\colhead{Standard model \tablenotemark{c}} &
\colhead{Reduced values \tablenotemark{c}}
}
\startdata
CO                         &  8   &  16  &  $<$12   &  19       &  19       \\
CO$_2$                     &  13  &  22  &  14      &  4.1(-3)  &  4.1(-3)  \\
CH$_4$                     &  1.5 &  2   &  2       &  22       &  2.2      \\
CH$_3$OH                   &  18  &  5   &  $<$4    &  15       &  1.5      \\
H$_2$CO                    &  6   &  4   &  $<$3    &  4.3      &  0.43     \\
HCOOH                      &  7   &  3   &  3       &  3.2(-6)  &  3.2(-6)  \\
NH$_3$                     &  15  &  13  &  20--30  &  25       &  25       \\

\enddata
\tablenotetext{a} { All values are expressed as a percentage of the H$_2$O value}
\tablenotetext{b} { See Gibb et al. 2000 for original references}
\tablenotetext{c} { X(H$_2$O)=2.3(-4)}
\end{deluxetable}

\begin{deluxetable}{lccccccccccl}
\rotate
\tabletypesize{\scriptsize}
\tablecaption{\label{tabx2} Peak gas-phase abundances and associated temperatures
for reduced ice composition run\tablenotemark{\dag}.}
\tablewidth{0pt}
\tablehead{
\colhead{Species} & \multicolumn{2}{c}{Model F(ice)} & \multicolumn{2}{c}{Model M(ice)} & \multicolumn{2}{c}{Model S(ice)} & \multicolumn{5}{c}{Sgr B2(N)} \\
\cline{8-12}
\colhead{ } &
\colhead{$n[i]/n_{H}$} & \colhead{T (K)} &
\colhead{$n[i]/n_{H}$} & \colhead{T (K)} &
\colhead{$n[i]/n_{H}$} & \colhead{T (K)} &
\colhead{$n[i]/n_{H}$} & \colhead{T$_{rot}$(K)} & \colhead{$\theta$} & \colhead{Ref.} & \colhead{Notes}}
\startdata
CO                   &
{\bf 4.4(-5)}  & {\bf 26} $^\ddag$    &    {\bf 4.1(-5)}  & {\bf 25} $^\ddag$    &    {\bf 3.4(-5)}  & 200 $^\ddag$    &   {\bf 1.6(-5)}  &  {\bf 50}   &  $23''$  &  (1)  &  From C$^{17}$O \\
H$_2$O               &
2.3(-4)  & 186   &    2.3(-4)  & 200   &    2.1(-4)  & 200   &   -   &   -   &  -  &     &   \\
CH$_4$               &
5.0(-6)  & 29 $^\ddag$   &    5.2(-6)  & 29 $^\ddag$   &    6.1(-6)  & 28 $^\ddag$   &   -   &   -   &  -  &     &   \\
NH$_3$               &
5.6(-5)  & 124   &    5.3(-5)  & 120   &    4.2(-5)  & 117   &   -   &  -    &  -  &     &   \\
HCO$^+$              &
{\bf 7.0(-10)} & {\bf 35}    &    {\bf 6.9(-10)} & {\bf 61}   &     {\bf 7.7(-10)} & {\bf 49}    &   {\bf 1.3(-10)}  &  {\bf 50}     &  $23''$  &  (1)  &  From H$^{13}$CO$^+$ \\
CH$_3$OH (early)            &
{\bf 3.3(-10)}  &  {\bf 44}  &    {\bf 7.0(-10)}  &  {\bf 42}  &    {\bf 1.1(-9)}  &  {\bf 40}  &   {\bf 1.7(-9)}    &  {\bf 45}     &  $23''$  &  (1)  &  (Halo)  \\
CH$_3$OH (late)             &
{\bf 3.2(-6)}  & 119   &    {\bf 2.6(-6)}  & 117   &    {\bf 1.4(-6)}  & 112   &   {\bf 8.3(-7)}    &  238    &  $2''$  &  (1)  &  (Core)  \\
H$_2$CO              &
1.1(-6)  & {\bf 42}    &    1.5(-6)  & {\bf 41}    &    1.9(-6)  & {\bf 40}    &   2.5(-10)    &  {\bf 50}     &  $23''$  &  (1)  &   \\
HCOOH (early)               &
9.9(-10)  &  {\bf 47}  &    1.3(-9)  &  {\bf 42}  &    1.5(-9)  & {\bf 40}    &   7.0(-11)    &  {\bf 74$^{+82}_{-30}$}     &  $23''$  &  (1)  &   \\
HCOOH (late)                &
{\bf 4.6(-9)}  & {\bf 124}   &    {\bf 4.7(-9)}  &{\bf  120}   &    {\bf 6.3(-10)}  &  {\bf 117}  &   {\bf 1.5(-9)}    &  {\bf 74$^{+82}_{-30}$}     &  $5''$   &  (1)  & Scaled from $23''$ data \\
HCOOCH$_3$           &
7.8(-11) & 84    &    1.0(-10) & 80    &    4.5(-10) & 78    &   1.9(-8)    &       &   $14'' \times 4''$ &  (2,3)  & (BIMA data) \\
CH$_3$OCH$_3$        &
{\bf 3.0(-9)}  & {\bf 200}   &    {\bf 6.3(-9)}  & {\bf 200}   &    {\bf 4.2(-9)}  & {\bf 149}   &   {\bf 1.3(-9)}    &   {\bf 197$^{+31}_{-22}$}    &  23'' &  (1)  &   \\
CH$_2$(OH)CHO        &
{\bf 8.8(-11)} & 124   &    {\bf 1.7(-10)} & 120   &    {\bf 3.1(-10)} & 117   &   {\bf 4.7(-11)}   &   50    &  $>1'$   &  (4)  &   \\
&  &    &     &    &     &                                   &   5.8(-11)   &   8     &  $55''$   &  (4)  &   \\
C$_2$H$_5$OH         &
{\bf 1.5(-9)}  & 124   &    {\bf 3.8(-9)}  & 120   &    {\bf 7.3(-9)}  & 117   &   7.0(-10)   &   73$^{+5}_{-4}$    &  23''  &  (1)  &   \\
&                               &    &      &    &      &    &   {\bf 1.5(-8)}    &   73$^{+5}_{-4}$    &  5''   &  (1)  &  Scaled from 23'' data \\
CH$_3$CHO (early)           &
2.8(-12) & {\bf 57}    &    {\bf 1.5(-10)} & {\bf 54}    &    {\bf 3.8(-10)} & {\bf 52}    &   {\bf 7.3(-11)}   &   {\bf 59$^{+22}_{-13}$}    &  23''  &  (1)  &  (a-type)  \\
CH$_3$CHO (late)            &
2.4(-11) & 200   &    1.9(-10) & 200   &    1.7(-9)  & 125   &   5.3(-9)    &   520$^{+1040}_{-220}$    &  23''  &  (1)  &  (b-type)  \\
& &    &    &   &    &                                       &   1.1(-7)    &   520$^{+1040}_{-220}$    &  5''  &  (1)  &  (b-type) Scaled from 23'' data  \\
CH$_2$CO             &
2.0(-9)  & 46    &    5.8(-9)  & 45    &    7.6(-9)  & 42    &   1.2(-10)   &   120$^{+68}_{-34}$    &  23''  &  (1)  &   \\
CH$_3$CN             &
2.0(-9)  & 96    &    {\bf 6.8(-9)}  & 94    &    {\bf 1.2(-8)}  & 103   &   1.1(-7)    &   400$^{+104}_{-86}$    &  2.7''  &  (1)  &  (Filling factor assumed) \\
             &                &     &      &     &      &    &   {\bf 3.1(-8)}    &   400$^{+104}_{-86}$    &  5''    &  (1)  &  Scaled from 2.7'' data \\
HCN (early)                 &
2.3(-8)  & {\bf 51}    &   5.9(-8)   &  {\bf 44}  &    1.9(-7)  & {\bf 41}    &   4.2(-10)    &   {\bf 50}    &  23''  &  (1)  &  From HC$^{15}$N \\
HCN (late)                 &
{\bf 2.1(-8)}  &   200  &    {\bf 6.8(-8)}  & 200   &    1.7(-7)  &  200   &   {\bf 9.0(-9)}   &  50  & 5''  & (1)  &  Scaled from 23'' data  \\
HNCO                 &
3.4(-11) & {\bf 200}   &    {\bf 1.1(-9)}  & {\bf 200}   &    {\bf 1.2(-8)}  & {\bf 163}   &   2.8(-10)   &   212$^{+46}_{-34}$    &  23''  &  (1)  &   \\
&                                &     &    &    &      &    &   {\bf 6.0(-9)}    &   {\bf 212$^{+46}_{-34}$}    &  5''   &  (1)  &  Scaled from 23'' data \\
NH$_2$CHO (early)           &
{\bf 3.0(-10)} & 45    &    {\bf 3.6(-10)} & 41    &    {\bf 4.3(-10)} & 41    &  {\bf 9.3(-11)} & 175$^{+53}_{-38}$  &  23''  &  (1)  &  (a-type)  \\
&                       &    &      &    &      &            &  2.0(-9)  & 175$^{+53}_{-38}$  &  5''   &  (1)  &  (a-type) Scaled from 23'' data  \\
NH$_2$CHO (late)            &
{\bf 1.0(-7)}  & {\bf 200}   &    {\bf 1.9(-7)}  & {\bf 200}   &    {\bf 1.4(-7)}  & 136   &  1.6(-9)  & 302$^{+126}_{-75}$ &  23''  &  (1)  &  (b-type)  \\
&                                 &    &     &   &      &    &  {\bf 3.3(-8)}  & {\bf 302$^{+126}_{-75}$} &  5''   &  (1)  &  (b-type) Scaled from 23'' data  \\
CH$_3$NH$_2$         &
1.1(-7)  & 124   &    {\bf 2.6(-7)}  & 120   &    1.8(-7)  & 117   &   {\bf 2.5(-6)}    &   230$^{+74}_{-46}$    &  1.1''  &  (1)  &  (Filling factor assumed) \\
NH$_2$OH             &
3.5(-7)  & 122   &    1.2(-6)  & 120   &    4.2(-6)  & 114   &   -    &   -    &  -  &    &   \\
(CH$_3$O)$_2$        &
6.3(-12) & 109   &    1.4(-11) & 105   &    1.2(-11) & 96    &   -    &   -    &  -  &    &   \\
(CH$_2$OH)$_2$       &
3.6(-11) & {\bf 166}   &    9.6(-11) & {\bf 171}   &    3.2(-10) & {\bf 175}   &    3.3(-10)   &   {\bf 200}    &  $84''$  &  (5)  &   \\
&                                 &    &     &    &     &    &    9.3(-8)    &   200    &  $5''$   &  (5)  & Scaled from $84''$ data \\
&                                 &    &     &    &     &    &    5.3(-11)   &   50     &  $84''$  &  (5)  &   \\
CH$_3$OCH$_2$OH      &
6.0(-11) & 127   &    2.5(-10) & 123   &    1.0(-9)  & 119   &   -    &   -    &  -  &    &   \\
CH$_3$ONH$_2$        &
3.7(-9)  & 124   &    1.6(-8)  & 123   &    6.4(-8)  & 117   &   -    &   -    &  -  &    &   \\
CH$_2$(OH)NH$_2$     &
6.4(-9)  & 148   &    2.4(-8)  & 147   &    6.8(-8)  & 146   &   -    &   -    &  -  &    &   \\
(CH$_3$)$_2$CO       &
1.7(-13) & 71    &    1.9(-12) & 69    &    9.5(-12) & 66    &     4.9(-9)   &   170    &  $12''.5 \times 5''.4$  &  (2)  & (BIMA data) \\
CH$_3$CONH$_2$       &
4.3(-10) & 122   &    2.2(-9)  & 120   &    2.0(-9)  & 114   &   3.0(-11)    &   8    &  Various  &  (6)  &   \\
CH$_3$COOH           &
4.9(-12) & 122   &    {\bf 9.3(-11)} & 120   &    {\bf 7.9(-10)} & 114   &    {\bf 1.0(-9)}    &       &  $11'' \times 4''$  &  (2,7)  & (BIMA data) \\
CH$_3$OCOCH$_3$      &
3.4(-14) & 98    &    4.0(-13) & 94    &    1.0(-12) & 90    &    -   &    -   &  -  &    &   \\
CH$_2$(OH)COCH$_3$   &
1.4(-13) & 122   &    3.2(-12) & 120   &    2.2(-11) & 114   &    -   &    -   &  -  &    &   \\
CH$_3$OCO            &
9.0(-13) & 200   &    3.6(-11) & 200   &    4.6(-10) & 183   &   -    &    -   &  -  &    &   \\
CH$_2$(OH)CO         &
5.0(-13) & 200   &    1.7(-11) & 200   &    6.0(-10) & 200   &   -    &    -   &  -  &    &   \\
(CH$_3$O)$_2$CO      &
6.5(-15) & 130   &    1.9(-13) & 125   &    2.9(-12) & 122   &   -    &   -    &  -  &    &   \\
(CH$_2$OH)$_2$CO     &
{\bf 1.3(-14)} & 170   &    {\bf 4.0(-13)} & 175   &    2.0(-11) & 179   &   {\bf $<$8.3(-12)}    &       &  $33''$  &  (8)  &   \\
CH$_3$OCOCH$_2$OH    &
2.6(-14) & 127   &    5.7(-13) & 123   &    2.2(-11) & 117   &   -    &   -    &  -  &    &   \\
CH$_3$OCONH$_2$      &
1.2(-11) & 124   &    1.4(-10) & 120   &    8.5(-10) & 117   &    -   &    -   &  -  &    &   \\
CH$_2$(OH)CONH$_2$   &
1.2(-11) & 152   &    1.4(-10) & 157   &    1.8(-9)  & 160   &    -   &    -   &  -  &    &   \\
(NH$_2$)$_2$CO       &
1.5(-9)  & 124   &    1.5(-8)  & 123   &    7.3(-8)  & 119   &    -   &    -   &  -  &    &   \\
NH$_2$COCHO          &
2.4(-11) & 124   &    8.4(-11) & 120   &    3.4(-10) & 117   &    -   &    -   &  -  &    &   \\
NH$_2$COOH           &
1.5(-9)  & 127   &    1.2(-8)  & 123   &    8.9(-8)  & 119   &    -   &    -   &  -  &    &   \\
CH$_3$OOH            &
2.1(-9)  & 116   &    6.5(-9)  & 115   &    1.9(-8)  & 109   &   -    &   -    &  -  &    &   \\
CH$_2$(OH)$_2$       &
5.5(-9)  & 124   &    1.9(-8)  & 120   &    6.8(-8)  & 117   &   -    &   -    &  -  &    &   \\
HOCOOH               &
6.5(-10) & 124   &    4.0(-9)  & 123   &    3.6(-8)  & 117   &   -    &   -    &  -  &    &   \\
CH$_3$OCOOH          &
3.8(-12) & 124   &    5.1(-11) & 123   &    6.3(-10) & 117   &   -    &   -    &  -  &    &   \\
CH$_2$(OH)COOH       &
6.9(-12) & 155   &    7.0(-11) & 160   &    1.8(-9)  & 117   &   -    &   -    &  -  &    &   \\
HCOCOCHO             &
5.8(-15) & 90    &    1.7(-14) & 87    &    6.9(-13) & 83    &   -    &   -    &  -  &    &   \\
CH$_3$COCHO          &
1.1(-13) & 79    &    3.1(-13) & 76    &    2.8(-12) & 75    &   -    &   -    &  -  &    &   \\
OHCCOOH              &
4.2(-13) & 127   &    5.7(-12) & 123   &    1.6(-10) & 119   &   -    &   -    &  -  &    &   \\
CH$_3$OCOCHO         &
4.2(-15) & 116   &    5.8(-14) & 112   &    9.8(-13) & 107   &   -    &   -    &  -  &    &   \\
CH$_2$(OH)COCHO      &
7.0(-15) & 130   &    1.3(-13) & 123   &    3.6(-12) & 119   &   -    &   -    &  -  &    &   \\
C$_2$H$_6$           &
1.4(-8)  & 94    &    5.2(-8)  & 92    &    1.6(-7)  & 88    &   -    &   -    &  -  &    &   \\
OHCCHO               &
1.9(-11) & 67    &    4.4(-11) & 65    &    8.0(-10) & 62    &   -    &   -    &  -  &    &   \\
H$_2$O$_2$           &
4.2(-7)  & 127   &    5.6(-7)  & 123   &    8.5(-7)  & 117   &   -    &   -    &  -  &    &   \\
(NH$_2$)$_2$         &
2.2(-7)  & 130   &    1.1(-6)  & 128   &    4.9(-6)  & 128   &   -    &   -    &  -  &    &   \\
COCHO                &
1.9(-12) & 200   &    2.3(-11) & 200   &    2.5(-10) & 191   &   -    &   -    &  -  &    &   \\
\enddata
\tablenotetext{\dag}{ $a(b)=a \times 10^{b}$}
\tablenotetext{\ddag}{ Modeled abundance values for CO and CH$_4$ are largely constant over a wide range of temperatures.}
\tablecomments{Bracketed comments in the `notes' column derive from the original reference. Iostopic ratios C$^{12}$/C$^{13}$=70, O$^{16}$/O$^{17}$=2044, and N$^{14}$/N$^{15}$=100 are assumed.}
\tablerefs{
(1) Nummelin et al. 2000; (2) Snyder et al. 2002 (3) Liu, Mehringer \& Snyder 2001;
(4) Hollis et al. 2004; (5) Hollis et al. 2002; (6) Hollis et al. 2006
(7) Remijan et al. 2002 (8) Apponi et al. 2006
}
\end{deluxetable}

\end{document}